\documentclass[12pt]{article}
\usepackage{amsthm,amsmath,amssymb,bbm,bm}
\usepackage{multirow}
\usepackage{xr-hyper}
\usepackage[pdftex]{graphicx}
\usepackage{subfigure}
\usepackage{wrapfig}
\usepackage{CJK}
\usepackage{array}
\usepackage{url}
\usepackage{booktabs}
\usepackage{algorithm}
\usepackage{algorithmic}
\usepackage{placeins}
\usepackage{mathtools}
\usepackage{romannum}
\usepackage{mathrsfs}
\usepackage{dsfont}
\usepackage{natbib}
\usepackage{relsize}
\usepackage{rotating}
\usepackage{enumitem}
\usepackage{setspace}
\usepackage{minitoc}
\usepackage{etoc}

\usepackage[colorlinks]{hyperref}
\hypersetup{
    colorlinks,
    linkcolor={red},
    filecolor={red}, 
    urlcolor={blue},
    citecolor={blue}
}

\makeatletter
\newcommand*{\addFileDependency}[1]{
  \typeout{(#1)}
  \@addtofilelist{#1}
  \IfFileExists{#1}{}{\typeout{No file #1.}}
}
\makeatother

\newcommand{\MC}[3]{\multicolumn{#1}{#2}{#3}}

\usepackage{etoolbox}
\newcommand{\zerodisplayskips}{%
  \setlength{\abovedisplayskip}{6.5pt}%
  \setlength{\belowdisplayskip}{6.5pt}%
  \setlength{\abovedisplayshortskip}{6.5pt}%
  \setlength{\belowdisplayshortskip}{6.5pt}}
\appto{\normalsize}{\zerodisplayskips}
\appto{\small}{\zerodisplayskips}
\appto{\footnotesize}{\zerodisplayskips}

\newcommand\numberthis{\addtocounter{equation}{1}\tag{\theequation}}

\usepackage[usenames,dvipsnames,svgnames,table]{xcolor}

\makeatletter
\newcommand{\vast}{\bBigg@{3}}
\makeatother

\usepackage{chngcntr}
\counterwithin{equation}{section}
\definecolor{ao(english)}{rgb}{0.0, 0.5, 0.0}

\theoremstyle{plain}
\newtheorem{definition}{Definition}[section]

\newtheorem{thm}{Theorem}

\counterwithin{thm}{section}
\newtheorem{proposition}{Proposition}[thm]
\counterwithin{proposition}{section}

\counterwithin{corollary}{section}

\counterwithin{theorem}{section}

\def\##1\#{\begin{align}#1\end{align}}
\def\$#1\${\begin{align*}#1\end{align*}}

\newtheorem{assump}{Assumption}

\def\beq#1\eeq{\begin{equation}#1\end{equation}}
\def\baa#1\eaa{\begin{eqnarray}#1\end{eqnarray}}
\def\bal#1\eal{\begin{align}#1\end{align}}

\DeclareMathOperator*{\argmin}{arg\,min}
\DeclareMathOperator*{\argmax}{arg\,max}

\def\T{\text{T}}

\newcommand{\blind}{1}

\usepackage[margin=0.8in]{geometry}
\begin{document}

\if1\blind
{ \title{\bf Estimating Optimal Infinite Horizon Dynamic Treatment Regimes via pT-Learning}

\author{
Wenzhuo Zhou\thanks{Department of Statistics, University of California Irvine, Irvine, CA; email: \href{mailto:wenzhuz3@uci.edu}{wenzhuz3@uci.edu};}, \,
Ruoqing Zhu\thanks{Department of Statistics, University of Illinois Urbana-Champaign, Champaign, IL; email: \href{mailto:rqzhu@illinois.edu}{rqzhu@illinois.edu};}, \,
and Annie Qu\thanks{Department of Statistics, University of California Irvine, Irvine, CA; email: \href{mailto:aqu2@uci.edu}{aqu2@uci.edu}.}
}

} \fi


\if1\blind
{
\title{\bf
Estimating Optimal Infinite Horizon Dynamic Treatment Regimes via pT-Learning
} \fi

\newpage

\date{}

\maketitle


\setcounter{page}{1}
\pagenumbering{arabic}

\vspace{-4mm}

\begin{abstract}

Recent advances in mobile health (mHealth) technology provide an effective way to monitor individuals' health statuses and deliver  just-in-time personalized interventions. However, the practical use of mHealth technology raises unique challenges to existing methodologies on learning an optimal dynamic treatment regime. Many mHealth applications involve decision-making with large numbers of intervention options and under an infinite time horizon setting where the number of decision stages diverges to infinity. In addition, temporary medication shortages may cause optimal treatments to be unavailable, while it is unclear what alternatives can be used. To address these challenges, we propose a \textbf{P}roximal \textbf{T}emporal consistency \textbf{Learning} (pT-Learning) framework to estimate an optimal regime that is adaptively adjusted between deterministic and stochastic sparse policy models. The resulting minimax estimator avoids the \textit{double sampling} issue in the existing algorithms. It can be further simplified and can easily incorporate off-policy data without mismatched distribution corrections. We study theoretical properties of the sparse policy and establish  finite-sample bounds on the excess risk and performance error. The proposed method is provided in our \texttt{proximalDTR} package and is evaluated through extensive simulation studies and the OhioT1DM mHealth dataset.

\end{abstract}


\section{Introduction}\label{introduction}

Mobile health (mHealth) technology has recently attracted much attention due to mobile devices such as smartphones or wearable devices for tracking physical activities and well-being. It makes real-time communications feasible between health providers and individuals \citep{sim2019mobile}. In addition, the mHealth technology can also be used to collect rich longitudinal data for exploring optimal dynamic treatment regimes, which are critical in delivering long-term personalized interventions \citep{nahum2018just}. For example, the OhioT1DM mHealth study \citep{marling2020ohiot1dm} collects eight weeks' worth of mHealth data for type 1 diabetes patients. All patients are equipped with mobile sensor bands for continuously measuring blood glucose level, insulin dose level, heart rate, carbohydrate intake, etc. This allows us to develop tailored dynamic treatment strategies to manage patients' blood glucose levels. However, current applications of mHealth technology in clinical use encounter some unique challenges. First, using mHealth technology involves data collection and requires decision-making over a very long period. This is often referred to as the infinite horizon setting. Secondly, most of the mHealth applications aim to provide multi-channel interventions with a large number of treatment combinations \citep{yang2019comparative}, or recommend continuous individualized dose levels \citep{marling2020ohiot1dm} for maximizing patients' clinical outcomes. Thirdly, there is an increasing demand for implementing robust dynamic treatment regimes to meet unexpected situations such as temporary shortage of medications or budget constraints in mHealth studies \citep{rehg2017mobile}. It is essential to design an optimal regime that can provide an alternative optimal or near-optimal treatment option as a backup choice. However, existing statistical methodologies are not well-developed for meeting the challenges mentioned above.


Although there is a rich body of literature on estimating dynamic treatment regimes \citep{murphy2003optimal, zhao2015new,  shi2018high} over a fixed period (finite horizon), only a limited number of statistical methodologies have been developed for the infinite horizon setting. \cite{ertefaie2018constructing} proposed a variant of Greedy GQ-learning to estimate optimal regimes. \cite{luckett2020estimating} proposed V-learning to search for an optimal policy over a pre-specified class of policies. \cite{xu2020latent} later extended V-learning to latent space models. Among other works, \cite{liao2020off}, \cite{uehara2020minimax} and \cite{shi2020statistical} focus on a target policy or value function evaluation instead of finding an optimal policy. In the computer science field, popular approaches include learning the optimal value function firstly, and then recovering the corresponding optimal policy \citep{antos2008learning, dai2018sbeed}. Other methods include the residual gradient algorithm \citep{baird1995residual} and PCL learning \citep{nachum2017bridging,chow2018path,nachum2018trust}, but these algorithms encounter the \textit{double sampling} problem \citep{sutton2018reinforcement}. Alternatively, entropy-augmented methods \citep{schulman2017equivalence, lee2018sparse, haarnoja2018soft} follow the principle of developing reinforcement learning algorithms with improved exploration and high robustness. However, their methods are not suitable for continuous state space or large numbers of treatment options. 

In this paper, we propose a novel \textbf{P}roximal \textbf{T}emporal consistency \textbf{Learning} (pT-Learning) framework for estimating the optimal infinite horizon treatment regime. Through revisiting the standard Bellman equation from an alternative perspective, we construct a proximal counterpart simultaneously addressing the non-smoothness issues and inducing an optimal sparse policy. This distinguishes our method from commonly used approaches in the existing literature. We utilize the path-wise consistency property of the constructed proximal Bellman operator to incorporate off-policy data, and further propose a consistent minimax sample estimator for the optimal policy via leveraging the idea of functional space embedding.

The pT-Learning framework enjoys several unique advantages. Firstly, it shows advantages when the number of treatment options is large and can be easily extended to a continuous treatment space. The induced optimal policy identifies treatments from a sparse subset of the treatment space, indicating that it only assigns (near-)optimal treatment options with nonzero probabilities. Also, this property adaptively adjusts the policy between stochastic and deterministic policy models through a data-driven sparsity parameter, hence bridging the two popular models together. In addition, the induced policy is robust to unexpected situations and guarantees the recommendation of near-optimal treatment alternatives when the optimal treatment is temporarily unavailable.
Secondly, the proposed minimax estimator can be directly optimized over the observed sample transition path without  the \textit{double sampling} required in existing algorithms. Thirdly, pT-Learning captures the optimal policy and value function jointly over any arbitrary state-action pair. This avoids the mismatched distributions adjustments, e.g., inverse propensity-score weighting in \cite{luckett2020estimating}, for off-policy data. Fourthly, our method intrinsically achieves flexibility in choosing the value function approximation class (including both linear and non-linear function approximation) without the risk of diverging from the optimal solution. In contrast, existing methods such as Q-learning \citep{watkins1992q}, Greedy GQ-learning \citep{ertefaie2018constructing}, and TD-learning \citep{dann2014policy} with its stabilized variant Emphatic-TD learning \citep{sutton2016emphatic,yu2016weak,yu2018generalized} 
may either diverge to infinity in off-policy training or only have guaranteed convergence under linear function approximation. Lastly, the proposed constrained minimax optimization problem can be reduced to an unconstrained minimization problem, which can be solved under a scalable and efficient unified \textit{actor-critic} framework. This greatly reduces the computational cost and improves the stability of the optimal policy and value function learning.

In addition to these unique advantages, our study makes important contributions to the fundamental problem of solving the Bellman equation when function approximation is used. We provide a substantial development for addressing the decade-long \textit{double sampling} problem in policy optimization. In addition, our approach draws a connection and provides an alternative understanding to the entropy-augmented Markov decision process problem \citep{schulman2017equivalence,lee2018sparse, haarnoja2018soft}.
Our method is motivated by addressing the non-smoothness issue of the Bellman equation while inducing policy sparsity. It is fundamentally different from the principles of the existing entropy-augmented methods, which focus on improving exploration ability and algorithmic robustness. In theory, we establish the first theoretical result for the adaptivity of sparse policy distributions. Moreover, we develop finite-sample upper bounds on both the excess risk and the performance error. To the best of our knowledge, this is the first non-asymptotic result to quantify the performance error on both deterministic and stochastic policy models jointly.

\section{Background and Notation}{\label{background}}

First, we introduce the background for the estimating dynamic treatment regime in infinite horizon settings, which can be modeled by the Markov decision process (MDP, \citealp{puterman2014markov}).  The MDP is denoted as a tuple $(\mathcal{S},\mathcal{A},\textbf{P},u)$, where $\mathcal{S}$ is a state space, $\mathcal{A}$ is an action (treatment) space, $\textbf{P}(\cdot|s,a)$ is an unknown Markov transition kernel, and $u$ is a unknown immediate utility function. The immediate utility at the time $t$ is defined as $R^{t} = u(S^{t+1},S^{t},A^{t}): \mathcal{S} \times  \mathcal{S} \times  \mathcal{A}  \rightarrow \mathbb{R}$. In this paper, we consider a finite action space, i.e., $|\mathcal{A}| < \infty$. A trajectory induced by the MDP can be written as $
\mathcal{D}=\left\{S^{1}, A^{1}, S^{2}, A^{2}, S^{3}, \ldots, \ldots, S^{T+1}\right\}$,
where $S^{t} \in \mathcal{S}$ is the patient's health state at $t$, $A^{t} \in \mathcal{A}$ is the action assignment at $t$, and $T$ denotes the length of trajectory assumed to be non-random for simplicity. The observed data $\mathcal{D}_{1:n} = \left\{\mathcal{D}_{i}\right\}_{i=1}^{n}$ comprises $n$ independent and identical distributed trajectories of $\mathcal{D}$. Here, the state evolves following the time-homogenous Markov process. For all $t \geq 1$, $S^{t+1}  \perp \left(S^{1}, A^{1},  \ldots, S^{t-1} , A^{t-1}\right) \mid \left(S^{t},A^{t}\right)$ and $\textbf{P}(S^{t+1} = s^{\prime} | S^{t} = s, A^{t} = a)= \textbf{P}(s^{\prime} | s,a)$.
A treatment regime (policy) $\pi: \mathcal{S} \rightarrow \mathcal{A}$ is a map from the state space $\mathcal{S}$ to the action space $\mathcal{A}$.

The discounted sum of utilities beyond the time $t$ is represented as $\sum_{k=1}^{\infty} \gamma^{k-1}  R^{t+k}$, where $\gamma \in  (0,1)$ is called discount factor. Our goal is to find a policy $\pi$  to maximize the expected discounted sum of utilities from time $t$ until death. The infinite-horizon value function is defined as $
V_{t}^{\pi}(s)=\mathbb{E}_{\pi}\left[\sum_{k=1}^{\infty} \gamma^{k-1}  R^{t+k} \mid S^{t}=s\right]$, where the expectation $\mathbb{E}_{\pi}$ is taken by assuming that the system follows a policy $\pi$. Accordingly, 
the infinite-horizon action-value function $
Q_{t}^{\pi}(s, a)=\mathbb{E}_{\pi}\left[\sum_{k=1}^{\infty} \gamma^{k-1} R^{t+k} \mid S^{t}=s, A^{t}=a\right]
$ can be similarly defined as $V_{t}^{\pi}(s)$, except that taking treatment $a$ given the state $s$ at time $t$ and then following $\pi$ till the end.
In a time-homogenous Markov process, $V_{t}^{\pi}(s)$ and $Q_{t}^{\pi}(s, a)$ would not depend on $t$ anymore \citep{sutton2018reinforcement}. And
 the optimal action-value function $Q^{\pi^{*}}(s,a) = \max_{\pi}Q^{\pi}(s, a)$ is \textit{unique}, which satisfies the Bellman optimality equation \citep{puterman2014markov}: $
Q^{\pi^{*}}\left(s, a\right)=\mathbb{E}_{{S}^{t+1}| s,a} [R^{t}+\gamma \max _{a^{\prime} \in \mathcal{A}} Q^{\pi^{*}}\left({{S}}^{t+1}, a^{\prime}\right) \mid S^{t}=s, A^{t}=a]
$,
where $\mathbb{E}_{{S}^{t+1}| s,a}$ is a short notation for $\mathbb{E}_{{S}^{t+1} \sim \textbf{P}(\cdot | s,a)}$. The policy $\pi^{*}$ is referred to as an optimal policy, but it might not be unique. An optimal policy $\pi^{*}$ can be obtained by taking the \textit{greedy} action of $Q^{\pi^{*}}(s,a)$, such that ${\pi}^{*}(s) = \arg \max_{a} Q^{\pi^{*}}(s,a)$.
Given a value function $V^{\pi}(s)$, the Bellman operator $\mathcal{B}$ is defined as
\#
 \mathcal{B}V^{\pi}(s) \coloneqq  \max _{a \in \mathcal{A}} \mathbb E_{{{S}^{t+1}| s,a}}  \left[R^{t}+\gamma  V^{\pi}({S}^{t+1}) \mid S^{t}=s, A^{t}=a \right]. 
 \label{greed_bell}
\#
Then $ \mathcal{B}V^{\pi^{*}}(s) =  V^{\pi^{*}}(s)$ for all $s \in \mathcal{S}$ where $V^{\pi^{*}}$ is the unique fixed point of the Bellman operator $\mathcal{B}$ \citep{bertsekas1997nonlinear}.

\section{Methodology}\label{proposed_method}

To develop the new framework, we introduce a proximal Bellman operator and the associated sparse and adaptive optimal policy in Section \ref{proximal_opt}. To estimate such optimal policy, we propose a minimax framework called proximal temporal consistency learning in Section \ref{proximal_learn}. The proposed method is able to address the practical challenges in mHealth applications, and also has some theoretical guarantees, e.g.,  the bounded performance error in Section \ref{theory}. 

The \textit{greedy} optimal policy $\pi^{*}(s) = \arg \max_{a} Q^{\pi^{*}}(s,a)$ is a deterministic policy, which means it suggests an action according to a deterministic rule without uncertainty. Several drawbacks exist for the deterministic policy class. First, the deterministic policy greedily takes an action at each decision stage, and thus fails to suggest a rule to pick up other (near-optimal) actions as back-up. This restriction leads the deterministic policy class to be non-robust for unexpected situations whenever the true optimal action is temporarily unavailable or restricted to implement. For example, the insufficient insulin or  patient's budget restricts him/her to adopt a sufficient dosage of insulin at each decision stage \citep{rehg2017mobile, marling2020ohiot1dm}. Second, the deterministic policy class is suboptimal due to poor exploration of environment, which is usually caused by taking greedy actions. The insufficient exploration impairs the performance of the algorithm, especially for the MDPs with large action or state spaces \citep{john1994best}. In contrast, the stochastic policy takes actions following a probability distribution, and therefore incorporates a certain degree of randomness, and successfully encourages exploration of the dynamic environment \citep{singh1994learning, schulman2015trust}. Also, the stochastic policy could suggest other actions with specific probabilities, which makes it more flexible and robust under unexpected situations. A more detailed comparison between deterministic and stochastic policy class is provided in Section 15, Appendix. The aforementioned advantages motivates many existing work to consider modeling a stochastic policy. In particular, one popular choice is using the Boltzmann distribution to model the optimal policy \citep{schulman2017equivalence, haarnoja2018soft, luckett2020estimating}. However, such policy distribution is prone to assigning nonzero probability mass to all actions, including non-optimal and dismissible ones. This could be problematic when the cardinality of the action space is large, which may cause the policy distribution to degenerate to a uniform distribution, and potentially fails to provide a desired action recommendation.

These challenges motivate us to develop an optimal policy following a distribution whose support set is a sparse subset of the action space containing only (near-)optimal actions. We refer to this class of policies as the \textit{sparse} policy class. Specially, the deterministic policy $\pi^{*}(s) = \arg \max_{a} Q^{\pi^{*}}(s,a)$ can be viewed as the most extreme case of the sparse policy and inherits some advantages of the sparse policy. However, it is greedy, resulting suboptimality issue and non-robustness to uncertainty. Therefore, we aim to design and estimate an optimal policy that enjoys a suitable and adaptive policy sparsity while inheriting the advantages of the stochastic policy model.

To start with, we first revisit the Bellman optimality equation in \eqref{greed_bell} from a policy-explicit view. Suppose the policy $\pi$ follows a stochastic distribution, then the Bellman optimality equation can be reformulated as
\#
\mathcal{B}V^{\pi^{*}}(s) \coloneqq \max _{\pi}  \mathbb{E}_{a \sim \pi(\cdot|s), \ {S}^{t+1} | s,a}\left[u(S^{t+1},s,a) + \gamma V^{\pi^{*}}({S}^{t+1})\right] = V^{\pi^{*}}(s),
\label{explicit_bellman}
\#
where $V^{\pi^{*}}$ is the unique fixed point of $\mathcal{B}$, and $\pi^{*}$ is the maximizer of $\max _{\pi}  \mathbb{E}_{a \sim \pi(\cdot|s), \ {S}^{t+1} | s,a} \linebreak \left[u(S^{t+1} ,s,a) + \gamma V^{\pi^{*}}({S}^{t+1})\right]$. We use this definition of $\pi^{*}$ for the rest of the paper. To solve the equation \eqref{explicit_bellman}, a natural idea is to jointly optimize $V^{\pi^{*}}$ and $\pi$ to minimize the discrepancy between the two sides of \eqref{explicit_bellman}. However, the equation \eqref{explicit_bellman} is nonlinear and contains a non-smooth \textit{max} operator. When either the state or action space is large, directly solving \eqref{explicit_bellman} often results in policies that are far from the optimal solution. In addition, the discontinuity and instability caused by the $\textit{max}$ operator make estimation very difficult without large amounts of samples. To address these issues, we need to consider a proximal counterpart of the Bellman equation \eqref{explicit_bellman}. 

\subsection{Proximal Bellman Operator}\label{proximal_opt}

In this section, we propose a proximal Bellman operator that circumvents the obstacles of solving \eqref{explicit_bellman}, while simultaneously inducing an adaptive and sparse optimal policy. The proposed framework bridges the gap between the deterministic and stochastic policy models and characterizes the intrinsic relationship between the two policy models.

Let $\mathcal{P}(\mathcal{A})$ be a convex probability simplex over $\mathcal{A}$, we reformulate the Bellman operator $\mathcal{B}$ under the Fenchel representation, i.e., 
\$
\mathcal{B}V^{\pi^{*}}(s) =  \max _{\pi \in \mathcal{P}(\mathcal{A})}  \sum_{a \in \mathcal{A}}\Big[  \mathbb{E}_{{S}^{t+1} | s,a}\big[u(S^{t+1},s,a) + \gamma V^{\pi^{*}}({S}^{t+1})\big] \cdot\pi(a|s) - \mu\left(\pi(a|s)\right) \Big],
\numberthis  \label{nesterov_structure}
\$
where $\langle \cdot,\cdot \rangle$ denotes the dot product and $\mu(\cdot)$ has to be convex and continuous. We take $\mu(\cdot)\equiv 0$, which satisfies the condition without introducing additional bias. The representation \eqref{nesterov_structure} provides a basis for constructing a proximity to $\mathcal{B}V^{\pi^{*}}(s)$. Specifically, we consider to add a strongly convex and continuous component, so-called proximity function, $d(\pi): \mathcal{P}(\mathcal{A}) \rightarrow \mathbb{R}$ to  \eqref{nesterov_structure}:
\$
\mathcal{B}_{\lambda}V^{\pi^{*}_{\lambda}}_{\lambda}(s) &=  \max _{\pi \in \mathcal{P}(\mathcal{A})}  \sum_{a \in \mathcal{A}} \Big[ \mathbb{E}_{{S}^{t+1}|s,a}\big[u(S^{t+1},s,a) + \gamma V^{\pi^{*}_{\lambda}}_{\lambda}({S}^{t+1})\big]\cdot\pi(a|s) + \lambda d\big(\pi(a|s)\big) \Big] \\
& =  \max _{\pi \in \mathcal{P}(\mathcal{A})}  \mathbb{E}_{a \sim \pi(\cdot |s)}\Big[Q^{\pi^{*}_{\lambda}}_{\lambda}(s,a) + \lambda \phi\big(\pi(a|s)\big)\Big],
\numberthis \label{bell_eq_smooth}
\$
where $\phi(x) = d(x)/x$ and $Q^{\pi^{*}_{\lambda}}_{\lambda}(s,a) \coloneqq  \mathbb{E}_{{S}^{t+1}| s,a}[u(S^{t+1},s,a) + \gamma V^{\pi^{*}_{\lambda}}_{\lambda}({S}^{t+1})]$
. Here, the proximal optimal value function  $V^{\pi^{*}_{\lambda}}_{\lambda}$ is the unique fixed point of $\mathcal{B}_{\lambda}$, i.e., $\mathcal{B}_{\lambda}V^{\pi^{*}_{\lambda}}_{\lambda}(s) = V^{\pi^{*}_{\lambda}}_{\lambda}(s)$, and the  policy $\pi^{*}_{\lambda}$ is the maximizer of \eqref{bell_eq_smooth}. The explicit forms of the proximal value functions are deferred in Section 12, Appendix.

    By the Fenchel transformation theorem, the proximity $\mathcal{B}_{\lambda}V^{\pi^{*}_{\lambda}}_{\lambda}(s)$ is a smooth approximation for $\mathcal{B}V^{\pi^{*}}(s)$ if some conditions on the proximity function $d(\cdot)$ are satisfied \citep{hiriart2012fundamentals}. In addition, we should notice that we are not satisfied with only achieving the smoothing purpose but aim to develop a sparse optimal policy. To achieve this goal, we define a class of proximity functions based on the $\kappa$-logarithm function \citep{korbel2019information}, i.e., $d(x) = x\phi(x) = -\frac{x}{2}\log_{\kappa}(x)$, where $
\log _{\kappa}(x)= \frac{1}{1-\kappa}\left(x^{1-\kappa}-1\right)
$
for $x > 0$ and $\kappa \neq 1$. In this paper, we consider a special case  $\kappa = 0$, and refer to the operator \eqref{bell_eq_smooth}  as the \textit{proximal Bellman operator}.

Accordingly, the proximal Bellman operator \eqref{bell_eq_smooth} has several unique properties. First, it is a valid approximation for the  Bellman operator $\mathcal{B}$, where the approximation bias is bounded in Theorem S.1 provided in Section K, Supplementary Materials. Second, it is a smooth substitute of $\mathcal{B}$ according to the closed form of $\mathcal{B}_{\lambda}$ in \eqref{sparsemax}.  Third, the proximal Bellman operator  $\mathcal{B}_{\lambda}$ induces a sparse optimal policy whose sparsity can be adjusted by the magnitude of $\lambda$.

Through verifying KKT conditions of the maximization in  \eqref{bell_eq_smooth}, the proximal Bellman operator $ \mathcal{B}_{\lambda}$ has a closed-form equivalence (Proposition S.1 in Section K, Supplementary Materials):
\#
\mathcal{B}_{\lambda}V^{\pi^{*}_{\lambda}}_{\lambda}(s) =   \frac{\lambda}{2} \left\{ 1 -  \sum_{a \in \mathcal{K}(s)}\left[ \left(\frac{\sum_{a^{\prime} \in \mathcal{K}(s)} \frac{Q^{\pi^{*}_{\lambda}}_{\lambda}(s, a^{\prime})}{\lambda}-1}{|\mathcal{K}(s)|} \right)^{2} - \left(\frac{Q^{\pi^{*}_{\lambda}}_{\lambda}(s,a)}{\lambda}\right)^{2}\right] \right\},
\label{sparsemax}
\#
where $\mathcal{K}(s) =\{a_{(i)} \in \mathcal{A}: Q^{\pi^{*}_{\lambda}}_{\lambda}(s, a_{(i)}) > \frac{1}{i} \sum_{j=1}^{i} Q^{\pi^{*}_{\lambda}}_{\lambda}(s, a_{(j)}) - \frac{\lambda}{i}\}$ represents the support action set at state $s$. Here $a_{(i)}$ is the action with the $i$-th largest state-action value, and $|\mathcal{K}(s)| \leq |\mathcal{A}|$ holds for any $s \in \mathcal{S}$. 

The proximal Bellman operator is  differentiable everywhere only except for few splitting points where the support set $\mathcal{K}(s)$ changes, and the degree of smoothness is determined by the magnitude of $\lambda$ according to Proposition S.1. We visualize this property in Figure \ref{fig:proxmax} under a binary action setting. As $\lambda$ increases, the proximal Bellman operator $\mathcal{B}_{\lambda}$ becomes a more smooth approximation of $ \mathcal{B}$ but the approximation bias increases accordingly, indicating that the parameter $\lambda$ controls the bias and smoothness trade-off. 
\begin{figure}[H]
	\centering
		\scalebox{0.515}[0.515]{\includegraphics{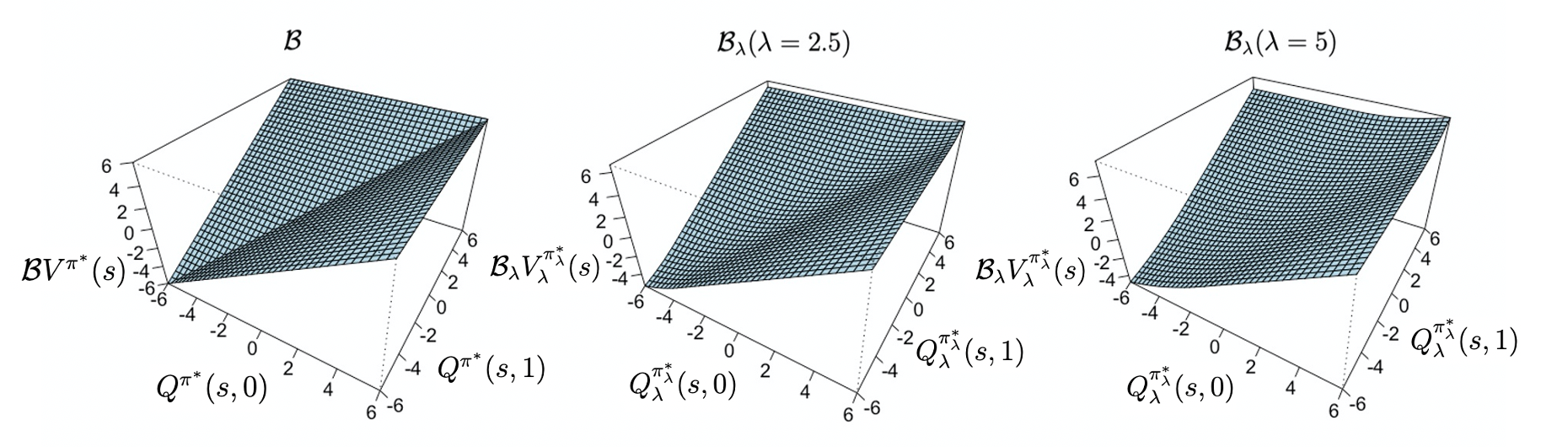}}
	\caption{A comparison of the standard and proximal Bellman operator in a binary action setting. 	}\label{fig:proxmax}
	\vspace{-5mm}
\end{figure}
In addition to the smoothness property, the proximal Bellman operator $\mathcal{B}_{\lambda}$ induces a sparse optimal policy distribution $\pi^{*}_{\lambda}(\cdot|s)$ whose support set is a sparse subset of the action space. We illustrate this point by presenting $\pi^{*}_{\lambda}(a|s)$  in terms of the state-action value function $Q^{\pi^{*}_{\lambda}}_{\lambda}(s, a)$ analytically (see the proof of Proposition S.1). That is
\#
 \pi^{*}_{\lambda}(a|s) = \Bigg(\frac{Q^{\pi^{*}_{\lambda}}_{\lambda}(s, a)}{\lambda}- \frac{\sum_{a^{\prime} \in \mathcal{K}(s)} \frac{Q^{\pi^{*}_{\lambda}}_{\lambda}(s, a^{\prime})}{\lambda}-1}{|\mathcal{K}(s)|}\Bigg)^{+},
 \label{sparse_pi}
\#
which is invariant to the location shift of the immediate utility. Given any state $s$, the equation \eqref{sparse_pi} defines a well-defined probability mass function in that $\sum_{a \in \mathcal{A}}  \pi^{*}_{\lambda}(a|s) = 1$. Also, the policy $\pi^{*}_{\lambda}$ is prone to assign a large probability to the action according to the rank of the state-action values. This ensures that the policy  $\pi^{*}_{\lambda}(a|s)$ can suggest the optimal action in a given state $s$.

Moreover, the sparsity is shown by analyzing the support set of $ \pi^{*}_{\lambda}(\cdot|s)$, i.e., $\{a \in \mathcal{A}: \pi^{*}_{\lambda}(a|s) > 0\}$, which satisfies
\#
\left\{a_{(i)} \in \mathcal{A}: Q^{\pi^{*}_{\lambda}}_{\lambda}(s, a_{(i)}) > \frac{1}{i} \sum_{j=1}^{i} Q^{\pi^{*}_{\lambda}}_{\lambda}(s, a_{(j)}) - \frac{\lambda}{i} \right\}.
\label{sparse_set}
\#
The inequality \eqref{sparse_set} indicates that the sparsity parameter $\lambda$ controls the margins between the smallest action value and the others included in the support set \eqref{sparse_set}.
In particular, the cardinality of the support set increases as $\lambda$ increases. Conversely, the support set of $\pi^{*}_{\lambda}(\cdot|s)$ shrinks as $\lambda$ becomes smaller. To illustrate how this mechanism works,  we use a ternary action setting as an example. By the inequality \eqref{sparse_set}, when $\lambda$ is sufficiently small in that $\lambda < Q^{\pi^{*}_{\lambda}}_{\lambda}(s, a_{(1)}) - Q^{\pi^{*}_{\lambda}}_{\lambda}(s, a_{(2)})$,  only $a_{(1)}$ is contained in the support set. The policy distribution becomes a deterministic rule and takes the greedy action with the largest action value $Q^{\pi^{*}_{\lambda}}_{\lambda}(s, a_{(1)})$. From this point of view, the Q-learning approach and its variants can be regarded as special cases in our framework. If $\lambda$ increases and falls into the range $\big[Q^{\pi^{*}_{\lambda}}_{\lambda}(s, a_{(1)}) - Q^{\pi^{*}_{\lambda}}_{\lambda}(s, a_{(2)}),  \sum^{2}_{i=1}Q^{\pi^{*}_{\lambda}}_{\lambda}(s, a_{(i)}) - 2  Q^{\pi^{*}_{\lambda}}_{\lambda}(s, a_{(3)}) \big]$,  the support set contains two actions $a_{(1)}$ and $a_{(2)}$ with the largest state-action values, but eliminates action $a_{(3)}$. This implies that the induced policy can be adaptively adjusted between the deterministic and stochastic policy models. In Section \ref{theory}, we  investigate the sparsity of the optimal policy distribution in more details.
\subsection{Proximal Temporal Consistency Learning}\label{proximal_learn}

Next, we introduce the proposed proximal temporal consistency learning framework for estimating the induced sparse policy $\pi^{*}_{\lambda}$. Our development mainly hinges on a path-wise property of the proximal Bellman operator $\mathcal{B}_{\lambda}$ and the functional space embedding \citep{gretton2012kernel}. The proposed minimax estimator is able to address the \textit{double sampling} issue and easily incorporate off-policy data, while intrinsically preserving the convergence in off-policy training with flexible function approximations.

First, we show that the proximal Bellman operator $\mathcal{B}_{\lambda}$ enjoys the temporal consistency property \citep{rawlik2013stochastic}. Most recently, many state-of-the-art methods based on this type of property have achieved great success in real-life applications \citep{nachum2017bridging, chow2018path,nachum2018trust}. The temporal consistency property connects the optimal policy and value function in one equation along with any arbitrary state-action pair, which provides an elegant way to incorporate off-policy data. We present the temporal consistency property for  $\mathcal{B}_{\lambda}$ in the following proposition.

\begin{proposition}\label{temporal_theory}
For any $s \in \mathcal{S}, a \in \mathcal{A}$ and $\lambda \in (0,+\infty)$, if $V^{\pi^{*}_{\lambda}}_{\lambda}$ is the fixed point of the proximal Bellman operator, i.e., $\mathcal{B}_{\lambda}V^{\pi^{*}_{\lambda}}_{\lambda} = V^{\pi^{*}_{\lambda}}_{\lambda} $, and $\pi^{*}_{\lambda}$ is the induced policy following the equation \eqref{sparse_pi}, then $(V^{\pi^{*}_{\lambda}}_{\lambda},\pi^{*}_{\lambda})$ is a solution of the following proximal temporal consistency equation:
\#
 \mathbb E_{S^{t+1}| s,a}\big[u(S^{t+1},s,a) + \gamma V^{\pi_{\lambda}}_{\lambda}(S^{t+1})\big]  - \lambda \phi^{\prime}( \pi_{\lambda}(a|s) )- {\Psi}(s)  + \psi(a|s)  - V^{\pi_{\lambda}}_{\lambda}(s) = 0,
\label{temproal_consistent}
\#
where ${\Psi}(s): \mathcal{S} \rightarrow [-\lambda/2,0]$ and $\psi(a|s) : \mathcal{S} \times \mathcal{A} \rightarrow \mathbb{R}^{+}$ are Lagrangian multipliers with $\psi(a|s) \cdot\pi_{\lambda}(a|s) = 0$, and  $\phi^{\prime}(x) = x-\frac{1}{2}$. Following, we call the discrepancy on the left-hand side of the equation \eqref{temproal_consistent} the proximal temporal consistency error (pT-error). Here, the ``temporal consistency'' characterizes that the equation holds for temporal transitions.  
\end{proposition}

Since the equation \eqref{temproal_consistent} holds for any arbitrary state-action pair $(s,a)$, the policy $\pi^{*}_{\lambda}$ can be estimated directly over the observed transition pairs. Utilizing this property lets us skip to adjust mismatched distributions using, e.g., inverse propensity-score weighting \citep{luckett2020estimating}, and avoids the curse of high variance \citep{jiang2016doubly} carried over from the data distribution corrections. To solve the equation \eqref{temproal_consistent},  an intuitive idea is to minimize the pT-error with the $L^2$ loss, that is
\vspace{-0.5cm}
\#
& \min_{V^{\pi_{\lambda}}_{\lambda}, \pi_{\lambda},{\Psi},\psi}  \mathbb E_{S^{t}, A^{t}}\left[\left(
\mathcal{T}_{\pi_\lambda}(S^{t},A^{t},V^{\pi_{\lambda}}_{\lambda}) - {\Psi}(S^{t})  + \psi(A^{t}|S^{t})  - V^{\pi_{\lambda}}_{\lambda}(S^{t}) \right)^2 \right]  \label{l2_loss} \\
  &  \qquad \text{s.t.} \quad \pi_{\lambda}  (a|s) \cdot\psi(a|s)=0, \; \psi(a|s)\geq0 \ \; \text{and} \;-\frac{\lambda}{2} \leq \Psi(s) \leq 0, \; \text{for all} \; s \in \mathcal{S} \; \text{and} \; a \in \mathcal{A}.
\label{l2_constrain}
\#
where $\mathcal{T}_{\pi_\lambda}(S^{t},A^{t},V^{\pi_{\lambda}}_{\lambda}) = \mathbb E_{S^{t+1} | S^{t},A^{t}} \big[ u(S^{t+1},S^{t},A^{t})  + \gamma V^{\pi_{\lambda}}_{\lambda}(S^{t+1})\big] -\lambda \phi^{\prime}( \pi_{\lambda}(A^{t}|S^{t}) ) $, and $ \mathbb E_{S^{t}, A^{t}}$ is a short notation for $\mathbb E_{\{S^{t}, A^{t}\}  \sim d_{\pi_{b}}}$ where $d_{\pi_{b}}$ is the behavior data distribution induced by the behavior policy  $\pi_{b}$. The behavior policy $\pi_{b}$ is the policy the decision maker follows in collecting the data.

However, directly minimizing the sample version of \eqref{l2_loss} is not realizable, as the conditional expectation $\mathbb E_{S^{t+1} | S^{t},A^{t}}$ in $\mathcal{T}_{\pi_\lambda}(S^{t},A^{t},V^{\pi_{\lambda}}_{\lambda})$ is unknown. To approximate this unknown expectation, bootstrapping can be applied but produces an inconsistent and biased sample estimator, where an extra conditional variance will be involved as a bias component \citep{fan2020theoretical}. This problem is usually referred to as the \textit{double-sampling} issue \citep{baird1995residual} and exists in many policy optimization approaches.

To avoid the \textit{double-sampling} bias, we consider embedding a Lebesgue measurable function class to the averaged pT-error. In particular, we define a \textit{critic} function $h \in \mathcal{H} = L^{2}(\mathcal{S} \times {\mathcal{A}})$ and formulate a novel embedding function $ \mathcal{L}_{\text{weight}}$ as follows,
\#
 \mathcal{L}_{\text{weight}}(V^{\pi_{\lambda}}_{\lambda},\pi_{\lambda},{\Psi},\psi,h) \coloneqq  \ \mathbb{E}_{S^{t}, A^{t}}  \big[h(S^{t},A^{t})\big(
{\mathcal{T}}_{\pi_\lambda}(S^{t},A^{t},V^{\pi_{\lambda}}_{\lambda})  - {\Psi}(S^{t})  + \psi(A^{t}|S^{t})  - V^{\pi_{\lambda}}_{\lambda}(S^{t})\big) \big].
\#
The \textit{critic} function is introduced here to fit the discrepancy of \eqref{temproal_consistent} and promotes the transition pairs with large pT-error. Unlike the naive $L^2$ loss \eqref{l2_loss}, the nested conditional expectation $\mathbb E_{S^{t+1}|S^{t}, A^{t}}$ in $
{\mathcal{T}}_{\pi_\lambda}(S^{t},A^{t},V^{\pi_{\lambda}}_{\lambda})$ is not inside the square function anymore. Intuitively, the $\mathcal{L}_{\text{weight}}$ circumvents the \textit{double sampling} issue since the second order moment of bootstrapping samples is not involved anymore and the extra condition variance vanishes. Fundamentally, the $\mathcal{L}_{\text{weight}}$ offers compensation for insufficient sampling on $\textbf{P}(s^{\prime}| s,a)$. The marginal information of the state-action pairs $d_{\pi_b}(s,a)$ and the transition kernel $\textbf{P}(s^{\prime}| s,a)$ can be aggregated as a joint distribution $p_{\pi_b}(s^{\prime}, s, a)$ by the linearity of the expectation. Instead of extracting the information from the transition kernel $\textbf{P}(s^{\prime}| s,a)$, i.e. approximating the conditional expectation $\mathbb E_{S^{t+1}|S^{t}, A^{t}}$, the $\mathcal{L}_{\text{weight}}$ can be approximated using the sample-path transition pairs drawn from the joint distribution $p_{\pi_b}(s^{\prime}, s, a)$. In essence, this explains why the $\mathcal{L}_{\text{weight}}$ can address the  \textit{double sampling} issue.

In the following theorem, we proceed to show $ \mathcal{L}_{\text{weight}}(V^{\pi_{\lambda}}_{\lambda},\pi_{\lambda},{\Psi},\psi,h)$ can identify the optimal value function and optimal policy $(V^{\pi^{*}_{\lambda}}_{\lambda}, \pi^{*}_{\lambda})$.

\begin{thm}\label{biconsistent}
Suppose $\mathcal{S} \times \mathcal{A}$ is Lebesgue measurable, and for any $h(\cdot)$ in a bounded $L^2$ space, i.e., $h(\cdot) \in \mathcal{H}^{\zeta}_{L^2} \coloneqq \{h: \|h\|_{L^2} \leq \zeta\}$ for $\zeta \in (0,+ \infty)$, then the loss  $\mathcal{L}_{\text{weight}}(V^{\pi^{*}_{\lambda}}_{\lambda},\pi^{*}_{\lambda} ,{\Psi},\psi,h) = 0$. Conversely, if there exists $({V}^{\tilde{\pi}_{\lambda}}_{\lambda},\tilde{\pi}_{\lambda})$ such that $\mathcal{L}_{\text{weight}}({V}^{\tilde{\pi}_{\lambda}}_{\lambda}, \tilde{\pi}_{\lambda},{\Psi},\psi,h) = 0$, then $({V}^{\tilde{\pi}_{\lambda}}_{\lambda},\tilde{\pi}_{\lambda})$ satisfies the proximal temporal  consistency equation \eqref{temproal_consistent}.
\end{thm}

As $\mathcal{L}_{\text{weight}}(V^{\pi^{*}_{\lambda}}_{\lambda},\pi^{*}_{\lambda} ,{\Psi},\psi,h) = 0$ holds for any $h \in \mathcal{H}^{\zeta}_{L^2}$, 
Theorem \ref{biconsistent} leads to a minimax optimization problem with a valid loss $\mathcal{L}^2_{\text{weight}}$,
\#
\min_{V^{\pi_{\lambda}}_{\lambda}, \pi_{\lambda},{\Psi},\psi} \max_{h \in
\mathcal{H}^{\zeta}_{L^2}} \ \mathcal{L}^2_{\text{weight}}(V^{\pi_{\lambda}}_{\lambda},\pi_{\lambda} ,{\Psi},\psi,h).
 \label{minimax_loss}
 \#
The minimax optimization of \eqref{minimax_loss} gives a clear direction to estimate $(V^{\pi^{*}_{\lambda}}_{\lambda}, \pi^{*}_{\lambda})$, but it still remains intractable. As the \textit{critic} function $h(\cdot)$ could be any arbitrary function in $\mathcal{H}^{\zeta}_{L^2}$, it makes infeasible to find a proper representation for $h(\cdot)$. Therefore, we introduce a tractable framework for  \eqref{minimax_loss}, where $h(\cdot)$ can be appropriately represented. Define $h^{*}: \mathcal{S} \times \mathcal{A} \rightarrow \mathbb{R}$ be the optimal \textit{critic} function if it satisfies
 \#
 h^{*}(\cdot) \in \argmax_{h \in
\mathcal{H}^{\zeta}_{L^2}} \ \mathcal{L}^2_{\text{weight}}(V^{\pi^{*}_{\lambda}}_{\lambda},\pi^{*}_{\lambda} ,{\Psi},\psi,h).
 \label{max_h_func}
 \#
In the following, we shows that $ h^{*}(\cdot)$ lies in a class of continuous functions under some regular continuity conditions which are easily satisfied in practice.
\begin{thm}\label{continuity_h_func}
Let $\mathbb{C}(\mathcal{S} \times \mathcal{A})$ be all continuous functions on $\mathcal{S} \times \mathcal{A}$. For any $(s,a) \in \mathcal{S} \times \mathcal{A}$ and $s^{\prime} \in \mathcal{S}$, the optimal \textit{critic} function $h^{*}(s,a)$ has the following properties: 

1. (\textit{Continuity}) Suppose the utility function $u(s^{\prime},s,a)$ and the transition kernel $\mathbf{P}(s^{\prime}|s,a)$ are continuous over $(s,a)$ for any $s^{\prime}$, then $h^{*} \in \mathcal{H}^{\zeta}_{L^2}  \cap \mathbb{C}(\mathcal{S} \times \mathcal{A})$ and is \textit{unique}.

2. (\textit{Lipschitz-continuity})  Suppose $u(s^{\prime},s,a)$ is uniformly $M_{u}$--Lipschitz continuous and $\mathbf{P}(s^{\prime}|s,a)$ is $M_{p}$--Lipschitz continuous over $(s,a)$  for any $s^{\prime}$, where $M_{u}$ and $M_{p}$ are some Lipschitz constants, then there must exist a Lipschitz constant $M_{h^{*}}$ such that $h^{*}(s,a)$ is $M_{h^{*}}$--Lipschitz continuous over $(s,a)$.
\end{thm}

Theorem \ref{continuity_h_func} states that the optimal \textit{critic} function  $h^{*}(s,a)$ is continuous over $(s,a)$ if only the utility function and the transition kernel are continuous. This continuity condition widely holds for precision medicine and reinforcement learning problems. As we mentioned, the $h^{*}$ could be any arbitrary function in $\mathcal{H}^{\zeta}_{L^2}$, which imposes exceptional difficulty in the representation of $h^{*}$. Theorem \ref{continuity_h_func} indicates that it is sufficient to represent $h$ in a bounded continuous function space which preserves the optimal solution of \eqref{minimax_loss}. This provides us a basis for constructing a tractable framework for solving \eqref{minimax_loss}.

We propose to represent $h^{*}$ in a bounded reproducing kernel Hilbert space (RKHS) such that $\mathcal{H}^{\zeta}_{\mathcal{K}} \coloneqq \{h: \| h \|^2_{ \mathcal{H}_{\text{RKHS}}} \leq \zeta \}$. When $\mathcal{H}^{\zeta}_{\mathcal{K}}$ is reproduced by a universal kernel, the error $\textit{err} \, (\zeta):= \sup _{f \in \mathbb{C}(\mathcal{S} \times \mathcal{A})} \inf _{h \in \mathcal{H}^{\zeta}_{\mathcal{K}}}\|f-h\|_{\infty}$ decreases as $\zeta$ increases and vanishes to zero as $\zeta$ goes to infinity \citep{bach2017breaking}. Therefore, any continuous function can be approximated by a function in $\mathcal{H}^{\zeta}_{\mathcal{K}}$ with arbitrarily small error. As a result, solving the inner maximization of \eqref{minimax_loss} over $h \in \mathcal{H}_{\mathcal{K}}^{\zeta}$ is feasible when $h^{*} \in \mathcal{H}_{L^{2}}^{\zeta} \cap \mathbb{C}(\mathcal{S} \times \mathcal{A})$, which also makes the optimization become tractable.

Although the optimization  \eqref{minimax_loss} is tractable with RKHS representation, solving the minimax optimization is still a challenging task. Int the following, we transform the minimax optimization to an easier solvable minimization problem, leveraging the idea of the kernel embedding \citep{gretton2012kernel}.

\begin{thm}\label{RKHS_minimax}
Suppose $h \in \mathcal{H}^{\zeta}_{\mathcal{K}}$ is reproduced by a universal kernel $K(\cdot,\cdot)$ in \eqref{minimax_loss} , then the minimax optimization \eqref{minimax_loss} can be decoupled to a single-stage minimization problem wherein
  \#
 \min_{V^{\pi_{\lambda}}_{\lambda},\pi_{\lambda},{\Psi},\psi} \ & \mathcal{L}_{U} :=  \mathbb E_{S^{t},\widetilde{S}^{t}, A^{t}, \widetilde{A}^{t},S^{t+1},\widetilde{S}^{t+1}} \bigg[ \Big(
\widetilde{\mathcal{T}}_{\pi_\lambda}(S^{t},A^{t},V^{\pi_{\lambda}}_{\lambda})   - {\Psi}(S^{t})  + \psi(A^{t}|S^{t})  - V^{\pi_{\lambda}}_{\lambda}(S^{t})\big) \notag \\
& \quad  \cdot \zeta  K\big( \{S^{t},A^{t}\}, \{ \widetilde{S}^{t},\widetilde{A}^{t}\}\big)  \big(\widetilde{\mathcal{T}}_{\pi_\lambda}(\widetilde{S^{t}},\widetilde{A}^{t},V^{\pi_{\lambda}}_{\lambda})   - {\Psi}(\widetilde{S^{t}})  + \psi(\widetilde{A}^{t}|\widetilde{S^{t}})  - V^{\pi_{\lambda}}_{\lambda}(\widetilde{S^{t}})\Big)  \bigg],
\label{loss_true}
  \#
  where $\widetilde{\mathcal{T}}_{\pi_\lambda}(S^{t},A^{t},V^{\pi_{\lambda}}_{\lambda}) =  u(S^{t+1},S^{t},A^{t}) + \gamma V^{\pi_{\lambda}}_{\lambda}(S^{t+1}) - \lambda \phi^{\prime}( \pi_{\lambda}(A^{t}|S^{t}) )$ and $(\widetilde{S}^{t},\widetilde{A}^{t},\widetilde{S}^{t+1} )$ is an independent copy of the transition pair $(S^{t},A^{t},S^{t+1})$.
  \end{thm}
Moreover, if the universal kernel $K(\cdot, \cdot)$ is strictly positive definite \citep{stewart1976positive}, 
 the loss $ \mathcal{L}_{U}$ is non-negative definite. This implies that $\mathcal{L}_{U} = 0 $ when $({V^{\pi_{\lambda}}_{\lambda},\pi_{\lambda}}) =  ({V^{\pi^{*}_{\lambda}}_{\lambda},\pi^{*}_{\lambda}})$, and thus the loss $\mathcal{L}_{U}$ is a valid loss can identify $({V^{\pi^{*}_{\lambda}}_{\lambda},\pi^{*}_{\lambda}})$.

Given the observed data $\mathcal{D}_{1:n}$ with the length of trajectory $T$, to minimize $\mathcal{L}_{U}$ under the constraints in \eqref{l2_constrain}, we propose a trajectory-based U-statistic estimator to capture the within-trajectory loss. Subsequently, the total loss $\mathcal{L}_{U}$ can be aggregated as the empirical mean of $n$ $i.i.d.$ within-trajectory loss. That is,
 \#
\min_{V^{\pi_{\lambda}}_{\lambda},\pi_{\lambda},{\Psi},\psi} \; \widehat{\mathcal{L}_{U}} = & \  \frac{1}{n}\sum_{i=1}^{n} \frac{2\zeta }{T(T-1)}  \sum_{ 1\leq j\neq k \leq T} \Big[\big(
\widetilde{\mathcal{T}}_{\pi_\lambda}(S_i^{j},A_i^{j},V^{\pi_{\lambda}}_{\lambda})   - {\Psi}(S_i^{j})  + \psi(A_i^{j}|S_i^{j})  - V^{\pi_{\lambda}}_{\lambda}(S_i^{j})\big) \notag \\
&  \quad \cdot  K\big( \{S_i^{j},A_i^{j}\}, \{ S_i^{k},{A}_i^{k}\}\big) \big(
\widetilde{\mathcal{T}}_{\pi_\lambda}(S_i^{k},A_i^{k},V^{\pi_{\lambda}}_{\lambda})   - {\Psi}(S_i^{k})  + \psi(A_i^{k}|S_i^{k})  - V^{\pi_{\lambda}}_{\lambda}(S_i^{t})\big)  \Big]. \notag \\
\quad  \text{s.t.} \quad \pi_{\lambda}  (a|s) & \cdot\psi(a|s)=0, \; \psi(a|s)\geq 0  \ \; \text{and} \;-\frac{\lambda}{2} \leq \Psi(s) \leq 0, \; \text{for all} \; s \in \mathcal{S} \; \text{and} \; a \in \mathcal{A}.
 \label{em_proposed_loss}
\#

\vspace{-3mm}

Unlike the inconsistent sample estimator in \eqref{l2_loss}, the proposed sample estimator $\widehat{\mathcal{L}_{U}}$ is consistent. The consistency is shown in Theorem \ref{concentration_ineq} in Section \ref{theory}, through examining the tail behavior of $\widehat{\mathcal{L}_{U}}$. In addition, the gradient of the proposed loss $\widehat{\mathcal{L}_{U}}$ can be approximated by the sampled transitions and optimized using any gradient descent type of algorithm. Hence, the proposed method achieves great flexibility in the function approximations of $(V^{\pi^{*}_{\lambda}}_{\lambda},  \pi^{*}_{\lambda})$, allowing both linear and nonlinear approximations without the risk of divergence from the optimal solution. In comparison, popular methods, including the Q-learning \citep{watkins1992q}, TD-learning \citep{dann2014policy} and Greedy GQ-learning \citep{maei2010toward, ertefaie2018constructing}, either diverge to infinity in off-policy training or have guaranteed convergence only in using linear function approximations. Also, we should note that the pT-Learning framework connects to the importance weighted variants of off-policy TD-learning algorithms when the \textit{critic} function $h(\cdot)$ is restricted into a bounded linear function space. We provide a discussion on this connection in Section 15, Appendix.

\section{Theory}\label{theory}
In this section, we establish the theoretical properties of the pT-Learning framework. Our study mainly focuses on three major parts. In the first part, we formally study the sparsity of the policy distribution $\pi^{*}_{\lambda}$. Theorem \ref{delta_sparse_theory} shows that the cardinality of the nonzero probability set for $\pi^{*}_{\lambda}$ is well-controlled by the magnitude of $\lambda$, which implies that $\pi^{*}_{\lambda}$ is adaptively adjusted between the deterministic and stochastic policy models. We believe this is the first theoretical result on investigating the sparsity of policy distributions. In the second part, we establish a convergence rate of the empirical risk  $\widehat{\mathcal{L}_{U}}$ towards the true risk ${\mathcal{L}_{U}}$. In particular,
Theorem \ref{concentration_ineq} provides a sharp exponential concentration bound, indicating that  $\widehat{\mathcal{L}_{U}}$ is a consistent estimate for ${\mathcal{L}_{U}}$ and  the \textit{double sample} issue is solved. In the last part, we measure the excess risk of  $\widehat{\mathcal{L}_{U}}$ which is applicable for both the parametric and nonparametric function spaces in Theorem \ref{risk_bound}. Furthermore, Theorem \ref{valuefunc_bound} provides a finite-sample upper bound on the performance of the estimated optimal value function. The bound indicates both the excess risk and the approximation bias of the proximal Bellman operator will affect the performance error.  To the best of our knowledge, this is the first non-asymptotic result to quantify the performance error on deterministic and stochastic policy models jointly.  In the following, we present our major assumptions and theorems. The proofs and additional theoretical results and assumptions are deferred to Supplementary Materials.

First, we define a notation called $\Delta_{\pi}$-sparsity to characterize the cardinality of nonzero probability actions for some policy $\pi$.
\begin{definition}\label{def_sparsity}
Denote the maximum cardinality of the support set for a given policy distribution $\pi(\cdot|s)$ over $s \in \mathcal{S}$ as $\Delta_{\pi}$, i.e., 
\$
\Delta_{\pi} \coloneqq \max_{s \in \mathcal{S}} | \{a \in \mathcal{A}: \pi(a|s) \neq 0 \} |,
\$
where $|\cdot|$ is a cardinality operator and $\Delta_{\pi} \leq |\mathcal{A}|$. Then, we call the policy distribution $\pi(\cdot|s)$ a $\Delta_{\pi}$-degree sparse policy distribution.
\end{definition}
Before we characterize the sparsity of the proposed sparse policy model, we first present a boundedness assumption. 

\begin{assump}\label{rbound_assum}
The immediate utility function $u(S^{t+1},S^{t},A^{t})$ is uniformly bounded by $R_{\max}$, i.e. $\|u\|_\infty  \leq R_{\max} < \infty $.
\end{assump}
Assumption \ref{rbound_assum} is a regular assumption to impose a boundedness condition on the MDP \citep{antos2008learning,liao2020off}.
 
\begin{thm}\label{delta_sparse_theory}
Under Assumption \ref{rbound_assum} and Assumptions S.1-S.2 provided in Supplementary Materials Section I, for all $s\in \mathcal{S}$, the sparse policy distribution $\pi^{*}_{\lambda}(\cdot|s)$ in \eqref{sparse_pi} has the following properties: the policy $\pi^{*}_{\lambda}$ is $\Delta_{\pi^{*}_{\lambda}}$-degree sparse where  $\Delta_{\pi^{*}_{\lambda}}  \rightarrow 1$ as $\lambda \rightarrow 0$, and $\Delta_{\pi_{\lambda}}  \rightarrow |\mathcal{A}|$ as $\lambda \rightarrow \infty$. In particular, for any $\epsilon >0$, there exists a positive number $\lambda_0$ such that for all $\lambda \geq \lambda_0$, the sparse policy distribution $\pi^{*}_{\lambda}(\cdot|s)$ approaches a  discrete uniform distribution $\mathcal{U}$ with a probability mass function $|\mathcal{A}|^{-1}$, i.e., $|\pi^{*}_{\lambda}(a|s) - |\mathcal{A}|^{-1}| < \epsilon$ for all $s \in \mathcal{S}$ and $a \in \mathcal{A}$.
\end{thm}

Theorem \ref{delta_sparse_theory} shows that the degree of sparseness, or the cardinality of the nonzero probability actions set, of the sparse policy $\pi^{*}_{\lambda}( \cdot | s)$ can be controlled by $\lambda$. In the most extreme case, when $\lambda \rightarrow 0$, the policy $\pi^{*}_{\lambda}$ becomes a deterministic policy. Alternatively, as $\lambda \rightarrow +\infty$, the induced policy $\pi^{*}_{\lambda}$ degenerates to a uniform distribution that assigns equal probabilities to all action arms.

Before we present the rest of the theoretical results, we need to introduce the stationarity and dependency of the stochastic process  $\{ (S^{t},A^{t}) \}_{t \geq 1}$. For each single patient trajectory, it is easily observed that the sequence $\{ (S^{t},A^{t}) \}_{t \geq 1}$  is a stationary Markov chain. In terms of sample dependency, \cite{farahmand2016regularized} and \cite{liao2020off} assume that within-trajectory samples are independent in order to reduce technical difficulties in the theoretical developments. However, this assumption might be too restrictive and often violated in practice. In our work, we consider the sample dependency and establish more rigorous theoretical results under the notion of the mixing process \citep{kosorok2008introduction}. Specifically,
for any stationary sequence of dependent random variables $\left\{S^{t},A^{t}\right\}_{t \geq 1}$, let $\mathcal{F}_{b}^{c}$ be the $\sigma$-field generated by $\left\{S^{b},A^{b}\right\}, \ldots, \left\{S^{c},A^{c}\right\}$ and define $\beta(k) =\mathbb{E}\big[\sup _{m \geq 1}\left\{\left|P\left(B \mid \mathcal{F}_{1}^{m}\right)-P(B)\right|: B \in \mathcal{F}_{m+k}^{\infty}\right\}\big]$, and we say that the process $\left\{S^{t},A^{t}\right\}_{t \geq 1}$ is $\beta$-mixing if $\beta(k) \rightarrow 0$ as $k \rightarrow \infty$. In the following, we assume a  mixing condition to quantify the within-trajectory sample dependency.

\begin{assump}\label{mixing}
For a strictly stationary sequence $\{ (S^{t},A^{t}) \}_{t \geq 1}$, there exists a constant $\delta_1 > 1$ such that the $\beta$-mixing coefficient corresponding to $\{ (S^{t},A^{t}) \}_{t \geq 1}$ satisfies $\beta(k) \lesssim \exp(-\delta_1 k)$ for $k \geq 1$.

\end{assump}

Assumption \ref{mixing} implies an exponential decay rate of the mixing process, which is typically for deriving a polynomial decay rate of the estimation error \citep{kosorok2008introduction}.  

\vspace{-1mm}

In Theorem \ref{concentration_ineq}, we study the tail behaviors of the empirical loss $\widehat{\mathcal{L}_{U}}$, and establish an exponential concentration inequality for $\widehat{\mathcal{L}_{U}}$ under an exponential rate of the $\beta$-mixing condition.
\begin{thm}\label{concentration_ineq}
For any sparsity parameter $\lambda < \infty$ and $\epsilon > 0$, under Assumptions \ref{rbound_assum}-\ref{mixing} and S.1-S.2 provided in Supplementary Materials Section I, we have $\epsilon$-divergence of $|\widehat{\mathcal{L}_{U}} - \mathcal{L}_{U}|$ bounded in probability for sufficiently large $T$, that is
\$
\mathbb P \left(|\widehat{\mathcal{L}_{U}} - \mathcal{L}_{U}|  > \epsilon  \right) \leq 2\Bigg[& \exp\left(-\frac{c_1{\epsilon}^2T/{4} - c_0c_1{\epsilon}U_{\max}^2\sqrt{T} }{ U_{\max}^2(\epsilon/2 - c_0 U_{\max}^2 /\sqrt{T})\log T \log\log4 T  +U_{\max}^4 }  \right )  \\
&+\exp\left (-\frac{n\epsilon^2}{2 U_{\max}^4}\right)  \Bigg],
\$
where $c_0,c_1$ are some constants depending on $\delta_1$; and $ U_{\max} = \frac{6R_{\max} +(5-4\gamma)\lambda}{2(1-\gamma)}$.
\end{thm}
Theorem \ref{concentration_ineq} shows that $\widehat{\mathcal{L}_{U}}$ is a consistent estimator for the loss ${\mathcal{L}_{U}}$ with a  high probability at an exponential rate. The concentration bound is sharper than the bound established in \cite{borisov2009exponential}. Here, we require an exponential decaying mixing rate, which is standard in the literature of deriving concentration inequalities for weakly dependent data \citep{borisov2009exponential, merlevede2011bernstein}. It should be possible to relax this mixing condition to a polynomial-decay rate of $\beta$-mixing with imposing an additional exponential $\alpha$-mixing condition. However, that is out of the scope of this paper.

We denote $\mathcal{L}_{U}^{*} = \inf_{\{V^{\pi_{\lambda}}_{\lambda},\pi_{\lambda},{\Psi},\psi\}}\mathcal{L}_{U}(V^{\pi_{\lambda}}_{\lambda},\pi_{\lambda},{\Psi},\psi)$ as the minimal risk, also called Bayes risk \citep{bartlett2006convexity}. Also, we define the empirical risk minimizer of $\widehat{\mathcal{L}_{U}}$ as
\$(\widehat {V}_{\lambda}^{\theta_1}, \widehat{\pi}_{\lambda}^{\theta_2}, \widehat \Psi^{\xi}, \widehat \psi^{\theta_3}) = \argmin_{(V_{\lambda}^{\theta_1},{\pi}_{\lambda}^{\theta_2},\Psi^{\xi}, \psi^{\theta_3}) \in \Theta_1 \times \Theta_2 \times \Xi \times \Theta_3}  \widehat{\mathcal{L}_{U}}(V_{\lambda}^{\theta_1},{\pi}_{\lambda}^{\theta_2},\Psi^{\xi}, \psi^{\theta_3}),
\$
where $\Theta_1, \Theta_2, \Xi$ and $\Theta_3$ are function spaces corresponding to $V_{\lambda}^{\theta_1}, {\pi}_{\lambda}^{\theta_2},\Psi^{\xi}$ and $\psi^{\theta_3}$, respectively. 

In Theorem \ref{risk_bound}, we establish the excess risk bound and convergence rate of $\mathcal{L}_{U}(\widehat {V}_{\lambda}^{\theta_1}, \widehat{\pi}_{\lambda}^{\theta_2}, \widehat \Psi^{\xi},\linebreak \widehat \psi^{\theta_3})  - \mathcal{L}^{*}_{U}$. The following assumption on the function space capacity is needed in order to develop the theoretical results.

 \begin{assump}\label{covering_num}
There exists a constant $\textbf{C} > 0$, and $q \in (0,2)$ such that for any $\varepsilon >0 $, $0 < \lambda \leq \lambda_{\max} < \infty $, $R_{\max}< \infty $, the following condition on metric entropy is satisfied,
\$
\log & \left( \max  \big\{ \mathcal{N}\left(\varepsilon, \Theta_1, \|\cdot\|_{\infty}\right) , \mathcal{N}\left(\varepsilon, \Theta_2, \|\cdot\|_{\infty}\right) ,  \mathcal{N}\left(\varepsilon, \Xi, \|\cdot\|_{\infty}\right), \mathcal{N}\left(\varepsilon, \Theta_3, \|\cdot\|_{\infty}\right) \big\} \right) \\
& \qquad \qquad \qquad  \qquad  \qquad  \qquad  \qquad  \leq  \textbf{C}\left(\frac{\max\{ \frac{4R_{\max} +(2-\gamma)\lambda_{\max}}{2-2\gamma},1\} }{\varepsilon}\right)^{q},
\$
where $\|\cdot\|_{\infty}$ denotes the supreme norm and $(V_{\lambda}^{\theta_1},{\pi}_{\lambda}^{\theta_2},\Psi^{\xi}, \psi^{\theta_3} ) \in \Theta_1 \times \Theta_2 \times \Xi \times \Theta_3$.
\end{assump}

Assumption \ref{covering_num} characterizes the complexity of the function spaces. In general, it is more difficult to estimate the functions as $\epsilon$ decreases. This assumption is satisfied in  the function spaces such as the RKHS and Sobolev space \citep{geer2000empirical,steinwart2008support}. Moreover, we compare this assumption with the assumption $3$ in \citep{antos2008learning} which considers parametric function spaces with finite effective dimension $D_{\mathcal{F}}$. Their assumption is equivalent to assuming $\log \mathcal{N}\left(\varepsilon, \Theta_1, \{\mathcal{D}_{i}\}^{n}_{i=1}\right) \leq D_{\mathcal{F}} \log \left({1}/{\varepsilon}\right)$, which is only able to account for the capacity of the finite dimension space. In contrast, our assumption is more general and allows our theorems can be applied to both finite and infinite dimension space.

\begin{thm}\label{risk_bound}
    Under Assumptions \ref{rbound_assum}-\ref{covering_num} and S.1-S.4 provided in Supplementary Materials Section I, for any $\delta \in (0,1]$, $q \in (0,2)$ and $\lambda \in (0,\lambda_{\max})$, the excess risk $\mathcal{E}(\mathcal{L}_{U}) \coloneqq \mathcal{L}_{U}(\widehat {V}_{\lambda}^{\theta_1},  \widehat{\pi}_{\lambda}^{\theta_2}, \widehat \Psi^{\xi}, \widehat \psi^{\theta_3}) - \mathcal{L}_{U}^{*} $ is upper bounded with probability at least $1-\delta$ for a sufficiently large $T$, that is
        \#
    \mathcal{E}(\mathcal{L}_{U}) & \leq \;  c_2n^{-1} +c_3 \sqrt{\log\left(\frac{2}{\delta}\right)}{n^{-\frac{1}{2}} }  +  J_1(\delta) \left( \frac{\log\log\left({2}^{-\frac{\delta_1}{1+\delta_1}}{c^{\frac{\delta_1}{1+\delta_1}}_{4}} {T}^{\frac{\delta_1}{1+\delta_1}}\right)}{{T}^{\frac{\delta_1}{1+\delta_1}}}  \right)^{1/2} \notag \\
 & \qquad  \qquad \qquad  \quad    +  J_2(\delta)\left({ {T}^{\frac{1-\delta_1}{(\delta_1+1)(2+q)}}}\right)  + J_3(\delta)  \left( \frac{\log\big({c_{5}}{T}^{\frac{2}{(1+\delta_1)}} \big) }{{T^{\frac{2}{2+q}}}} \right)^{1/2},
 \label{risk_detail}
    \#
    where the terms $\{J_1(\delta), J_2(\delta), J_3(\delta)\}$ are functions of $\delta, \delta_1, \textbf{C}, c_4, q, U_{\max}, \lambda_{\max}, R_{\max},\gamma$ and $\zeta$; additionally, the constants terms $\{c_2,c_3,c_4,c_5\}$ depend on $\delta, \delta_1, \textbf{C}, U_{\max}, \lambda_{\max}, R_{\max}$ and $\gamma$.
    
    If the constant and logarithmic terms are omitted, the excess risk can be simplified and achieves the rate
        \$
    \mathcal{E}(\mathcal{L}_{U})  \asymp \; \mathcal{O}\left( n^{-\frac{1}{2}}   \boldsymbol{\vee}  \left(	T^{\frac{\delta_1 - 1}{\delta_1 + 1}}\right)^{-\frac{1}{2+q}}    \boldsymbol{\vee} 
    \left(\frac{T^{\frac{\delta_1}{1+\delta_1}}}{\log\log(T^{\frac{\delta_1}{1+\delta_1}})}  \right)^{-\frac{1}{2}} \right).
    \$
\end{thm}
The first two terms in \eqref{risk_detail} control the statistical error from the empirical estimation over $n$ i.i.d. trajectories. The third and fourth terms bound the trajectory-based stochastic error from the variability inherent in weakly dependent within-trajectory samples. It observes that the third and fourth terms depends on the complexity of the function spaces $\Theta_1 \times \Theta_2 \times \Xi \times \Theta_3$. In addition, the last term is a remainder term due to using the \textit{block devices} in dealing with the sample dependency. In general, this remainder term converges to zero much faster than other main terms, especially when the sample dependency is weak in that $\delta_1$ is large. Hence, the remainder term does not affect the established convergence rate.

Theorem \ref{risk_bound} provides a more powerful result than just a consistency of estimation as it establishes a finite-sample upper bound guarantee for the risk.
In particular, if the process $\{ (S^{t},A^{t}) \}_{t \geq 1}$ forgets its past history sufficiently fast, i.e. $\delta_1 \rightarrow \infty$, then $T^{({\delta_1 - 1})/({\delta_1 + 1})}$ and $T^{{\delta_1}/{(1+\delta_1)}}$ both converge to $T$. Since the term $\log\log(T^{{\delta_1}/{(1+\delta_1)}})$ is a negligible to $\mathcal{O}(T^{{\delta_1}/{(1+\delta_1)}})$, we may achieve the optimal sample complexity upper bound $\mathcal{O}(n^{-{1}/{(2+q)}}) $  if $T$ is the same order of $n$. In particular, when the state space $\mathcal{S}$ is an open Euclidean ball in $\mathbb{R}^{d}$, and for a second-order Sobolev space $\mathbb{W}^{j,2}$ with $j > d/2$, one can choose $q = d/j$ to obtain the risk upper bound of the rate $\mathcal{O}(n^{-d/2(j + d)})$ if $T$ is the same order of $n$.
Besides establishing the risk bound in infinite dimensional function spaces,  Theorem \ref{risk_bound} can be also applied for the finite dimensional function spaces. That is, Theorem \ref{risk_bound} recovers the best upper error bound $n^{-1/2}$  when $q \rightarrow 0$.

In the following, we provide a finite sample upper bound on the performance error of the estimated optimal value function. Let $K(\cdot,\cdot)$ be a continuous integral strictly positive definite  kernel on the compact metric space $\mathcal{S} \times \mathcal{A}$, and there exists an orthonormal basis $e_1, e_2,...,$ for $L^2(\mathcal{S} \times \mathcal{A})$. Also, we let $\kappa_1,\kappa_2,...,$ be corresponding eigenvalues such that
$
K(\{s,a\}, \{\widetilde{s},\widetilde{a}\}) =\sum_{j=1}^{\infty} \kappa_{j} e_{j}(\{s,a\}) e_{j}( \{\widetilde{s},\widetilde{a}\})
$,
where $\{s,a\},\{\widetilde{s},\widetilde{a}\} \in \mathcal{S} \times \mathcal{A}$. The $L^2$ norm that  $\sqrt{\int f^{2}(s) \mathrm{d} d_{\pi_b}(s)}$ for the observed data distribution $d_{\pi_b}$ over $\mathcal{S}$ is denoted by  $\|f\|_{L^2}$ .

\begin{thm}\label{valuefunc_bound}
    Under Assumptions \ref{rbound_assum}-\ref{covering_num} and S.1-S.4 provided in Supplementary Materials Section I, for $\lambda \in (0,\lambda_{\max})$ and the finite minimum eigenvalue $\kappa_{\min} = \min\{\kappa_1,\kappa_2,...,\} < \infty$, the performance error between the estimated optimal value function $\widehat {V}_{\lambda}^{\theta_1}$ with respect to the proximal Bellman operator $\mathcal{B}_{\lambda}$ and the optimal value function ${V}^{\pi^{*}}$ with respect to the  standard Bellman operator $\mathcal{B}$ is upper bounded by
    \$
    \|  \widehat {V}_{\lambda}^{\theta_1} - {V}^{\pi^{*}} \|^2_{L^{2}} \leq \frac{c_{6}\mathcal{E}(\mathcal{L}_{U})}{\kappa_{\min}(1-\gamma)^2}  +  \frac{c_{7}\lambda^2}{(1-\gamma)^2}|\mathcal{A}| +  \frac{c_{8}\lambda^2|\mathcal{A}| + \lambda^2c_{9}}{(1-\gamma)^2|\mathcal{A}|},
    \$
    where $c_{6},c_{7},c_{8},c_{9}$ are some terms of $\delta_1, \textbf{C}, c_2,c_3,c_4,c_5, J_1(\delta),J_2(\delta), J_3(\delta), q, U_{\max},  \zeta, \lambda_{\max}, R_{\max}$ and $\gamma$ in which $c_{7} > c_{8}$.
\end{thm}

The above bound provides an insight regarding the performance error of the proposed method. Specifically, under the regularity conditions, the $L^2$ distance between the estimated optimal proximal value function $\widehat{V}_{\lambda}^{\theta_{1}}$ and the optimal value function ${V}^{\pi^{*}}$ is upper bounded, and this gap diminishes with the growth of sample size $n$ and time length $T$, and with the decay of the smoothing parameter $\lambda$.
Note that the last two terms is approaching to zero as $\lambda$ is sufficiently small. This implies that only a small smoothing bias is involved in the finite sample bound of $\| \widehat {V}_{\lambda}^{\theta_1} - {V}^{\pi^{*}} \|^2_{L_{2}}$. Moreover, combined with the sample bound for an excess risk $\mathcal{E}(\mathcal{L}_{U})$ in Theorem \ref{risk_bound}, Theorem \ref{valuefunc_bound} indicates that the behavior of the upper bound as a function of samples $\mathcal{O}(n^{-{1}/{(2+q)}})$ is the best if $T$ is the same order of $n$. In general,  a large $\lambda$ increases the smoothing error but decreases the approximation error as the solution function space is better behaved due to stronger smoothness. However, the approximation error is not reflected in our sample bound because we make the zero approximation error assumption, i.e., Assumption S.4 in Supplementary Materials.

\section{Implementation and Algorithm}\label{implement_algo}

For optimizing \eqref{em_proposed_loss}, the functions $(V^{\pi_{\lambda}}_{\lambda},\pi_{\lambda},{\Psi},\psi)$ are required to be parameterized by a class of models for practical implementations. One may parameterize $(V^{\pi_{\lambda}}_{\lambda},\pi_{\lambda},{\Psi},\psi)$ by $\{V^{\pi_{\lambda}}_{\lambda}(\cdot;\beta), \pi_{\lambda}(\cdot;\theta), \linebreak {\Psi}(\cdot;\omega), \psi(\cdot; \xi)\}$, where $(\theta, \beta,\omega,\xi)$ are associated parameters. Note that the policy (actor) $\pi_{\lambda}(\cdot;\theta)$ and the value function (critic) $V^{\pi_{\lambda}}_{\lambda}(\cdot; \beta)$ are separately represented by the two sets of parameters $(\theta,\beta)$. This parametrization representation is aligned with the \textit{actor-critic} paradigm in estimating the policy, which requires assistance and evaluation from the critic \citep{mnih2016asynchronous}. In our algorithm development, we introduce a unified  \textit{actor-critic} paradigm where the actor itself can play a critic role such that the actor can be self-supervised during the training process. We allow both $V^{\pi_{\lambda}}_{\lambda}$ and $\pi_{\lambda}$ to be represented in terms of $Q^{\pi_{\lambda}}_{\lambda}$ based on the connections in \eqref{bell_eq_smooth} and \eqref{sparse_pi}, respectively. In other words, it is sufficient to only parametrize $Q^{\pi_{\lambda}}_{\lambda}$ for  $V^{\pi_{\lambda}}_{\lambda}$ and  $\pi_{\lambda}$. Specifically, if we parametrize $Q^{\pi_{\lambda}}_{\lambda}$ by $ Q^{\pi_{\lambda}}_{\lambda}(\cdot; \theta)$, then $\{V^{\pi_{\lambda}}_{\lambda}, \pi_{\lambda}\}$ can be parametrized by the same set of parameter $\theta$, i.e. $\{ V^{\pi_{\lambda}}_{\lambda}(\cdot;\theta), \pi_{\lambda}(\cdot;\theta)\}$. One advantage of the new diagram is to reduce the parameter space, form two sets of parameters to one set, and hence relaxing the computational intensity. Another key advantage is that the new diagram only need to track the target policy $\pi_{\lambda}(\cdot;\theta)$ instead of tracking the non-stationary target $V^{\pi_{\lambda}}_{\lambda}(\cdot;\theta)$ which usually results in divergence issues.

For any state-action pair $(s,a)$, we follow the new unified \textit{actor-critic} framework to approximate $Q^{\pi_{\lambda}}_{\lambda}(s,a)$ using basis function approximations. The state-action function $Q^{\pi_{\lambda}}_{\lambda}(s,a; \theta)$ is represented by a linear combination of basis functions $\theta^\T \varphi(s,a)$, where $\theta$ is a $p_{Q}$-dimensional weight vector and $\varphi(s, a)$ is an column vector of non-linear basis functions computed at $(s,a)$. For $m = p_{Q}/|\mathcal{A}|$, the vector $\varphi(s,a)$ sets the basis function value $\varphi(s) = \big(\varphi_1(s), \varphi_2(s),...,\varphi_{m}(s)\big)$ in the corresponding slot for a specific action $a$,  while the values of basis function for the rest of the actions are set to be zero. That is 
$$
{\setstretch{1.25}
\varphi\left(s, a_{1}\right)=\left[\begin{array}{c}
\varphi_1(s) \\
... \\
\varphi_{m}(s) \\
\textbf{0}_{m(|\mathcal{A}|-1) \times1}
\end{array}\right],...,  \varphi\left(s, a_{k}\right)=\left[\begin{array}{c}
\textbf{0}_{m(k-1) \times1}\\
\varphi_1(s) \\
... \\
\varphi_{m}(s) \\
\textbf{0}_{m(|\mathcal{A}|-k) \times1}
\end{array}\right],..., \varphi\left(s, a_{|\mathcal{A}|}\right)=\left[\begin{array}{c}
\textbf{0}_{m(|\mathcal{A}|-1) \times1}\\
\varphi_1(s) \\
... \\
\varphi_{m}(s)
\end{array}\right],
}
$$
where $\textbf{0}$ is a zero vector. Similarly, we model the two Lagrangian functions as ${\Psi}(s;\omega) = \omega^\T\varphi(s)$ and $\psi(a|s; \xi) = \xi^\T\varphi(s, a)$. The flexibility of these working models can be achieved by selecting different $\varphi(\cdot)$, such as B-splines, radial basis and Fourier basis functions. Also note that $Q^{\pi_{\lambda}}_{\lambda}(s,a; \theta)$ can be parameterized by a non-linear approximation architecture and the optimization convergence is guaranteed in the pT-learning framework.

In the following, we reformulate the empirical risk minimization \eqref{em_proposed_loss} to a nonlinear programming problem for which the objective is converted to a quadratic term with a nonlinear equality constraint and two nonlinear inequality constraint,
\#
 &\min_{\theta,\omega,\xi} \;  \widehat{\mathcal{L}_{U}}(\theta,\omega,\xi) =  \  \frac{\zeta}{n}\sum_{i=1}^{n}  \big(D_{i}(\theta) -  P_{i}(\theta)  - Z_{i}(\omega)  + W_{i}(\xi)\big)^\T \Omega_i  \big(D_{i}(\theta) -  P_{i}(\theta)  - Z_{i}(\omega)  + W_{i}(\xi)\big) \notag \\
& \ \text{s.t.} \ \pi_{\lambda}(a|s;\theta) \cdot\psi(a|s; \xi)=0, \; \psi(a|s; \xi)\geq 0  \ \text{and} \ -\frac{\lambda}{2} \leq {\Psi}(s;\omega) \leq 0, \; \text{for all} \; s \in \mathcal{S}, a \in \mathcal{A},
\label{linear_working}
\#
where $\Omega_i \in  \mathbb R^{T \times T}$ is a weight matrix with $\{\Omega_i\}_{jk} = \, \big\{2\big( K( \{S_i^{j},A_i^{j}\}, \{ S_i^{k},{A}_i^{k}\}) - \mathds{1}(j=k)\big)/T(T-1)\big\}_{jk}$, and $D_{i}(\theta) ,  P_{i}(\theta), Z_{i}(\omega), W_{i}(\xi)$ are provided in Section 11, Appendix.

To solve this quadratic programming problem with non-linearity constraints, one may consider applying sequential quadratic programming \citep{fletcher2010sequential} to optimize the objective function with linearized constraints. However, this requires several  derivatives which are required to be solved analytically before iteration. Therefore, it can be quite cumbersome in practical implementations. Another approach is to apply exterior penalty methods \citep{boyd2004convex}, which convert a constrained problem to a series of unconstrained optimization problems. However, the size of constraints is $\mathcal{O}(|\mathcal{S}||\mathcal{A}|)$, and thus the computation is intensive if either $|\mathcal{S}|$ or $|\mathcal{A}|$ is large.

Alternatively, we propose a computationally more efficient algorithm, solving an unconstrained optimization via imposing certain restrictions on the representation of $\{Z_{i}(\omega), W_{i}(\xi)\}$ such that the two Lagrangian functions satisfy the constraints automatically. Although this re-parametrization may sacrifice certain model flexibility, it gains computational advantages as it solves a simpler quadratic optimization problem. More specifically, we parametrize $Z_{i}(\omega)$ by flipping a sigmoid function:
${\Psi}(s;\omega) = -({\lambda}/{2})\cdot \left({1 + \exp({-k_0({\omega^\T s - b_0)}})}\right)^{-1} \in  [-\lambda/2, 0],
$
where $b_0$ is the sigmoid's midpoint and $k_0$ is the logistic growth rate. Obviously, the first inequality constraint $-{\lambda}/{2} \leq {\Psi}(s;\omega) \leq 0$ is automatically satisfied under this parameterization. Observe that the $\pi_{\lambda}(a|s;\theta) > 0$ when $a \in \mathcal{K}(s)$, and $\pi_{\lambda}(a|s;\theta) =0$ otherwise. Therefore, we can further reduce our parameter space via modeling $\psi(a|s; \theta)$ as a function of $\theta$ instead of $\xi$. That is,
$
\psi(a|s; \theta) = \Big ( \frac{\sum_{a \in \mathcal{K}(s)} \theta^\T \varphi(s,a) - \lambda}{\lambda| \mathcal{K}(s)|} - \frac{ \theta^\T \varphi(s,a)}{\lambda}  \Big)^{+},
$
where the constraints $\pi_{\lambda}(a|s;\theta) \cdot\psi(a|s; \theta)=0$ and $\psi(a|s; \theta) \geq 0$ are always satisfied, and $W_{i}(\cdot)$ in \eqref{linear_working} are changed to  $W_{i}(\theta)$. The parameter $\xi$ is replaced by $\theta$, and hence the computational complexity is further reduced. The representation of $\widehat{\mathcal{L}_{U}}(\theta,\omega,\xi)$ under ${\Psi}(s;\omega)$ and $\psi(a|s; \theta)$, i.e., $\widehat{\mathcal{L}_{U}}(\theta,\omega)$, and its derivatives are provided in Section 11, Appendix.

\begin{algorithm}[H]
\setstretch{1.1}
	\caption{pT-Learning with Stochastic Gradient Descent}
	\label{SGD Algorithm}
	\begin{algorithmic}[1]
			\STATE \textbf{Input} observed data $\mathcal{D}_{1:n}$ as the transition pairs format $\{(S_i^{t},A_i^{t},R_i^{t},S_i^{t+1}): t=1,...,T\}^n_{i=1}$.
			\STATE \textbf{Initialize} the primary and auxiliary parameters $(\theta,\omega) = (\theta_0,\omega_0)$, the mini-batch size $n_0$, the learning rates $\alpha_{\theta} = \alpha_{\theta}^0 , \alpha_{\omega} = \alpha_{\omega}^0$, the scale parameter $\zeta =\zeta_0$, the factors $(\kappa_{e}=1,\kappa_{\pi}=1,\kappa_{\alpha}=1 )$, the sparsity parameter $\lambda = \lambda_0$, the bandwidth $\textit{bw}$ = $\textit{bw}_0$, and the stopping criterion $\varepsilon$.
			\STATE \textbf{For} $k=1$ to $k = \textit{max.iter}$
						\STATE \; \ Randomly sample a mini-batch $\{(S_i^{t},A_i^{t},R_i^{t},S_i^{t+1}): t=1,...,T\}^{n_0}_{i=1}$.
						\STATE \; \ Compute the gradient w.r.t. $\theta$ as
				\begin{center}
	 $
		\bar{\Delta}_{\theta} = \zeta_0\mathbb{P}_{n_0} [D_{i}(\theta) - P_{i}(\theta) + W_{i}(\theta) -Z_{i}(\omega)]^\T \Omega_i [\kappa_{e} \nabla_{\theta} D_{i}(\theta)  + \kappa_{\pi}  \nabla_{\theta} P_{i}(\theta) + \kappa_{\alpha} \nabla_{\theta} W_{i}(\theta) ].
					$
						\end{center}
				\STATE \; \  Compute the gradient w.r.t. $\omega$,  
							$		\bar{\Delta}_{\omega} = \zeta_0\mathbb{P}_{n_0}  [D_{i}(\theta) - P_{i}(\theta) + W_{i}(\theta) -Z_{i}(\omega)]^\T \Omega_i \nabla_{\omega} Z_{i}(\omega)$.				\STATE \;  \ Decay the learning rate $\alpha_{\theta}^{k} = \mathcal{O}(k^{-1/2})$, $\alpha_{\omega}^{k} = \mathcal{O}(k^{-1})$.
				\STATE \;  \ Update the parameters of interest as
									$\theta^{k} \leftarrow \theta^{k-1} - \alpha_{\theta}^{k}\bar{\Delta}_{\theta}, \,\omega^{k} \leftarrow \omega^{k-1} - \alpha_{\omega}^{k}\bar{\Delta}_{\omega}.$
									\STATE \;  \ Stop if $\|\theta^{k} - \theta^{k-1} \| \leq \varepsilon$.
			  \STATE \textbf{Return}	 $\widehat \theta = \theta^{k} $.
				\end{algorithmic}
\end{algorithm}

The minimization of $\widehat{\mathcal{L}_{U}}(\theta,\omega)$ requires $\mathcal{O}(nqT^2)$ time complexity in calculating the exact gradients. Here, we implement a stochastic gradient descent (SGD) algorithm, where the training is on mini-batch datasets and is faster than a vanilla gradient descent or BFGS algorithm. We summarize details of the proposed algorithm in Algorithm \ref{SGD Algorithm}, where the derivations of the gradients are provided in Section 11, Appendix.
In addition, the time complexity of calculating the gradient requires $\mathcal{O}(|\mathcal{A}|\log(|\mathcal{A}|)Tqn)$ by a naive sorting algorithm, and we can further improve the time complexity to $\mathcal{O}(|\mathcal{A}|Tqn)$ by utilizing the bucket-sorting algorithm \citep{blum1973time}.

\section{Simulation Studies}\label{simulation}

We conduct two numerical experiments to evaluate the finite sample performance of the proposed method. In the first experiment, we consider a binary treatment setting following a benchmark generative model \citep{luckett2020estimating, liao2020off}. In the second experiment, we mimic an mHealth cohort study aiming to deliver personalized interventions with $12$ choices for managing an individual's glucose level. In both experiments, we compare our approach to state-of-the-art methods including linear, polynomial and Gaussian V-learning \citep{luckett2020estimating}, and Greedy GQ-learning \citep{ertefaie2018constructing}. The proposed method is available in our \texttt{proximalDTR} R package.

In the first example, we let the current state ${S}_{i}^{t}=\left(S_{i, 1}^{t}, S_{i, 2}^{t}\right)^{T}$ be a two-dimensional vector and the current action $A_{i}^{t} \in \{1,0\}$. The next state ${S}_{i}^{t+1}$ is generated according to
$
{S}_{i, 1}^{t+1}=(3 / 4)(2 A_{i}^{t}-1) S_{i, 1}^{t}+(1 / 4) S_{i, 1}^{t} S_{i, 2}^{t}+\epsilon_{i,1}^{t}$ and $
S_{i, 2}^{t+1}=(3 / 4)(1-2 A_{i}^{t}) S_{i, 2}^{t}+(1 / 4) S_{i, 1}^{t} S_{i, 2}^{t}+\epsilon_{i,2}^{t},
$
where the random noises $\epsilon_{i,1}^{t+1}$ and $\epsilon_{i,2}^{t+1}$ follow an independent Gaussian distribution $N(0,0.5^2)$. Note that assigning $A_{i}^{t} = 1$ imposes a positive effect on ${S}_{i, 1}^{t+1}$ but has a negative effect on ${S}_{i, 2}^{t+1}$. We define a nonlinear utility function $u(S^{t+1}_{i}, S^{t}_{i},A^{t}_{i})$ such that
\$
R_{i}^{t} =  u(S^{t+1}_{i}, S^{t}_{i}, A^{t}_{i}) =  (1/4)(S_{i, 1}^{t+1})^3 +  2S_{i, 1}^{t+1} +  (1/2)(S_{i, 2}^{t+1})^3 + S_{i, 1}^{t+1} +(1/4)(2A_{i}^{t}-1).
\$
The initial state $S^{1}_{i}$ follows a Gaussian distribution $N(0,I_{2\times 2})$ and  $A_{i}^{t}$ is randomly assigned with an equal probability for each treatment arm as in micro-randomized trials.

We consider different scenarios in which the number of patients $n = \{25, 50,100\}$ and the follow-up time length $T = \{24,36,48\}$. The discount factor $\gamma$ is set to be $0.9$, focusing on long-term benefits. To specify the basis function $\varphi(s)$ mentioned in Section \ref{implement_algo}, we consider a cubic spline containing $6$ knots located in equal space of interval $[0,1]$, and then apply it to the state variables normalized between $0$ and $1$.

The objective function $\widehat{\mathcal{L}_{U}}(\theta,\omega)$ may not be convex with respect to both $\theta$ and $\omega$. Therefore, we adopt a multiple initialization method to determine an appropriate initial point. Specifically, we choose an initial point with the smallest objective value among $50$ randomly generated initial points.  In our two numerical experiments, we consider a fixed $\lambda=0.1$. Alternatively, we can also use a $k$-fold cross validation procedure to select sparsity parameter $\lambda$. We choose an optimal $\lambda$ which maximizes the lower bound of the empirical discounted sum of utilities, i.e., 
$
\widehat \lambda =   \argmax_{\lambda} \frac{1}{k}\sum^{k}_{r=1} \mathbb{P}_{n(r)} {\widehat{V^{\pi_{\lambda}}_{\lambda}}}^{(r)}(S^{1}) - \frac{\lambda \phi(0)}{{1-\gamma}},
$
where $\widehat{V^{\pi_{\lambda}}_{\lambda}}^{(r)}$ is the proximal value function estimator on the $r$th training set, and  $\mathbb{P}_{n(r)}$ is the empirical measure on the initial state $S_1$ for the $r$th validation set. Note that the cross-validation procedure may help to select a better $\lambda$ but it is computationally expensive. The cross-validation results, the tuning parameters set-up, and sensitivity analyses on different choices of $\lambda$ for model performance are provided in Section 13, Appendix.
To evaluate the model performance, we use the mean utility under the estimated policy as an evaluation criterion. Specifically, after obtaining the estimated policy, we simulate $100$ independent individual patients following this estimated policy over $100$ stages, and calculate the mean utility. Therefore, a larger value of the mean utility indicates a better policy. In Table \ref{simu1}, we report the average and standard deviations of the mean utilities over $50$ simulations.

\vspace{-3mm}

\begin{table}  [H]
\scriptsize	
  \caption{{Example 1: the average and standard deviation (in parentheses) of the mean utilities under the estimated optimal policy based on $50$ simulation runs.
     \vspace{0.2cm}} }\label{simu1}
		\renewcommand{\arraystretch}{1.3}
		\centering
\begin{tabular}{cc|ccccc|c}
\hline$n$ & $T$ & Proposed & Greedy-GQL & Linear VL & Poly VL & Gauss VL & Observed \\
\hline 25 & 24 & $0.3827(0.121)$ &$0.0787(0.175)$  &$0.2561(0.011)$ &$0.2564(0.011)$ & $0.2561(0.011)$ & $0.0033$ \\
& 36 & $0.4153(0.065)$ & $0.0716(0.234)$ & $0.2560(0.013)$& $0.2561(0.011)$ & $0.2558(0.014)$ & $0.0025$ \\
& 48 & $0.4001(0.078)$ & $0.0840(0.213)$& $0.2578(0.012)$ & $0.2578(0.014)$ &$0.2575(0.012)$ & $0.0033$\\
\hline 50 & 24 & $0.4154(0.080)$ & $0.0844(0.209)$  &$0.2564(0.012)$ & $0.2569(0.012)$ & $0.2564(0.012)$ & $0.0092$ \\
& 36 & $0.4029(0.093)$ & $0.0829(0.235)$  & $0.2570(0.010)$ & $0.2574(0.012)$ & $0.2569(0.011)$ &$0.0080$ \\
& 48 & $0.3888(0.084)$ & $0.0836(0.236)$& $0.2564(0.012)$ & $0.2570(0.012)$ & $0.2530(0.026)$ & $0.0036$\\
\hline 100 & 24 & $0.4290(0.059)$ & $0.0875(0.240)$ & $0.2529(0.016)$ & $0.2551(0.011)$ & $0.2547(0.011)$ & $0.0041$ \\
& 36 &  $0.4050(0.062)$ & $0.0780(0.266)$ & $0.2538(0.010)$ & $0.2542(0.010)$ & $0.2535(0.010)$  & $0.0043$ \\
& 48 & $ 0.4112(0.052)$ & $0.0945(0.254)$ & $0.2548(0.011)$ & $0.2552(0.015)$ & $0.2543(0.017)$  & $0.0090$ \\
\hline
\end{tabular}
\end{table}

\vspace{-3mm}

Table \ref{simu1} shows that the proposed method outperforms the competing methods in all scenarios. When $n=100$ and  $T= 24$, compared to the baseline, i.e.,  the observed mean utility, the proposed method improves $0.4249$ while the polynomial V-learning and Greedy GQ-learning only improve $0.2510$ and $0.0834$, respectively. This is mainly because the proposed method does not impose restrictions on the class of policies as in V-learning. Also, the induced policy can be automatically adjusted between deterministic and stochastic policy models. In addition, the proposed method utilizes the smoothed proximal Bellman operator, and thus avoids discontinuity and instability. Moreover, the proposed method has advantages in off-policy training due to utilizing the proximal temporal consistency property. In contrast, Greedy GQ-learning tends to solve the hard Bellman optimality equation, which could be difficult without large amounts of samples. In terms of computations, pT-Learning also achieves great gains over V-learning. The actual computation time tables are provided in Section 13 of the Appendix.

In the second example, we simulate cohorts of patients with type 1 diabetes to mimic the mobile health study \citep{maahs2012outpatient}. The study aims to achieve long-term glycemic control and maintain the rest of the health index in the desired ranges. Specifically, our hypothetical mHealth study targets to searching the optimal policy for suggesting multi-channel interventions; the insulin injection (IN), physical activity (PA), and dietary intake (DI), to control the blood glucose (BG) level in the desired range while maintaining healthy levels of Adiponectin (AD) and blood pressure (BP)  \citep{fidler2011hypoglycemia}. Hence, the immediate utility $R_{i}^{t}$ is defined by a weighted summation of patients' health status as following,
\vspace{-4mm}
   \$
 	R_{i}^{t} =& \, \alpha_1\mathbbm{1}(  70 \leq \text{BG}_{i}^{t+1} \leq 120 )
+ \alpha_2\mathbbm{1}(\text{BG}_{i}^{t+1} < 70 \ \text{or} \ \text{BG}_{i}^{t+1} > 150 )  +\alpha_3\mathbbm{1}( 120 \leq \text{BG}_{i}^{t+1} \leq 150 ) \\
   &+  \alpha_4\mathbbm{1}(5 \leq \text{AD}_{i}^{t+1} \leq 23) + \alpha_5\mathbbm{1}(\text{AD}_{i}^{t+1} < 5 \ \text{or} \ \text{AD}_{i}^{t+1} >23) +    \alpha_6\mathbbm{1}(66 \leq \text{BP}_{i}^{t+1} \leq 80) \\
   & + \alpha_7\mathbbm{1}(\text{BP}_{i}^{t+1} < 66 \ \text{or} \ \text{BP}_{i}^{t+1} > 80),
 \$
 
 \vspace{-4mm}
 
 \noindent where $(\alpha_1,...,\alpha_7) = (3,-3,-1,2,-1,2,-1)$ are weights reflecting the clinical consequences. 

The patients are assigned treatment from a combination of the insulin injections (Yes/No), physical activity (No/Moderate/Strong) and dietary intake (Yes/No). Note that there are total $12$ different treatment choices, i.e., $\mathcal{A} = \{1,...,12\}$. The details of the $12$ different treatment combinations are enumerated in Section 13, Appendix. At each stage, the treatment $A_i^{t} \in \mathcal{A}$ is randomly assigned with equal probability for each treatment arm. The patient's state $(\text{BG}_{i}^{t},\text{AD}_{i}^{t},\text{BP}_{i}^{t})$ evolves according to the given dynamic model:
$
\text{BG}_{i}^{t+1} = \gamma_{11}\mu_{\text{BG}} +  \gamma_{12}\text{BG}_{i}^{t} - \gamma_{13} \text{AD}_{i}^{t}  + \sum_{a \in \mathcal{A}} \mu_{1a} \mathbbm{1}(A_i^t  = a)+ \epsilon_{1i}^{t+1};
\text{AD}_{i}^{t+1} =  \gamma_{21}\mu_{\text{AD}} + \gamma_{22} \text{AD}_{i}^{t} + \gamma_{23} \text{BG}_{i}^{t}   + \sum_{a \in \mathcal{A}} \mu_{2a} \mathbbm{1}(A_i^t  = a) + \epsilon_{2i}^{t+1};
\text{BP}_{i}^{t+1} =  \gamma_{31}\mu_{\text{BP}} + \gamma_{32} \text{BP}_{i}^{t} + \sum_{a \in \mathcal{A}} \mu_{3a} \mathbbm{1}(A_i^t  = a)  + \epsilon_{3i}^{t+1},
$
where $\epsilon_{1i},\epsilon_{2i}$ and $\epsilon_{3i}$ are individual-level Gaussian random noises. The values of coefficients in the generative model are provided in Section 13, Appendix.

  \begin{table}  [H]
  	\vspace{-2mm}
\scriptsize	
  \caption{{Example 2 (mHealth study): the average and standard deviation (in parentheses) of the mean utilities under the estimated optimal policy based on $50$ simulation runs.
     \vspace{0.2cm}} }\label{simu2}
		\renewcommand{\arraystretch}{1.3}
		\centering
\begin{tabular}{cc|ccccc|c}
\hline$n$ & $T$ & Proposed & Greedy-GQL & Linear VL & Poly VL & Gauss VL  & Observed \\
\hline 25 & 24 & $2.489(0.631)$ &$1.561(1.034)$& $1.743(0.748)$ & $1.555(0.652)$& $1.787(0.784)$ & $1.481$ \\
& 36 & $ 2.878(0.568)$ &$1.500(1.244)$ &$1.694(0.702)$ & $1.725(0.705)$ & $1.770(0.674)$ & $1.473$\\
\hline 75 & 24 & $2.685(0.268)$ & $1.588(1.310)$& $1.667(0.482)$ & $1.724(0.862)$ & $1.885(0.898)$ & $1.464$ \\
& 36 & $ 3.247(0.510)$ &$1.772(1.277)$ & $1.713(0.423)$ & $2.162(0.954)$ & $1.918(0.894)$ & $1.478$\\
\hline 100 & 24 & $2.785(0.401)$ &$1.720(1.232)$ & $1.561(0.317)$ & $1.849(0.904)$ & $1.981(0.573)$ & $1.471$ \\
& 36 & $3.409(0.429)$ & $1.854(1.453)$&$1.768(0.622)$ & $2.105(0.912)$ & $2.177(1.086)$ & $1.473$ \\
\hline
\end{tabular}
\vspace{-4mm}
 \end{table}
 
The cohort data is generated under different scenarios with a follow-up time length $T=\{24,36\}$ and sample size $n=\{25, 75,100\}$. The discount factor $\gamma$ is set to be $0.9$. We use the same evaluation criterion as in Simulation Example 1 except that the testing set is simulated over $36$ stages. The results are summarized in Table \ref{simu2}. The pT-Learning approach achieves the best performance. For example, when $n=100$ and $T=36$, pT-Learning achieves $56.7\%$ and $83.8\%$ improvements compared to Gaussian V-learning and Greedy GQ-learning, respectively. The improvement of pT-Learning compared to V-learning is due to the sparse property of pT-Learning. For a better illustration, we visualize the sparsity and compare the estimated policy distribution of pT-Learning and V-learning in Section 13 of the Appendix. On the other hand, pT-Learning is better than Greedy GQ-learning mainly because Greedy GQ-learning requires modeling the entire data-generating process, and hence produces a large over-estimation error. Worst of all, the Greedy GQ-learning approach is designed to take the greedy action based on the fitted model. This again magnifies the over-estimation error and ultimately leads to a huge sub-optimality. In contrast, pT-Learning only requires modeling the optimal policy and optimal value function. Additionally, pT-Learning does not take a greedy action, but follows an adaptive and sparse stochastic policy model.

\section{Application to Ohio Type 1 Diabetes Data}\label{real_data}

We apply the proposed method to two cohorts of individuals from the OhioT1DM dataset \citep{marling2020ohiot1dm}, which was used to study the long-term blood glucose management via the just-in-time interventions on the type 1 diabetes patients. Each cohort contains $6$ individuals of age ranging within 20 to 60. All patients were on insulin pump therapy with continuous glucose monitoring (CGM) sensors, and the life-event data is collected via a mobile phone app.  The physiological data of the first cohorts was collected by \textit{Basis} sensor bands, and the second cohort used \textit{Empatica} sensor bands. Specifically, the dataset includes CGM blood glucose levels measured per 5 minutes, insulin dose levels delivered to the patient, meal intakes, and corresponding carbohydrate estimates. Note that the heart rate measured per 5 minutes is available only for patients using the \textit{Basis} sensor band (the first cohort), while the magnitude of acceleration aggregated per minute is only available with the \textit{Empatica} sensor band (the second cohort). The data also include other features such as the self-reported times of work, sleep and stress, etc. Based on the preliminary investigation, each patient has distinct blood glucose dynamics. Therefore, we follow the pre-processing strategy used in \cite{zhu2020causal}, and treat the data of each patient as a single dataset. Then we estimate the optimal policy by treating each day as an independent sample. For the first cohort, we consider a binary intervention setting, i.e., whether or not to provide the insulin injection; and for the second cohort, we study the individualized dose-finding problem by discretizing the continuous dose level into $14$ disjoint intervals for intervention options. We compare the performance of pT-Learning with the same competing methods as in Section \ref{simulation}.

We summarize the collected measurements over $60$-min intervals such that the length of each trajectory is $T=24$. After removing missing samples and outliers, each dataset contains $n=15$ trajectories on average. For the first six patients who wore the \textit{Basis} sensor band, the patient's states at each stage include the average blood glucose levels $S^{t}_{i, 1}$, the average heart rate $S^{t}_{i, 2}$ and the total carbohydrates $S^{t}_{i, 3}$ intake from time $t-1$ and $t$. For others equipped with the \textit{Empatica} sensor band, the states are the same as the first six patients, except that the average heart rate is substituted by the average magnitude of acceleration. Here, the utility is defined as the average of the index of glycemic control \citep{rodbard2009interpretation} between time $t-1$ and $t$, measuring the health status of the patient's glucose level. That is
$
{R^{t}_{i}}=-\frac{\mathbbm{1}({S^{t}_{i,1}}>140)|{S^{t}_{i,1}}-140|^{1.35} + \mathbbm{1}({S^{t}_{i,1}}<80)({S^{t}_{i,1}} - 80)^{2}}{30},
$
where $R^{t}_{i}$ is non-positive and a larger value is preferred. Our goal is to maximize the expected discounted sum of utilities $\mathbb E_{\pi} \sum_{t\geq 1}\gamma^{t-1}R^{t}$. In the first cohort study, the treatment is binary, i.e. $A^t_{i} \in \{0,1\}$. In the second study, we provide the optimal insulin dose suggestion via the uniform discretization of the continuous dose level, i.e., $A^{t}_{i} \in \left\{0=A_{(1)}<\ldots<A_{\left(14\right)}=\max({A})\right\}$.

Since the data-generating process is unknown, it is hard to use the mean utility under the estimated policy as the evaluation criterion as in Section \ref{simulation}. Instead, we follow \cite{luckett2020estimating} to utilize the Monte Carlo approximation of the expected discounted sum of utilities for evaluating the model performance, i.e. $\mathbb{P}_n\widehat{V^{\pi}}(S^{1}_{i})$, where $S^{1}_i$ is the initial state for the $i$th trajectory. In the competing methods, the quantity $\widehat{V^{\pi}}(\cdot)$ represents the estimated value function. In our method, we consider the lower bound of our estimated value function $\widehat{V^{\pi_{\lambda}}_{\widehat{\lambda}}}(\cdot)$, i.e., $\mathbb{P}_n\widehat{V^{\pi_{\lambda}}_{\widehat{\lambda}}}(S^{1}_{i}) - (1-\gamma)^{-1}\widehat{\lambda} \phi(0)$, to mitigate the effects of the sparsity parameter $\widehat{\lambda}$ on the utilities. This quantity can be also interpreted as the worst case of the discounted sum of utilities produced by pT-Learning. The discounted sum of observed utilities, i.e, $\mathbb{P}_n \sum_{t \geq 1}\gamma^{t-1}R^{t}_i$, is used as the baseline.

\begin{figure}[h]
	\centering
		\scalebox{0.28}[0.26]{\includegraphics{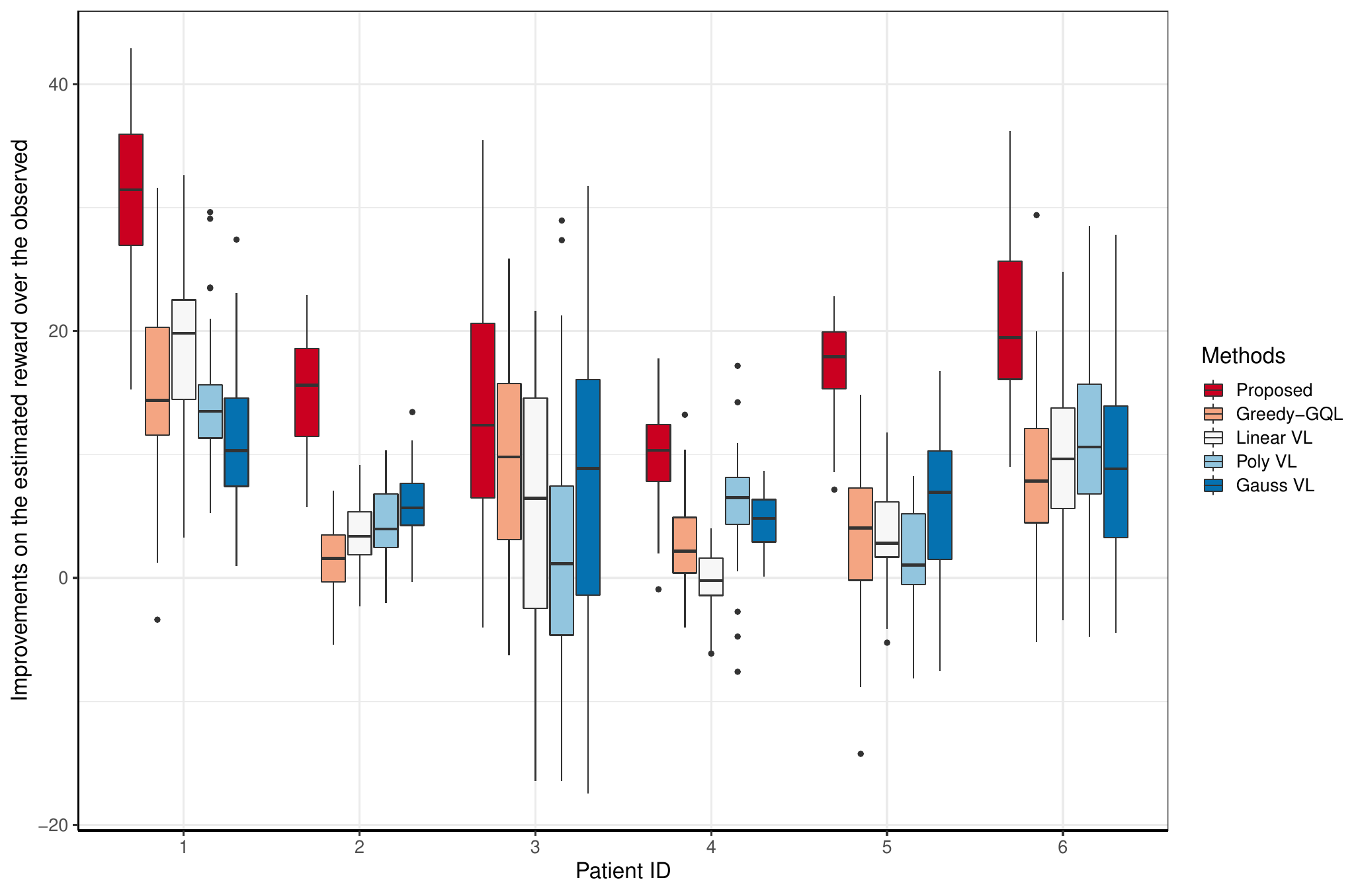}}
		\vspace{-2mm}
	\caption{The first cohort patients: boxplots of the improvements on the discounted sum of utilities under estimated policy over $50$ simulation runs, with $|\mathcal{A}|=2$ and $\gamma=0.9$.}\label{binary09}
		\vspace{-4mm}
\end{figure}

\begin{figure}[h]
\vspace{-4mm}
	\centering
\scalebox{0.28}[0.26]{\includegraphics{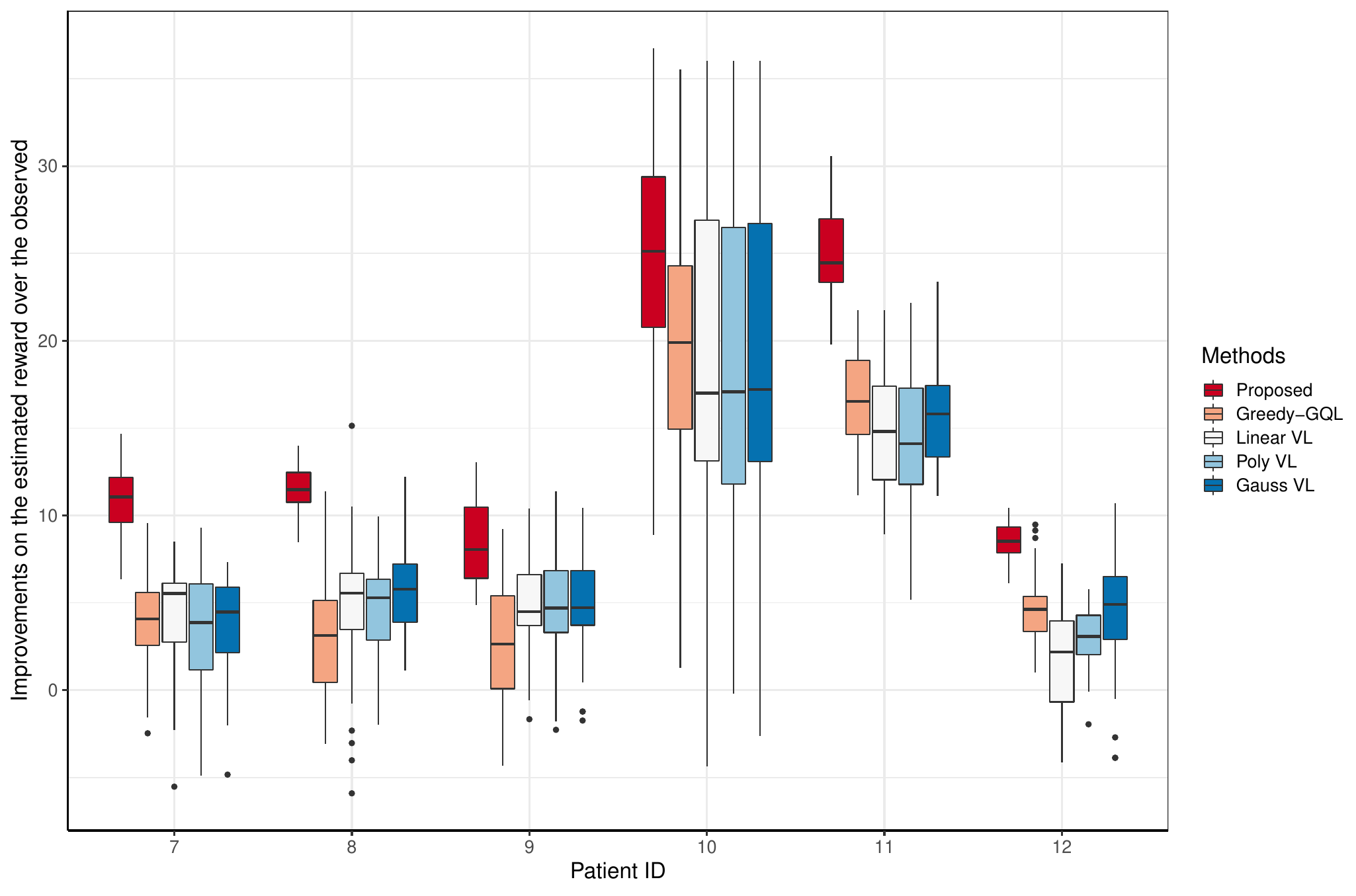}}
			\vspace{-2mm}
	\caption{The second cohort patients: boxplots of the improvements on the discounted sum of utilities under estimated policy over $50$ simulation runs, with $|\mathcal{A}|=14$ and $\gamma=0.9$.}\label{multi09}
	\vspace{-4mm}
\end{figure}

In our experiments, we choose two discount factors, $\gamma = 0.9$ and $\gamma = 0.8$. For the $\gamma = 0.9$ setting, the boxplots of the relative improvements of utilities are provided in Figure \ref{binary09}-\ref{multi09} for the first and second cohort data, respectively. Additional results are presented in Section 13, Appendix. Figure \ref{binary09} indicates that pT-Learning achieves better improvements over competing methods across all patients. For example, for the Patient $1$, the proposed method has $98.0\%$ and $51.2\%$ improvement rates compared to Greedy GQ-learning and the linear V-learning, respectively. This is mainly because pT-Learning does not impose restrictions on the class of policies and thus is more flexible. The standard deviation of the proposed method is the smallest among all approaches, reflecting that pT-Learning is relatively stable due to the smoothness property of the proximal Bellman operator and the implemented unified \textit{actor-critic} framework. For the second cohort study with a large cardinality treatment space, pT-Learning substantially outperforms the competing methods in Figure \ref{multi09}. For the $7$th patient, the improvement of pT-Learning to Greedy GQ-learning and Gaussian V-learning attains $175.4\%$ and $187.0\%$, respectively. This result shows that pT-Learning achieves high efficiency in the continuous treatment space, and our sparse policy estimation has a clear benefit with large numbers of treatment options.

\section{Discussion}\label{discussion}

In this paper, we propose a novel proximal temporal consistency learning framework for estimating the optimal dynamic treatment regime in infinite time horizon settings. The constructed proximal Bellman operator directly leads to a smoothed Bellman optimality equation, while simultaneously inducing a sparse optimal policy. The proposed minimax policy estimator resolves the double sampling issue and can be easily optimized by a scalable and efficient SGD algorithm.

Several improvements and extensions are worth exploring in the future. First, we may extend our algorithm to deal with strong temporal dependency. The idea of the experience-replay \citep{mnih2015human} might be useful. It is shown that the gradients calculated by the experience-replay algorithm are $1-\mathcal{O}(n^{-1})$-nearly independent. Second, it is interesting to extend pT-Learning to the long-term average reward setting \citep{murphy2016batch}. Third, developing statistical inference methods for quantifying uncertainty of the policy and value function is also important. 

Under the non-stationary learning setting where the environment varies over time, a relatively high exploration is typically preferred, as it may lead to a better policy estimation in the long run. Therefore, the V-learning method potentially gains more benefits than pT-Learning in policy estimations due to a higher exploration rate. To improve the exploration ability of pT-Learning, one may consider to adopt the $\varepsilon$-greedy strategy \citep{sutton1998introduction}. Other future directions include developing a rigorous extension to continuous action space with theoretical justifications, and constructing a state-varying $\lambda(s)$ which results in a group-wise smoothness and sparsity. The detailed discussion of the aforementioned extensions is provided in Appendix.

\section{Acknowledgements}
The authors thank the Editor,  Associate Editor and the anonymous reviewers for their insightful suggestions and helpful feedback which improved the paper significantly. The authors declare no financial or non-financial interest that has arisen from the direct applications of this research. This work is supported by NSF grants DMS 2210640, DMS 1952406 and DMS 2210657.

\section{Supplementary Materials}
The online supplementary materials provide all technical proofs of main theorems as well as the additional theoretical results. 

\bibliographystyle{asa}
\bibliography{pT_learning}

\newpage 

\begin{center} 
\huge{``Appendix''}
\end{center}

\section{Implementation Details and Gradients in Algorithm}\label{supp_implement}
In this section, we provide some details in the implemented algorithm in Section \ref{implement_algo} of the main text. 
\$
\widehat{\mathcal{L}_{U}}(\theta,\omega, \xi) \coloneqq  \frac{\zeta}{n}\sum_{i=1}^{n}  \big(D_{i}(\theta) -  P_{i}(\theta)  - Z_{i}(\omega)  + W_{i}(\xi)\big)^\T \Omega_i  \big(D_{i}(\theta) -  P_{i}(\theta)  - Z_{i}(\omega)  + W_{i}(\xi)\big),
\$
where 
\begin{align*}
& \big\{\Omega_i\big\}_{jk} = \, \bigg\{2\big( K( \{S_i^{j},A_i^{j}\}, \{ S_i^{k},{A}_i^{k}\}) - \mathbbm{1}(j=k)\big)/T(T-1)\bigg\}_{jk}, \, \text{for} \, j,k=1,...,T. \\ 
&D_{i}(\theta) = \, \Bigg \{R^{j}_{i} + \frac{1}{2\lambda} \bigg[ \theta^\T \bigg(\sum_{a^{\prime}  \in \mathcal{K}(S_i^{j+1})} \gamma \varphi(S_i^{j+1},a^{\prime} ) \varphi(S_i^{j+1},a^{\prime} )^{\T} -  \sum_{a^{\prime} \in \mathcal{K}(S_i^{j})} \varphi(S_i^{j},a^{\prime} ) \varphi(S_i^{j},a)^{\T}\bigg) \theta  \\
& +   \frac{\bigg(\sum_{a^{\prime}  \in \mathcal{K}(S_i^{j})} \theta^\T \varphi(S_i^{j},a^{\prime} ) - \lambda \bigg)^2}{|\mathcal{K}(S_i^{j})|} -  \frac{\gamma \bigg(\sum_{a^{\prime}  \in \mathcal{K}(S_i^{j+1})} \theta^\T \varphi(S_i^{j+1},a^{\prime} ) - \lambda \bigg)^2}{|\mathcal{K}(S_i^{j+1})|} + (\gamma - 1)\lambda^2\bigg] \Bigg \}_{j=1}^{T} \in \mathbb R^{T \times 1}, \\
 &P_{i}(\theta) = \,  \Bigg \{ \bigg (\frac{\theta^\T \varphi(S_i^{j},A_i^{j})}{\lambda} - \frac{\sum_{a \in \mathcal{K}(S_i^{j})} \theta^\T \varphi(S_i^{j},a^{\prime} ) - \lambda}{\lambda| \mathcal{K}(S_i^{j})|} \bigg)^{+} \Bigg \}_{j=1}^{T} \in \mathbb R^{T \times 1}, \, \\
&W_{i}(\xi) = \Bigg \{ \xi^\T\varphi(S_i^{j},A_i^{j}) \Bigg \}_{j=1}^{T}, \\
&Z_{i}(\omega) = \left \{\omega^\T\varphi(S_i^{j}) \right\}_{j=1}^{T} .
\end{align*}
Following the proposed computational efficient algorithm and representations in the main text, we aim to solve an unconstrained minimization problem with the objective function
\$
\widehat{\mathcal{L}_{U}}(\theta,\omega) \coloneqq  \frac{\zeta}{n}\sum_{i=1}^{n}  \big(D_{i}(\theta) -  P_{i}(\theta)  - Z_{i}(\omega)  + W_{i}(\theta)\big)^\T \Omega_i  \big(D_{i}(\theta) -  P_{i}(\theta)  - Z_{i}(\omega)  + W_{i}(\theta)\big),
\$
where 
\begin{align*}
& \big\{\Omega_i\big\}_{jk} = \, \bigg\{2\big( K( \{S_i^{j},A_i^{j}\}, \{ S_i^{k},{A}_i^{k}\}) - \mathbbm{1}(j=k)\big)/T(T-1)\bigg\}_{jk}, \, \text{for} \, j,k=1,...,T. \\ 
&D_{i}(\theta) = \, \Bigg \{R^{j}_{i} + \frac{1}{2\lambda} \bigg[ \theta^\T \bigg(\sum_{a^{\prime}  \in \mathcal{K}(S_i^{j+1})} \gamma \varphi(S_i^{j+1},a^{\prime} ) \varphi(S_i^{j+1},a^{\prime} )^{\T} -  \sum_{a^{\prime} \in \mathcal{K}(S_i^{j})} \varphi(S_i^{j},a^{\prime} ) \varphi(S_i^{j},a)^{\T}\bigg) \theta  \\
& +   \frac{\bigg(\sum_{a^{\prime}  \in \mathcal{K}(S_i^{j})} \theta^\T \varphi(S_i^{j},a^{\prime} ) - \lambda \bigg)^2}{|\mathcal{K}(S_i^{j})|} -  \frac{\gamma \bigg(\sum_{a^{\prime}  \in \mathcal{K}(S_i^{j+1})} \theta^\T \varphi(S_i^{j+1},a^{\prime} ) - \lambda \bigg)^2}{|\mathcal{K}(S_i^{j+1})|} + (\gamma - 1)\lambda^2\bigg] \Bigg \}_{j=1}^{T} \in \mathbb R^{T \times 1}, \\
 &P_{i}(\theta) = \,  \Bigg \{ \bigg (\frac{\theta^\T \varphi(S_i^{j},A_i^{j})}{\lambda} - \frac{\sum_{a \in \mathcal{K}(S_i^{j})} \theta^\T \varphi(S_i^{j},a^{\prime} ) - \lambda}{\lambda| \mathcal{K}(S_i^{j})|} \bigg)^{+} \Bigg \}_{j=1}^{T} \in \mathbb R^{T \times 1}, \, \\
&W_{i}(\theta) = \Bigg \{ \bigg ( \frac{\sum_{a^{\prime} \in \mathcal{K}(S_i^{j})} \theta^\T \varphi(S_i^{j},a^{\prime}) - \lambda}{\lambda| \mathcal{K}(S_i^{j})|} - \frac{ \theta^\T \varphi(S_i^{j},A_i^{j})}{\lambda}  \bigg)^{+}  \Bigg \}_{j=1}^{T}  \in \mathbb {R^{+}}^{T \times 1}, \\
&Z_{i}(\omega) = \left \{\frac{ -{\lambda}/{2} }{1 + \exp\left({-k_0({\omega^\T S_i^{j} - b_0)}}\right)} \right\}_{j=1}^{T} \in  \left[-\frac{\lambda}{2}, 0\right]^{T \times 1}. \\
& \, \text{Here} \ b_0 \ \text{is the sigmoid’s midpoint and} \ k_0 \ \text{is the logistic growth rate.}
\end{align*}
First, we derive the row vector of the gradient of $\widehat{\mathcal{L}_{U}}(\theta,\omega)$ over $\theta$, i.e., $\nabla_{\theta}\widehat{\mathcal{L}_{U}}(\theta,\omega) $. Taking the derivative, we have that 
\#
\nabla_{\theta}\widehat{\mathcal{L}_{U}}(\theta,\omega) &=   \frac{\zeta}{n}\sum_{i=1}^{n} \big(D_{i}(\theta) -  P_{i}(\theta)  - Z_{i}(\omega)  + W_{i}(\theta)\big)^\T \Omega_i \nabla_{\theta}\big\{D_{i}(\theta) -  P_{i}(\theta)  - Z_{i}(\omega)  + W_{i}(\theta)\big\} \notag \\ 
& = \frac{\zeta}{n}\sum_{i=1}^{n} \big(D_{i}(\theta) -  P_{i}(\theta)  - Z_{i}(\omega)  + W_{i}(\theta)\big)^\T \Omega_i \big(\nabla_{\theta}D_{i}(\theta) -  \nabla_{\theta}P_{i}(\theta) +\nabla_{\theta} W_{i}(\theta)\big).
\label{theta_eq}
\#
Next, we derive the exact forms for $\nabla_{\theta}D^{j}_{i}(\theta),   \nabla_{\theta}P^{j}_{i}(\theta)$ and $\nabla_{\theta} W^{j}_{i}(\theta)$ ( $j=1,...,T$) separately. 
\$
\underbrace{\nabla_{\theta}D^{j}_{i}(\theta)}_{1\times p_{\theta}} & = \frac{1}{\lambda} \bigg[\theta^\T \bigg(\sum_{a^{\prime}  \in \mathcal{K}(S_i^{j+1})} \gamma \varphi(S_i^{j+1},a^{\prime} ) \varphi(S_i^{j+1},a^{\prime} )^{\T} -  \sum_{a^{\prime} \in \mathcal{K}(S_i^{j})} \varphi(S_i^{j},a^{\prime} ) \varphi(S_i^{j},a)^{\T}\bigg)  \\
& +   \frac{\big(\sum_{a^{\prime}  \in \mathcal{K}(S_i^{j})} \theta^\T \varphi(S_i^{j},a^{\prime} ) - \lambda \big)\big(\sum_{a^{\prime}  \in \mathcal{K}(S_i^{j})}\varphi(S_i^{j},a^{\prime} )^{\T} \big)}{|\mathcal{K}(S_i^{j})|} \\ 
& -\frac{\gamma \big(\sum_{a^{\prime}  \in \mathcal{K}(S_i^{j+1})} \theta^\T \varphi(S_i^{j+1},a^{\prime} ) - \lambda \big)\big(\sum_{a^{\prime}  \in \mathcal{K}(S_i^{j+1})}\varphi(S_i^{j+1},a^{\prime} )^{\T}   \big)}{|\mathcal{K}(S_i^{j+1})|} \bigg],
\$
where $p_{\theta}$ is the dimension of $\theta$ and $\mathcal{K}(s;\theta) = \{a_{(i)} \in \mathcal{A}: \theta^{\T}\varphi(s,a_{(i)} ) > \frac{1}{i}\sum^{i}_{m=1}\theta^{\T}\varphi(s,a_{(m)} ) - \frac{\lambda}{i} \}$ is the support set. Note that $\nabla D_{i}(\theta)$ is differential everywhere (the derivative exists) except for (a small number of) splitting points where $\mathcal{K}(s;\theta) \neq \mathcal{K}(s;\theta+\boldsymbol{\varepsilon})$ for some  infinitesimally $p_{\theta}$-dimensional $\boldsymbol{\varepsilon}$. In the numerical implementation,  we use the current $\theta$ to evaluate the support set for calculating the derivative on those splitting points. Alternatively, we can also take an arbitrary matrix in the set of generalized Clarke’s Jacobians \citep{clarke1990optimization} on the splitting points. As the number of splitting points is few, using the current $\theta$ to evaluate the support set does not affect the stability of the optimization very much. Therefore, we implement the proposed method following the first approach for simplicity and leave the second approach as an alternative. \\
Next, we proceed to derive the gradient $\nabla_{\theta}P_{i}(\theta)$: 
\$
\underbrace{\nabla_{\theta}P^{j}_{i}(\theta)}_{1\times p_{\theta}} =\left\{
\begin{aligned}
\frac{\varphi^{\T}(S_i^{j},A_i^{j})}{\lambda} - \frac{\sum_{a^{\prime}  \in \mathcal{K}(S_i^{j})}\varphi^{\T}(S_i^{j},a^{\prime})}{\lambda|\mathcal{K}(S_i^{j}) |} & , &  \frac{ \theta^{\T}\varphi(S_i^{j},A_i^{j})}{\lambda} > \frac{\sum_{a^{\prime}  \in \mathcal{K}(S_i^{j})}\theta^{\T}\varphi(S_i^{j},a) - \lambda}{\lambda|\mathcal{K}(S_i^{j}) |} ; \\
0\quad\quad\quad\quad \quad \quad \quad \quad & , &\frac{ \theta^{\T}\varphi(S_i^{j},A_i^{j})}{\lambda} \leq  \frac{\sum_{a^{\prime}  \in \mathcal{K}(S_i^{j})}\theta^{\T}\varphi(S_i^{j},a) - \lambda}{\lambda|\mathcal{K}(S_i^{j}) |}.
\end{aligned}
\right.
\$
According to $\nabla_{\theta}P^{j}_{i}(\theta)$, it is easy to derive $\nabla_{\theta} W^{j}_{i}(\theta)$:
\$
\underbrace{\nabla_{\theta}W^{j}_{i}(\theta)}_{1\times p_{\theta}} =\left\{
\begin{aligned}
 \frac{\sum_{a^{\prime}  \in \mathcal{K}(S_i^{j})}\varphi^{\T}(S_i^{j},a^{\prime})}{\lambda|\mathcal{K}(S_i^{j}) |} - \frac{\varphi^{\T}(S_i^{j},A_i^{j})}{\lambda} & , &   \frac{\sum_{a^{\prime}  \in \mathcal{K}(S_i^{j})}\theta^{\T}\varphi(S_i^{j},a) - \lambda}{\lambda|\mathcal{K}(S_i^{j}) |} > \frac{ \theta^{\T}\varphi(S_i^{j},A_i^{j})}{\lambda} ; \\
0\quad\quad\quad\quad \quad \quad \quad \quad & , & \frac{\sum_{a^{\prime}  \in \mathcal{K}(S_i^{j})}\theta^{\T}\varphi(S_i^{j},a) - \lambda}{\lambda|\mathcal{K}(S_i^{j}) |} \leq \frac{ \theta^{\T}\varphi(S_i^{j},A_i^{j})}{\lambda}.
\end{aligned}
\right.
\$
We then calculate the stochastic gradient
\$
\underbrace{\bar{\Delta}_{\theta}}_{1\times p_{\theta}} = \frac{\zeta_0}{n_0}\sum\limits^{n_0}_{i=1} \underbrace{[D_{i}(\theta) - P_{i}(\theta) + W_{i}(\theta) -Z_{i}(\omega)]^\T}_{1\times T} \underbrace{\Omega_i}_{T \times T} \underbrace{[\kappa_{e} \nabla_{\theta} D_{i}(\theta)  + \kappa_{\pi}  \nabla_{\theta} P_{i}(\theta) + \kappa_{\alpha} \nabla_{\theta} W_{i}(\theta) ]}_{T \times p_{\theta}}
\$ 
by taking all the components derived above into Equation \eqref{theta_eq}. In the next step, we derive the row vector of the gradient of $\widehat{\mathcal{L}_{U}}(\theta,\omega)$ over $\omega$, i.e., $\nabla_{\omega}\widehat{\mathcal{L}_{U}}(\theta,\omega) $
\#
\bar{\Delta}_{\omega}  = \nabla_{\omega}\widehat{\mathcal{L}_{U}}(\theta,\omega)  =  \frac{\zeta_0}{n_0}\sum^{n_0}_{i=1}  [D_{i}(\theta) - P_{i}(\theta) + W_{i}(\theta) -Z_{i}(\omega)]^\T \Omega_i \nabla_{\omega} Z_{i}(\omega).
\label{omega_eq}
\#
Therefore, we focus on deriving $\nabla_{\omega} Z^{j}_{i}(\omega)$  ($j=1,...,T$) as follows:
\$
\underbrace{\nabla_{\omega} Z^{j}_{i}(\omega)}_{1\times p_{\omega}} & = \frac{\lambda}{2}\frac{1}{ \big(1+\exp(-k_0(\omega^{\T}S^{j}_{i}-b_0))\big)^2} \nabla_{\omega}\big(1+\exp(-k_0(\omega^{\T}S^{j}_{i}-b_0))\big) \\ 
& = \frac{\lambda}{2} \frac{1}{ \big(1+\exp(-k_0(\omega^{\T}S^{j}_{i}-b_0))\big)^2} \exp(-k_0(\omega^{\T}S^{j}_{i}-b_0))(-k_0 {S^{j}_{i}}^{\T}) \\
& = \frac{-\lambda k_0{S^{j}_{i}}^{\T}}{2}\bigg[\frac{1}{1+\exp(-k_0(\omega^{\T}S^{j}_{i}-b_0))} - \frac{1}{\big(1+\exp(-k_0(\omega^{\T}S^{j}_{i}-b_0))\big)^2} \bigg] \\ 
& = \underbrace{\frac{-\lambda k_0{S^{j}_{i}}^{\T}}{2}}_{1\times p_{\omega}}\underbrace{\bigg[\frac{1}{1+\exp(-k_0(\omega^{\T}S^{j}_{i}-b_0))}\bigg]\bigg[1- \frac{1}{1+\exp(-k_0(\omega^{\T}S^{j}_{i}-b_0))}\bigg]}_{\text{scalar}}, 
\$
where $p_{\omega}$ is the dimension of $\omega$. Taking all the components into Equation \eqref{omega_eq}, we obtain the stochastic gradient $\bar{\Delta}_{\omega}$.

\section{Explicit Form of the Proximal Value Function}\label{supp_explicit_v}

The proximal value function is intrinsically defined by the proximal Bellman operator in our framework. In the following, we derive an explicit form of $V^{\pi}_{\lambda}$ under the policy $\pi$ in \eqref{explict_proximal_value}, motivated from the proximal Bellman operator in Equation \eqref{bell_eq_smooth}. Specifically, 
\begin{align}
\mathcal{B}_{\lambda}V^{\pi^{*}_{\lambda}}_{\lambda}(s) &=  \max _{\pi \in \mathcal{P}(\mathcal{A})}  \sum_{a \in \mathcal{A}} \Big[ \mathbb{E}_{{S}^{t+1}|s,a}\big[u(S^{t+1},s,a) + \gamma V^{\pi^{*}_{\lambda}}_{\lambda}({S}^{t+1})\big]\cdot \pi(a|s) + \lambda d\big(\pi(a|s)\big) \Big] \\
& =  \max _{\pi \in \mathcal{P}(\mathcal{A})} \mathbb{E}_{a \sim \pi(\cdot|s), \ {S}^{t+1} | s,a}\left[u(S^{t+1},s,a) + \gamma V^{\pi^{*}_{\lambda}}_\lambda({S}^{t+1}) + \lambda \phi(\pi(a|s)) \right] \notag \\
& =  \max _{\pi \in \mathcal{P}(\mathcal{A})} \mathbb{E}_{a \sim \pi(\cdot|s), \ {S}^{t+1} | s,a}\left[\underbrace{\left\{u(S^{t+1},s,a) + \lambda \phi(\pi(a|s))\right\}}_{u_{\text{augmented}}(S^{t+1},s,a,\pi)} + \gamma V^{\pi^{*}_{\lambda}}_\lambda({S}^{t+1}) \right] \notag \\
& = \max _{\pi \in \mathcal{P}(\mathcal{A})} \mathbb{E}_{a \sim \pi(\cdot|s), \ {S}^{t+1} | s,a}\left[u_{\text{augmented}}(S^{t+1},s,a,\pi) + \gamma V^{\pi^{*}_{\lambda}}_\lambda({S}^{t+1}) \right] 
\label{new_aug}
\end{align}
where $u_{\text{augmented}}(S^{t+1},s,a,\pi) \coloneqq u(S^{t+1},s,a) + \lambda \phi(\pi(a|s))$ is a newly defined composite utility function. By aligning the equation \eqref{new_aug} with the standard Bellman operator 
\$
\mathcal{B}V^{\pi^{*}}(s) = \max _{\pi \in \mathcal{P}(\mathcal{A})} \mathbb{E}_{a \sim \pi(\cdot|s), \ {S}^{t+1} | s,a}\left[u(S^{t+1},s,a) + \gamma V^{\pi^{*}}({S}^{t+1}) \right],
\$
and combining with the definition of the standard value function 
\$
V^{\pi}(s)=\mathbb{E}_{\pi}\left[\sum_{k=1}^{\infty}\gamma^{k-1} R^{t+k} \mid S^{t}=s\right],
\$
where $R^{t} = u(S^{t+1},S^{t},A^{t})$, we  obtain an explicit form for the proximal value function $V^{\pi}_{\lambda}$ under any policy $\pi$, i.e., 
\#
V^{\pi}_{\lambda}(s)=\mathbb{E}_{\pi}\left[\sum_{k=1}^{\infty} \gamma^{k-1} R_{\text{augmented},\pi}^{t+k} \mid S^{t}=s\right], 
\label{explict_proximal_value}
\#
where $R^{t}_{\text{augmented},\pi} = u_{\text{augmented}}(S^{t+1},S^{t},A^{t},\pi)$. Consequently, this gives the explicit representation of the proximal value function $V^{\pi}_{\lambda}$. The interpretation of the proximal value function $V^{\pi}_{\lambda}(s)$ is a discounted sum of augmented utilities given state $s$.

\section{Additional Numerical Results and Details}\label{supp_numerical}

\subsection{Actual Computation Time for Simulation Experiments}

We provide actual computation time tables of simulation experiments as follows:

\begin{table}[H]
\small
  \caption{{Example 1: The computation time in seconds for a single-experiment using one \texttt{Intel i7-10700K} CPU
     \vspace{0.2cm}} }\label{simu1_time}
		\renewcommand{\arraystretch}{1.4}
		\centering
\begin{tabular}{cc|ccccc}
\hline$n$ & $T$ & Proposed & Greedy-GQL & Linear VL & Poly VL & Gauss VL    \\
\hline 25 & 24 & 1.25 &1.17  & 11.03 &12.22 & 16.25    \\
& 36 & 1.40 & 1.48 & 12.90& 14.89  & 17.45  \\
& 48 & 2.96 &2.13& 13.67 & 14.74 &19.85\\
\hline 50 & 24 & 3.32 & 1.96 &13.88  & 14.82 & 18.02 \\
& 36 & 4.04 &2.84  & 15.68 & 17.66  & 20.96 \\
& 48 &  5.10& 4.32&  17.49 & 19.54 & 23.58  \\
\hline 100 & 24 & 4.76& 4.18  &  17.76 &19.43 &24.77    \\
& 36 & 5.67& 3.17  & 21.82 & 23.98 & 31.09\\
& 48 &5.93 &4.87 & 26.71 & 29.65 & 33.06 \\
\hline
\end{tabular}
\end{table}

\begin{table}  [H]
  	\vspace{-2mm}
\small
  \caption{Example 2: The computation time in seconds for a single-experiment using one \texttt{Intel i7-10700K} CPU
     \vspace{0.2cm}}\label{simu2_time}
		\renewcommand{\arraystretch}{1.4}
		\centering
\begin{tabular}{cc|ccccc}
\hline$n$ & $T$ & Proposed & Greedy-GQL  & Linear VL & Poly VL & Gauss VL  \\
\hline 25 & 24 & $21.69$ &$18.46$& $41.08$ & $48.23$& $64.42$  \\
& 36 & $ 25.19$ &$20.71$ &$62.51$ & $68.33$ & $81.38$ \\
\hline 75 & 24 & $36.05$ & $29.80$& $95.35$ & $97.76$ & $142.68$  \\
& 36 & $ 42.23 $ &$44.56$ & $139.74$ & $152.29$ &$208.34$ \\
\hline 100 & 24 & $39.64$ &$35.71$ & $121.14$ & $141.74$ & $197.45$  \\
& 36 & $48.29$ & $47.72$&$178.61$ & $201.68$ & $249.26$ \\
\hline
\end{tabular}
\vspace{-2mm}
 \end{table}
 
As a result, the proposed method is computationally competitive to Greedy GQ-learning \citep{ertefaie2018constructing} which is based on solving estimating equations. In comparison to V-learning \citep{luckett2020estimating}, the proposed method is computationally faster than the V-learning method; especially in the Simulation Example 2 which has a large action space.

\subsection{The Generative Model Setting in Simulation Example 2}

In this section, we give details for coefficients of the generative model in Simulation Example 2. The patient's state $(\text{BG}_{i}^{t},\text{AD}_{i}^{t},\text{BP}_{i}^{t})$ evolves according to the dynamic model:
\$
\text{BG}_{i}^{t+1} &= \gamma_{11}\mu_{\text{BG}} +  \gamma_{12}\text{BG}_{i}^{t} - \gamma_{13} \text{AD}_{i}^{t}  + \sum_{a \in \mathcal{A}}\mu_{1a} \mathbbm{1}(A_i^t  = a)+ \epsilon_{1i}^{t+1}; \\
\text{AD}_{i}^{t+1} &=  \gamma_{21}\mu_{\text{AD}} + \gamma_{22} \text{AD}_{i}^{t} + \gamma_{23} \text{BG}_{i}^{t}   + \sum_{a \in \mathcal{A}}\mu_{2a} \mathbbm{1}(A_i^t  = a) + \epsilon_{2i}^{t+1}; \\
\text{BP}_{i}^{t+1} &=  \gamma_{31}\mu_{\text{BP}} + \gamma_{32} \text{BP}_{i}^{t} + \sum_{a \in \mathcal{A}}\mu_{3a} \mathbbm{1}(A_i^t  = a)  + \epsilon_{3i}^{t+1},
\$
 where $\epsilon_{1i},\epsilon_{2i}$ and $\epsilon_{3i}$ are individual-level Gaussian random noises, and the value of coefficients $\mu_{ja}$, $j=1,2,3$, are defined in the following. 
   \begin{table}  [H]
  \caption{{Example 2: $ \text{Coefficients table for} \ \mu_{1a},\mu_{2a},\mu_{3a}$. } }\label{table1}
     \center
		\renewcommand{\arraystretch}{1.4}
\begin{tabular}{lcccccccccccc}
\hline$\mu_{ia}$ & \it{1} & \it{2} &\it{3}  & \it{4} & \it{5}  & \it{6} & \it{7} & \it{8}&\it{9}&\it{10}&\it{11}&\it{12} \\
\hline \it{1} & 0 & $-4$ &$-6$  & 6 & 4  & 10 & 8 & 12 & 14 &2 &4 & 2 \\
\hline \it{2} & 0 & 2 & $-2$  & $3$ & $1$  & $1$ & 0 & 2&$-2$&$3$ &$1$&$1$ \\
\hline \it{3} & 0 & 4 &8  &0 & 12  & 4 & 0 & 4&8&0 &12&4 \\
\hline
\end{tabular}
 \end{table}
 
The definition of the coefficients on non-treatment effects $\gamma_{ij}$ and average effects $\mu_{\text{BG}}, \mu_{\text{AD}}$ and $\mu_{\text{BP}}$ are provided as follows: 
\$
& \gamma_{11} = 0.15, \ \gamma_{21} = 0.15 \ \text{and} \ \gamma_{31} = 0.1. \\
\mu_{\text{BG}} = 100 \,& \text{mg/dL}, \ \mu_{\text{AD}} = 12 \, \text{mg/L} \ \text{and} \  \mu_{\text{BP}} = 70 \, \text{mmHg}.
\$

\subsection{The Action Space in Simulation Example 2}

In Simulation Example 2, patients are assigned by treatments from a combination of insulin injections (IN; Yes/No), physical activity (PA; No/Moderate/Strong) and dietary intake (DI; Yes/No). There are total $12$ different treatment choices, i.e., $\mathcal{A} = \{1,...,12\}$. We enumerate the treatment combinations as follows.
 \begin{table}[H]
 \center
   \caption{{Simulation Example 2: The combinations of potential actions patient received at each time point
     \vspace{0.3cm}} }\label{combine}
		\renewcommand{\arraystretch}{1.5}
 \begin{tabular}{lc}
Action ID &  Actions \\
\hline     
1 & No IN + No PA + No DI \\
2 & No IN + Moderate PA + No DI  \\
3 & No IN + Strong PA + No DI \\
4 & No IN + Moderate PA + Yes DI  \\
5 & No IN + Strong PA + Yes DI \\
6 & No IN + No PA + Yes DI \\
7 & Yes IN + No PA + No DI \\
8 & Yes IN + Moderate PA + No DI  \\
9 & Yes IN + Strong PA + No DI \\
10 & Yes IN + Moderate PA + Yes DI  \\
11 & Yes IN + Strong PA + Yes DI \\
12 & Yes IN + No PA + Yes DI \\
\hline
\end{tabular}
 \end{table}

\subsection{Sensitivity Analyses on $\lambda$ in Simulation Studies}

We conduct sensitivity analyses on examining the effect of different $\lambda$ values over model performance. The following figures illustrate the results of the sensitivity analyses in Simulation Examples 1 and 2, respectively, where the x-axis represents the different choices of $\lambda$ values. The set for $\lambda$ is $\{0.025,0.05,0.075,0.1,0.15,0.25,0.75,1, 2,5,10\}$. The y-axis represents the means of the estimated rewards over $30$ simulation experiments.

\begin{figure}[H]
\caption{Example 1: The $\lambda$ sensitivity analysis on performance over $30$ repeated simulation experiments}
\centering
\includegraphics[scale=0.52]{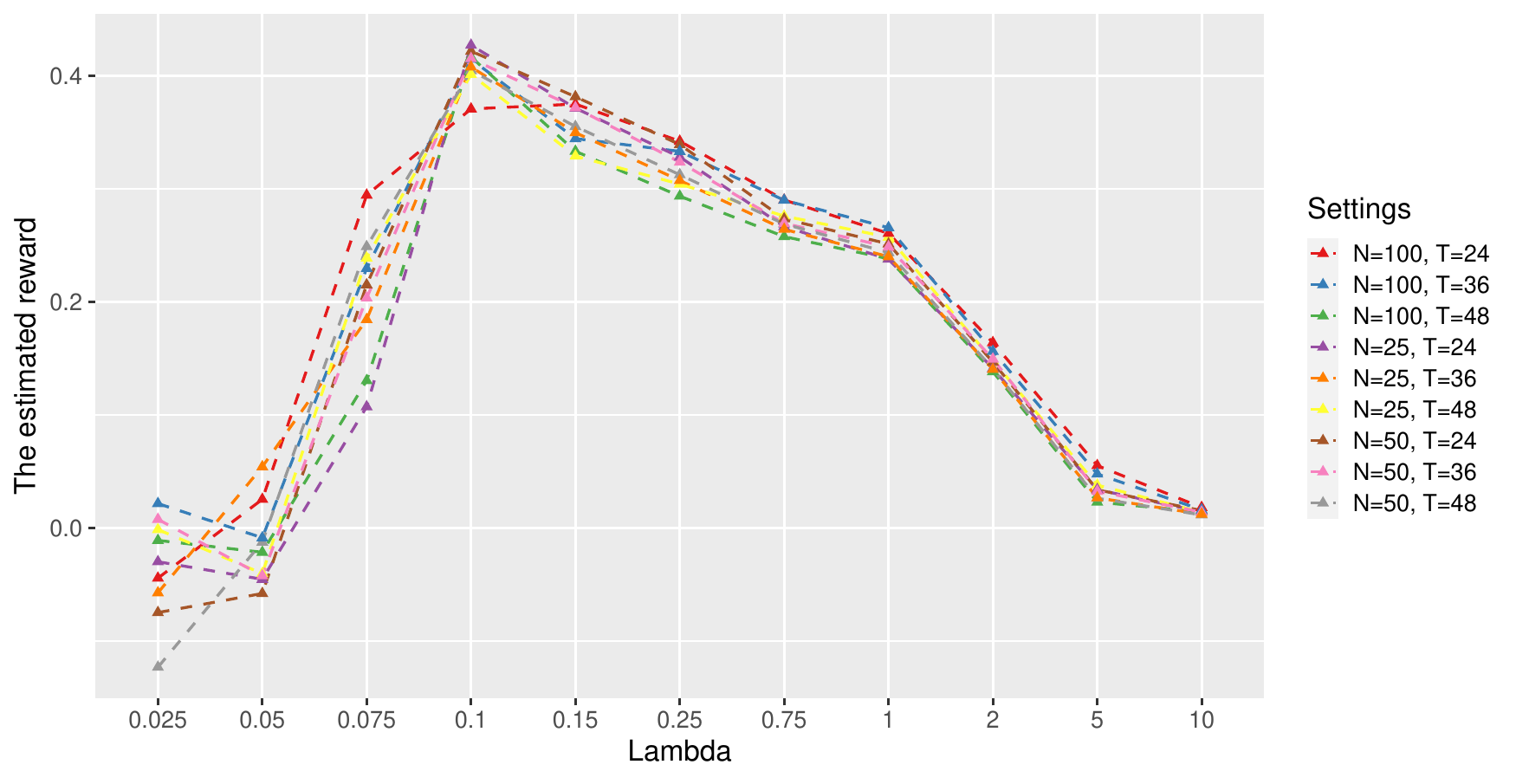}
\label{ex1_sensi}
\end{figure}



\begin{figure}[H]
\caption{Example 2: The $\lambda$ sensitivity analysis on performance over $30$ repeated simulation experiments}
\centering
\includegraphics[scale=0.52]{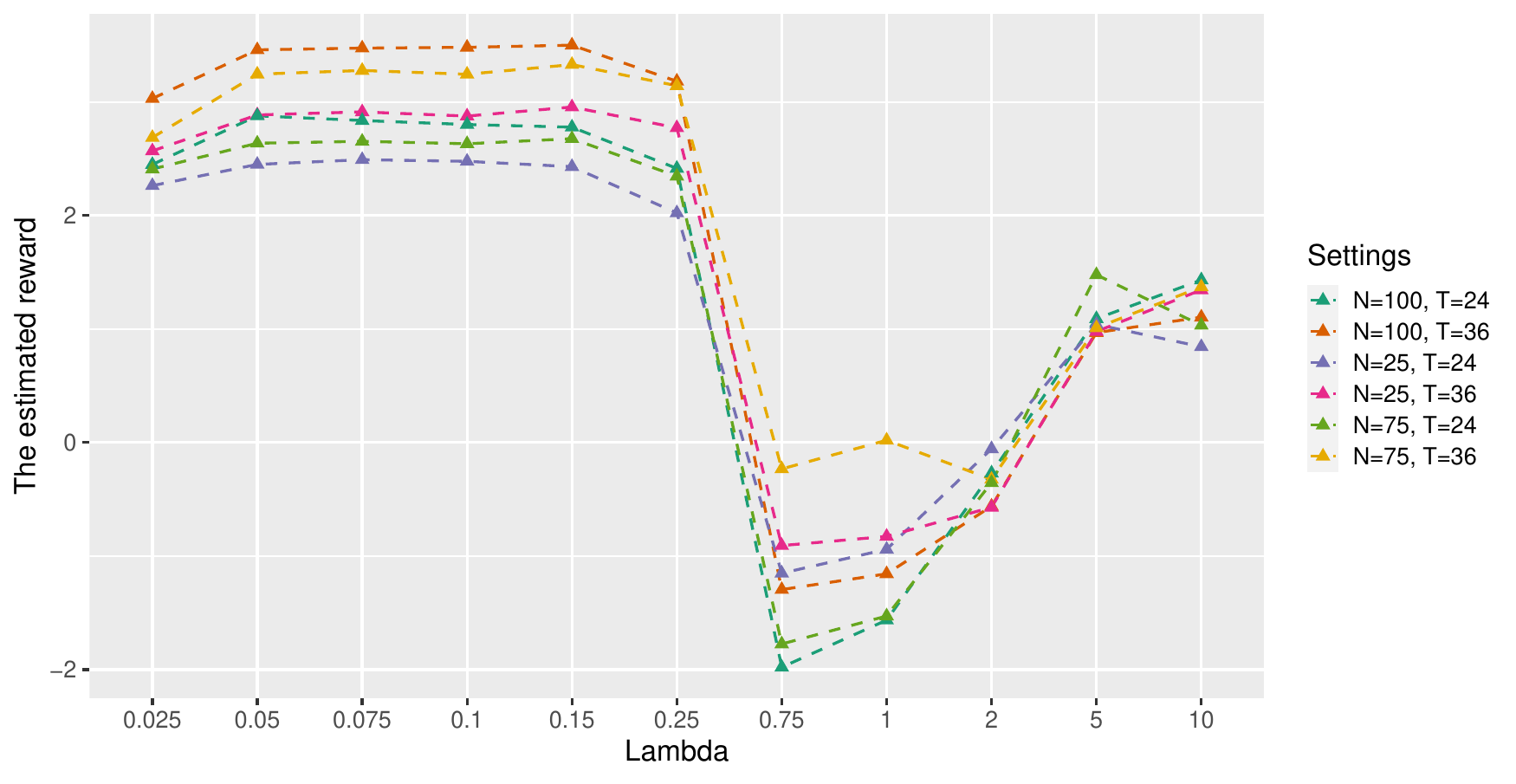}
\label{ex2_sensi}
\end{figure}

In Figure \ref{ex1_sensi}, the best performance of the lambda is within the region $[0.1,0.15]$, and the performance is negatively affected by specifying too small or too large $\lambda$. 
 In Figure \ref{ex2_sensi}, the best region of $\lambda$ falls into $[0.05,0.15]$, and the performance for $\lambda$s outside the region is not competitive. Figure \ref{ex2_sensi} also indicates that a large $\lambda$ may lead to a uniform estimated policy distribution (the behavior policy used to generate the simulation data). For example, when $\lambda=10$, all the curves become more flat and the estimated reward is close to $1.5$, which matches the observed reward. This phenomenon aligns with our Theorem 4.1.

\subsection{Cross-validation Results for Simulation Example 1}

In Section 6 of the maintext, we provide a cross-validation procedure for $\lambda$ tuning. Here, we provide the cross-validation results of Simulation Example 1 for illustrations. Specifically, we choose the cross-validation set
$
\{0.1,0.25, 0.5,0.75,1,2\}
$ 
for $\lambda$ tuning. The results are presented in Table \ref{cv_whole_table}. Based on our numerical experience, the cross-validation tuning on $\lambda$ potentially leads to a better model performance. On the other hand, the cross-validation procedure is computationally expensive.

\begin{table}[H]
\scriptsize
  \caption{{Example 1: The average and standard deviation of the mean utilities under the estimated optimal policy based on $50$ simulation runs. A higher mean utility is better.} }\label{simu1}
		\renewcommand{\arraystretch}{1.4}
		\centering
\begin{tabular}{cc|ccccc|c}
\hline$n$ & $T$ & Proposed (\textbf{CV-lambda}) & Greedy-GQL & Linear VL & Poly VL & Gauss VL & Observed \\
\hline 25 & 24 & $0.3785(0.119)$ &$0.0787(0.175)$  &$0.2533(0.021)$ &$0.2532(0.010)$ & $0.2470(0.015)$ & $0.0033$ \\
& 36 & $0.4183(0.061)$ & $0.0716(0.234)$ & $0.2553(0.013)$& $0.2558(0.016)$ & $0.2522(0.021)$ & $0.0025$ \\
& 48 & $0.3900( 0.119)$ & $0.0840(0.213)$& $0.2584(0.011)$ & $0.2584(0.014)$ &$0.2570(0.014)$ & $0.0033$\\
\hline 50 & 24 & $0.4112(0.082)$ & $0.0844(0.209)$  &$0.2497(0.012)$ & $0.2488(0.011)$ & $0.2483(0.014)$ & $0.0092$ \\
& 36 & $0.4052(0.091)$ & $0.0829(0.235)$  & $0.2533(0.017)$ & $0.2533(0.020)$ & $0.2566(0.013)$ &$0.0080$ \\
& 48 & $0.3888(0.084)$ & $0.0836(0.236)$& $0.2515(0.014)$ & $0.2518(0.013)$ & $0.2523(0.019)$ & $0.0036$\\
\hline 100 & 24 & $0.4186(0.072)$ & $0.0875(0.240)$ & $0.2441(0.014)$ & $0.2450(0.012)$ & $0.2444(0.013)$ & $0.0041$ \\
& 36 &  $0.4024(0.061)$ & $0.0780(0.266)$ & $0.2489(0.013)$ & $0.2494(0.013)$ & $0.2517(0.014)$ & $0.0043$ \\
& 48 & $0.4064(0.052)$ & $0.0945(0.254)$ & $0.2507(0.014)$ & $0.2507(0.014)$ & $0.2484(0.028)$ & $0.0090$ \\
\hline
\end{tabular}
\label{cv_whole_table}
\end{table}

\subsection{Training Details in Numerical Experiments}

In this section, we provide  details for model training in the numerical experiments. In Simulation Example 1, we set the kernel bandwidth to be $\text{bw}_0=0.5$. The parameter $\lambda$  is fixed to be $0.1$ and the discounted factor is $0.9$. The initial parameters for $\omega$ are $(0,0,0,0,0,0)$ and the initial parameters for $\theta$ are selected from $50$ initial points generated by following the uniform distribution $\text{Unif}[0,2]$ with dimension $p_{\theta} = 38$. The learning rate for parameter $\theta$ is $\alpha_\theta^0 = 0.15 $, and  the learning rate for parameter $\omega$ is $\alpha_\omega^0 = 0.01$. In Simulation Example 2, we set the kernel bandwidth to be $\text{bw}_0=0.5$. The parameter $\lambda$ is fixed to be $0.1$ and the discounted factor is $0.9$. The initial parameters for $\omega$ are $(0,0,0,0,0,0)$ and the initial parameters for $\theta$ are selected from $200$ initial points generated by following the uniform distribution $\text{Unif}[-2,2]$ with dimension $p_{\theta} = 336$. The learning rate for parameter $\theta$ is $\alpha_\theta^0 = 0.05$, and the learning rate for parameter $\omega$ is $\alpha_\omega^0 = 0.01$. In real data analysis, we set the kernel bandwidth to be $\text{bw}_0=0.5$. The parameter $\lambda$ is selected by the cross-validation procedure over tuning set $(0.075,0.1,0.15)$. The discounted factors are set to be $0.8$ and $0.9$ for different scenarios. The initial parameters for $\omega$ are $(0,0,0,0,0,0)$. The initial parameters for $\theta$ are selected from $200$ initial points generated by following uniform distributions with dimension $p_{\theta} = 56$ for the first cohort experiment and $p_{\theta} = 392$ for the second cohort experiment. The learning rate for parameter $\theta$ is $\alpha_\theta^0 = 0.05$, and the learning rate for parameter $\omega$ is $\alpha_\omega^0 = 0.01$.

\subsection{Additional Numerical Plots and Results}

In the following, we visualize and compare an estimated policy distribution of pT-Learning and V-learning in Figure \ref{sparse_simu} for Simulation Example 2. The pT-Learning estimated policy successfully allocates the optimal treatment by assigning large probabilities to the optimal treatment $7$ and the near-optimal treatment $4$. However, V-learning gives non-negligible probabilities to all non-optimal treatments, which leads to an inferior treatment recommendation. 

 \begin{figure}[H]
	\centering
	\scalebox{0.7}[0.7]{\includegraphics{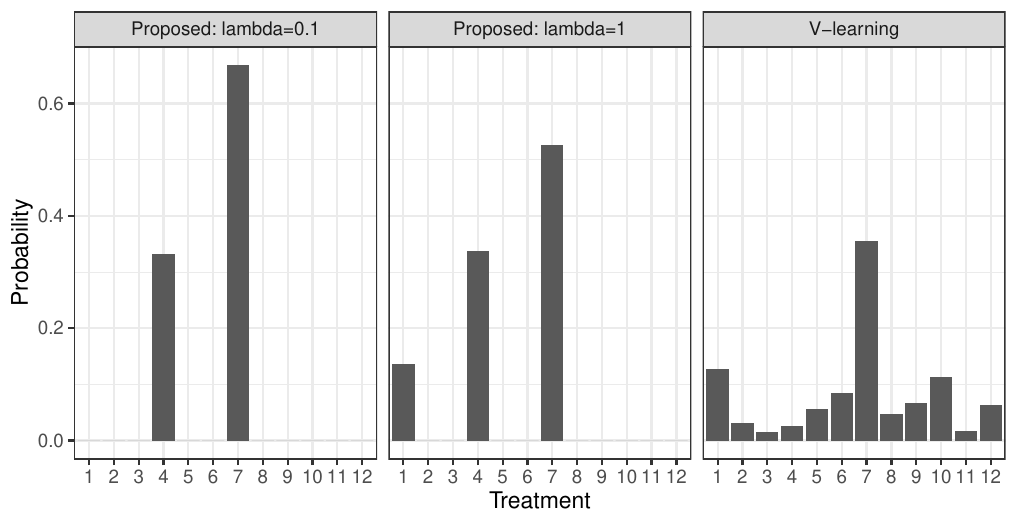}}
	\caption{The estimated optimal policy distribution of pT-Learning and V-learning for one patient at a specific time point. The $7$th treatment is the optimal treatment, and the $4$th treatment is a near-optimal treatment; other treatments are sub-optimal or non-optimal.
	}\label{ex2_data}
	\label{sparse_simu}
\end{figure}

 In Real Data Analysis , the boxplots (for two cohorts) of the improvements on the Monte Carlo discounted sum of utilities relative to the baseline for $\gamma = 0.8$ are presented in the following. 
 
\begin{figure}[H]
	\centering
		\scalebox{0.40}[0.37]{\includegraphics{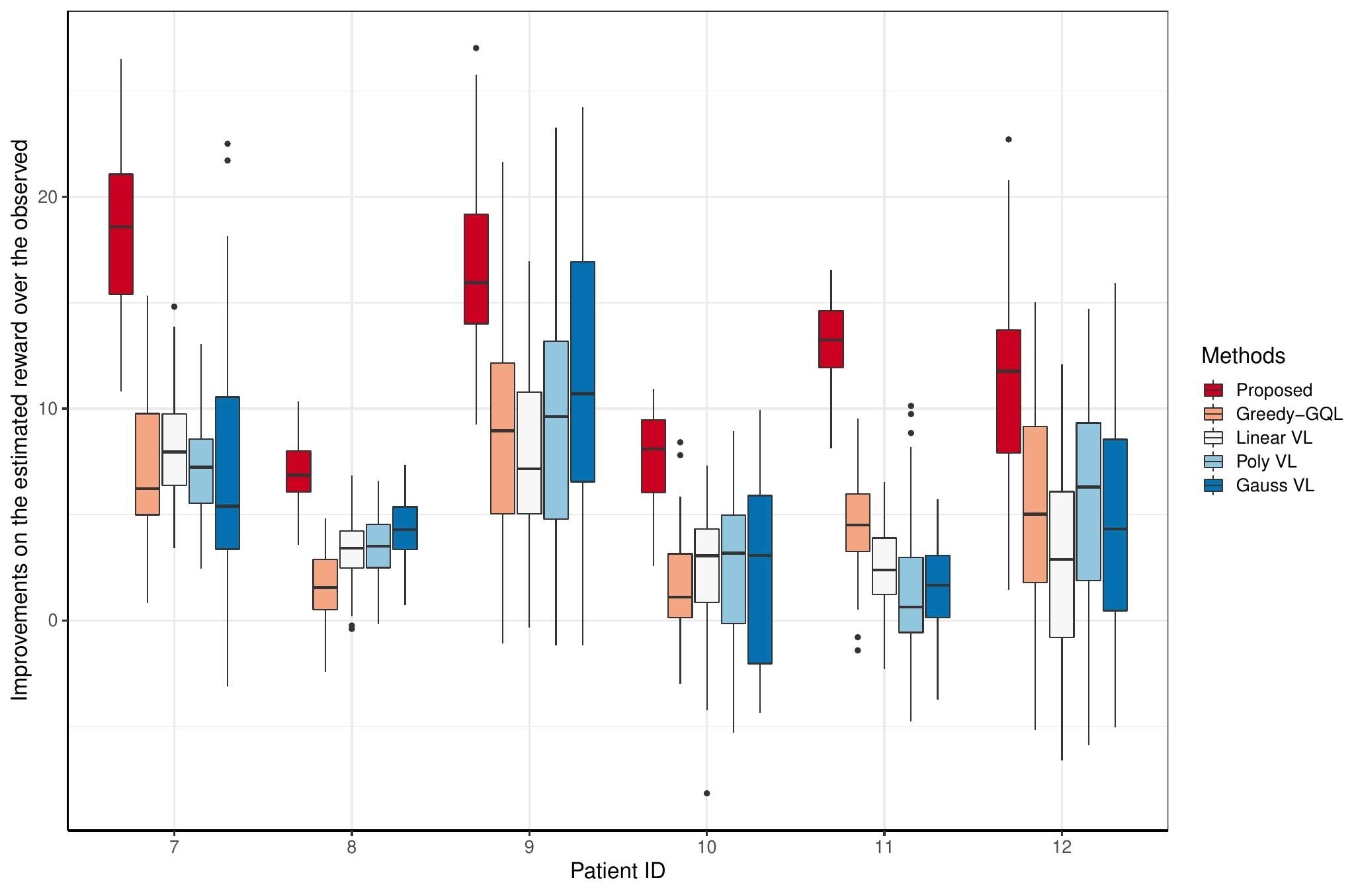}}
	\caption{The first cohort of patients who wore the \textit{Basis} sensor bands: Boxplots of the improvements on the Monto Carlo discounted sum of utilities relative to the baseline discounted sum of observed utilities. The results are summarized over $50$ simulation runs when the  treatment is binary and discounted factor is $\gamma=0.8$.}\label{binary08}
	\vspace{-5mm}
\end{figure}

\begin{figure}[H]
	\centering
		\scalebox{0.40}[0.37]{\includegraphics{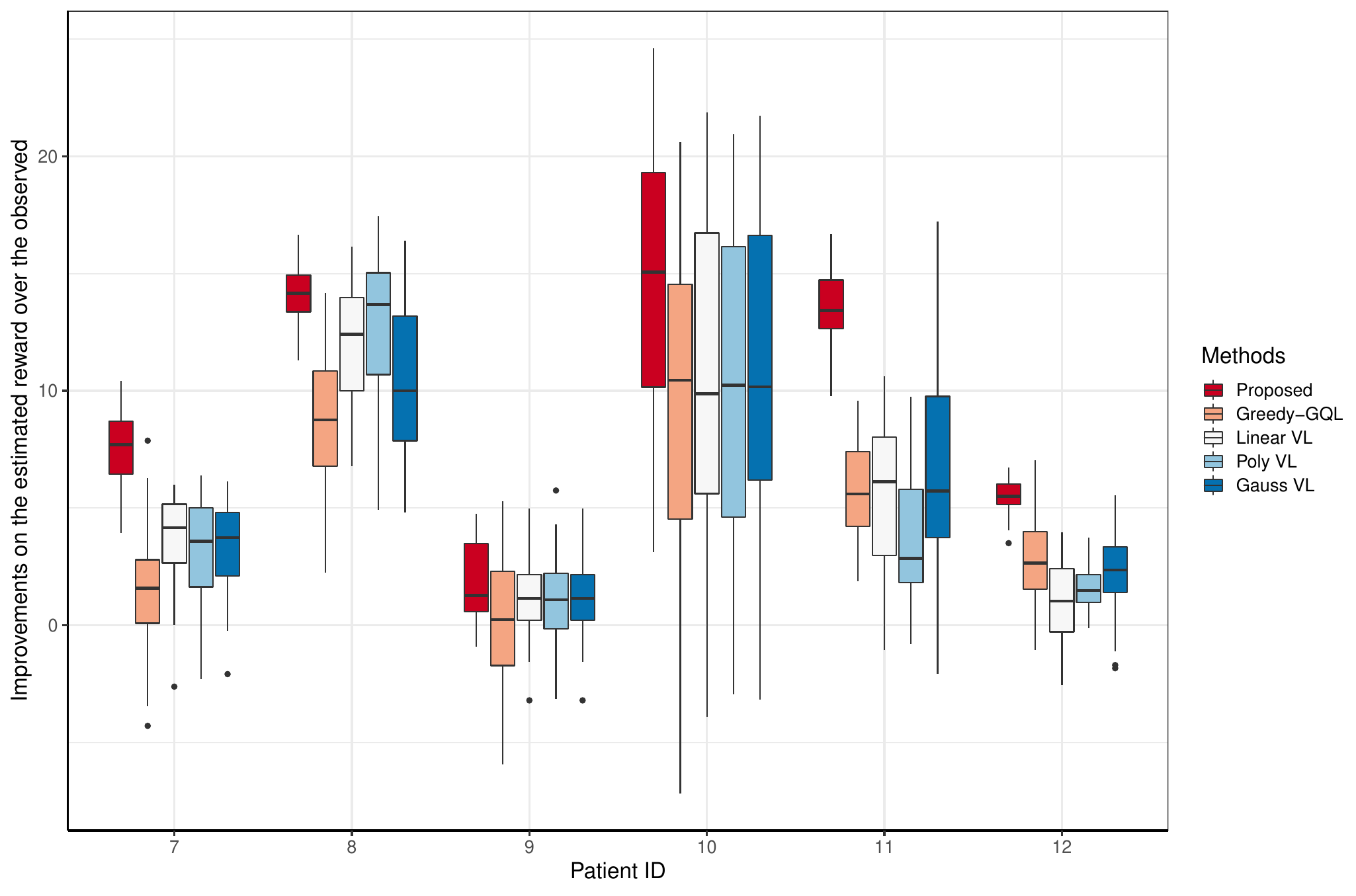}}

	\caption{The second cohort of patients who wore the \textit{Empatica} sensor bands: Boxplots of the improvements on the Monto Carlo discounted sum of utilities relative to the baseline discounted sum of observed utilities. The results are summarized over $50$ simulation runs when the cardinality of the treatment $|\mathcal{A}|=14$ and discounted factor is $\gamma=0.8$.}\label{multi08}
	\vspace{-5mm}
\end{figure}

 \begin{table}  [H]
 \small
\begin{center}
\caption{{The first cohort of patients who wore the \textit{Basis} sensor bands: The average and standard deviation of the Monto Carlo discounted sum of utilities. The results are summarized over $50$ simulation runs when the treatment is binary.} }  \label{real_cohort1}
\vspace{0.3cm}
\renewcommand{\arraystretch}{1.15} \tabcolsep 0.09in 
\begin{tabular}{c|rrrrr|r}  \hline  \rule{0pt}{0.75\normalbaselineskip}
   Patient ID & Proposed & Greedy-GQL & Linear VL & Poly VL & Gauss VL & Observed \\  \hline  \rule{0pt}{0.75\normalbaselineskip}
         &  \MC{6}{c}{{$\gamma=0.9$}}  \\
Patient $1$     & -34.44(5.11) & -49.77(8.09) &-44.93(9.10) & -47.63(8.83)& -51.25(9.68) & -65.41(4.75)   \\
Patient $2$     &-12.03(3.36) & -25.91(2.08) & -23.79(1.64) & -22.98(1.96) & -21.21(2.18) & -27.23(2.20) \\
Patient $3$    & -77.33(6.75) & -81.99(4.59) &-85.46(8.34) & -89.47(7.46) & -83.76(11.05) & -91.17(6.19)\\
Patient $4$       &  -26.29(3.16) &-33.49(4.54) & -36.40(2.44) & -32.36(6.12) & -31.71(2.60) & -36.28(3.03)\\
Patient $5$       & -32.47(1.58) & -46.54(5.26) & -46.30(2.92) & -48.02(2.28) & -44.10(4.41) & -49.74(3.20)\\
Patient $6$       &   -33.86(2.76) & -46.20(3.20) & -44.58(3.89) &-43.80(4.06) &-45.33(5.38) & -54.77(5.90) \\
\hline
         &  \MC{6}{c}{{$\gamma = 0.8$}}  \\
Patient $1$     & -15.33(3.06) & -26.67(2.06) &-25.91(2.87) &  -26.69(2.91)& -26.59(5.65) & -33.95(4.02)   \\
Patient $2$     &-7.10(0.38) & -12.47(1.13) & -10.68(0.90) & -10.50(0.97) & -9.61(0.95) & -13.99(1.44) \\
Patient $3$    & -33.39(3.43) & -41.46(3.71)&-42.04(2.19) & -40.72(4.66) & -38.56(6.06) & -50.10(4.05)\\
Patient $4$       &  -8.83(2.50) &-14.96(2.37) & -14.26(2.65) & -13.98(3.44) & -14.20(4.40) & -16.52(1.60)\\
Patient $5$       & -12.29(0.67) & -21.02(1.02) & -23.07(1.45) & -23.99(2.75) & -23.77(1.18) & -25.43(1.84)\\
Patient $6$       &  -16.69(1.28) & -23.21(1.88) & -25.81(1.73) &-22.62(1.45) & -23.93(2.54) & -28.65(4.76) \\
\hline
\end{tabular}
\end{center}
\end{table}

 \begin{table}  [H]
 \small
\begin{center}
\caption{{The second cohort of patients who wore the \textit{Empatica} sensor bands: The average and standard deviation of the Monto Carlo discounted sum of utilities. The results are summarized over $50$ simulation runs when the cardinality of the treatment $|\mathcal{A}|=14$. } }  \label{real_cohort2}
\vspace{0.3cm}
\renewcommand{\arraystretch}{1.15} \tabcolsep 0.09in 
\begin{tabular}{c|rrrrr|r}  \hline  \rule{0pt}{0.75\normalbaselineskip}
   Patient ID & Proposed & Greedy-GQL & Linear VL & Poly VL & Gauss VL & Observed \\  \hline  \rule{0pt}{0.75\normalbaselineskip}
         &  \MC{6}{c}{{$\gamma=0.9$}}  \\
Patient $7$     & -6.68(0.17) & -13.59(2.02) &-12.99(2.34) &  -13.99(2.87)& -13.75(2.27) & -17.53(2.16)   \\
Patient $8$     &-4.26(0.75) & -12.16(3.10) & -11.11(3.52) & -11.04(2.74) & -10.01(2.44) & -15.77(1.13) \\
Patient $9$    &  -3.45(0.55) & -9.29(1.85) &-7.22(1.98) & -7.32(2.60) &  -7.02(2.12) & -12.04(2.27)\\
Patient $10$       &  -4.20(0.81) &-9.56(2.62) &-10.68(3.71) & -11.20(3.73) & -10.54(3.50) & -28.98(8.45)\\
Patient $11$       & -5.51(0.11) &-13.89(1.98) & -15.84(1.59) &-16.34(2.39) & -14.90(1.76) & -30.59(2.42)\\
Patient $12$       &  -4.34(0.25) & -8.21(1.57) & -11.24(2.55) &-9.86(1.54) &-8.44(2.75) & -12.88(0.98) \\
\hline
         &  \MC{6}{c}{{$\gamma = 0.8$}}  \\
Patient $7$     &-2.67(0.14) & -8.75(2.28) &-6.54(1.18) &   -7.06(1.44)& -6.9(1.22) & -10.29(1.60)   \\
Patient $8$     &-1.62(0.31) & -7.31(2.94) & -3.72(2.37) & -3.40(3.29) &-5.33(3.33) & -15.77(1.13) \\
Patient $9$    & -4.08(0.54) & -5.63(1.82) &-4.74(0.79) & -4.95(0.99) & -4.74(0.79) & -5.91(1.63)\\
Patient $10$       & -2.81(0.57) &-8.48(2.89) & -7.48(1.46) & -8.09(1.74) &-7.23(1.21) & -17.65(6.60)\\
Patient $11$       &-3.56(0.13) & -11.38(1.85) &-11.68(2.03) & -13.33(2.18) & -10.58(3.67) & -17.19(1.49)\\
Patient $12$       &  -1.11(0.13) & -3.82(1.46) & -5.60(1.32) &-4.99(0.67) & -4.42(1.36) & -6.58(0.70) \\
\hline
\end{tabular}
\end{center}
\end{table}

\section{Comparisons Between Stochastic and Non-stochastic Policy Class}\label{supp_deterministic}

In this section, we discuss our motivation for considering stochastic policy classes over non-stochastic (deterministic) policy classes. The motivation is two-fold: suboptimality and robustness. In the following, we discuss them using several illustration examples.

\medskip

Deterministic policies could be suboptimal. A deterministic policy class is often suboptimal due to the poor exploration of the environment by taking greedy actions, thus impairing its performance \citep{john1994best}. In contrast, a stochastic policy class takes actions with certain randomness, encouraging the exploration of the dynamic environment \citep{singh1994learning, sutton1999policy, schulman2015trust, liao2020batch}. The policy with relatively high exploration power is desirable for large Markov decision process problems (consisting of large action or state space) \citep{sutton2018reinforcement}.

In the following, we illustrate the effect of exploration in Simulation Example 2 by performing additional numerical experiments. Note that the competing method, Greedy GQ-learning \citep{ertefaie2018constructing} estimates a deterministic policy by taking a greedy action derived from the Q-function. Instead, the proposed pT-Learning method is constructed under a (sparse) stochastic policy class.

\begin{figure}[H] 
\centering
\includegraphics[scale=0.4]{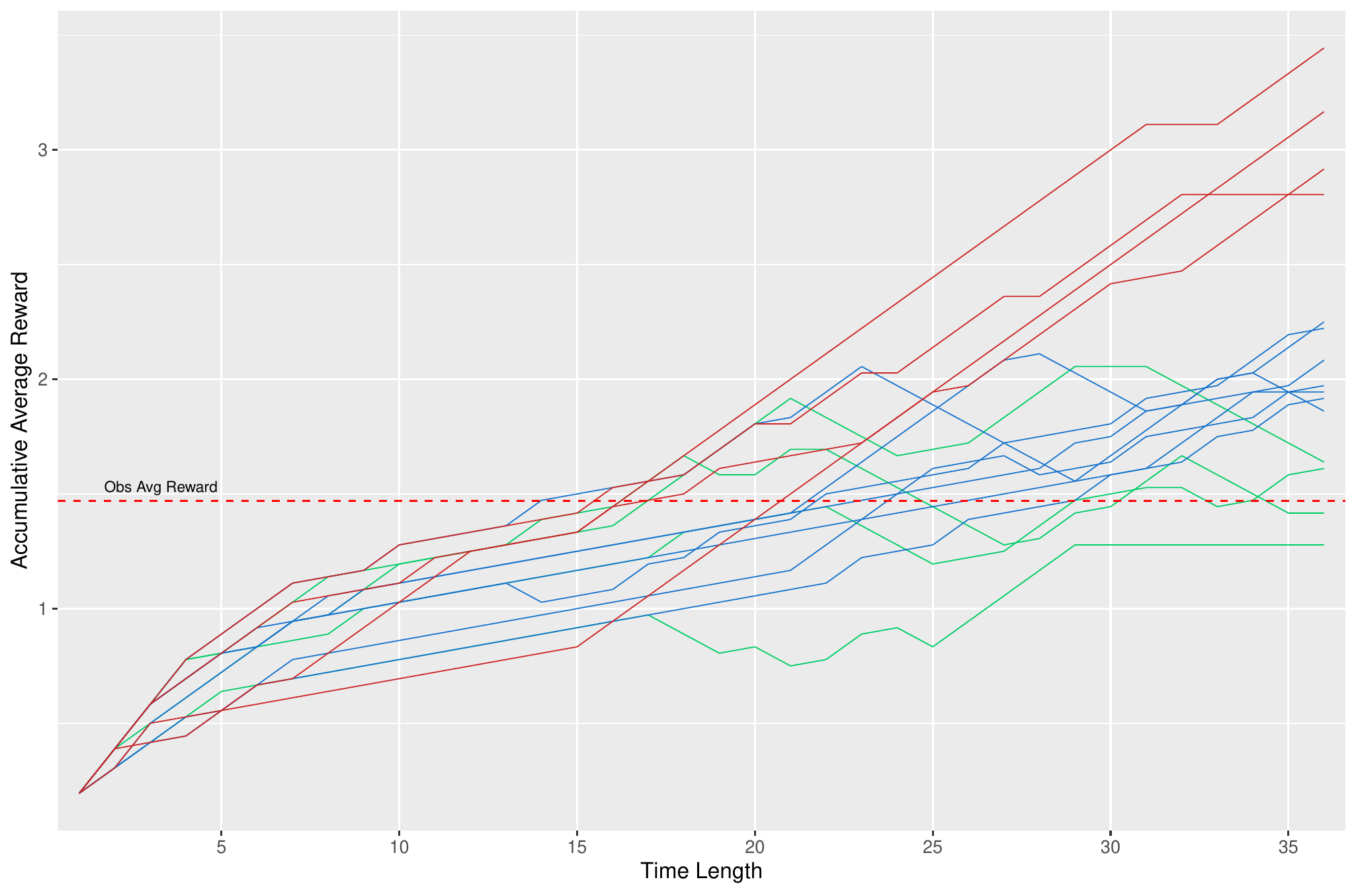}
\caption{Greedy GQ-learning: the accumulative average reward on the testing set over the time horizon. The red dashed line is the observed average reward (under a randomized policy). We may identify three groups of curves in the plot. The red curves have a strong increasing tendency, with a large average reward (compared to the observed average reward) at the end of follow-up time ($T=36$). Therefore, they can be categorized as the optimal ones. The blue curves achieve improved performance compared to the observed average reward. These blue curves are suboptimal compared with the red curves. Therefore, the blue curves are regarded as suboptimality. The green curves' performance is not significantly different from the observed average rewards, so they are categorized as non-optimalities.}
\label{gcq_avg}
\end{figure}
\begin{figure}[H]
\centering
\includegraphics[scale=0.4]{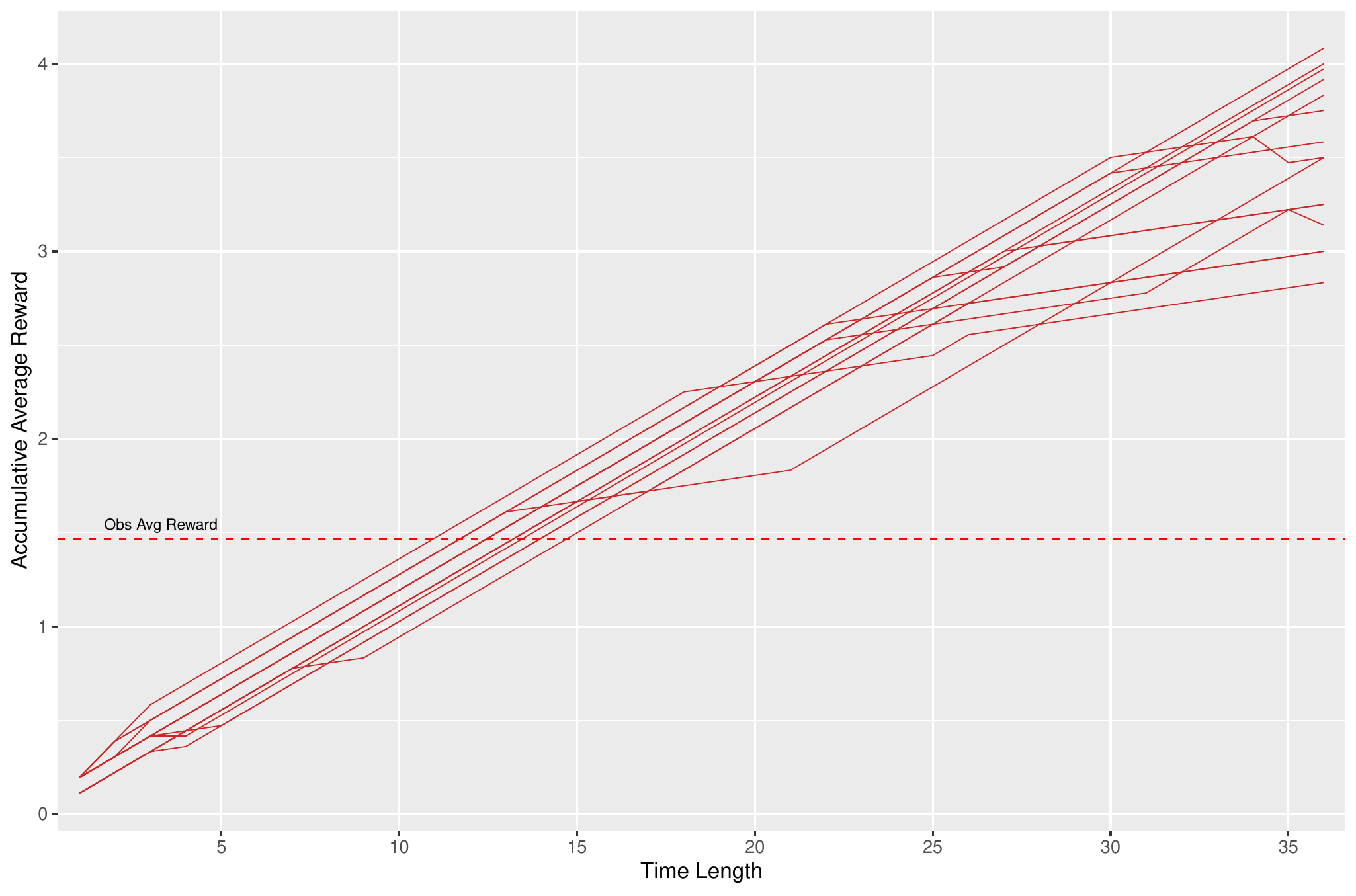}
\caption{pT-Learning: the accumulative average reward on the testing set over the time horizon. The red dashed line is the observed average reward. The red curves belong to the optimalities because all have a strong increasing tendency and a large average reward (compared to the observed average reward) at the end of follow-up time ($T=36$).}
\label{pt_avg}
\end{figure}
\begin{figure}[H]
\minipage{0.45\textwidth}
  \includegraphics[width=\linewidth]{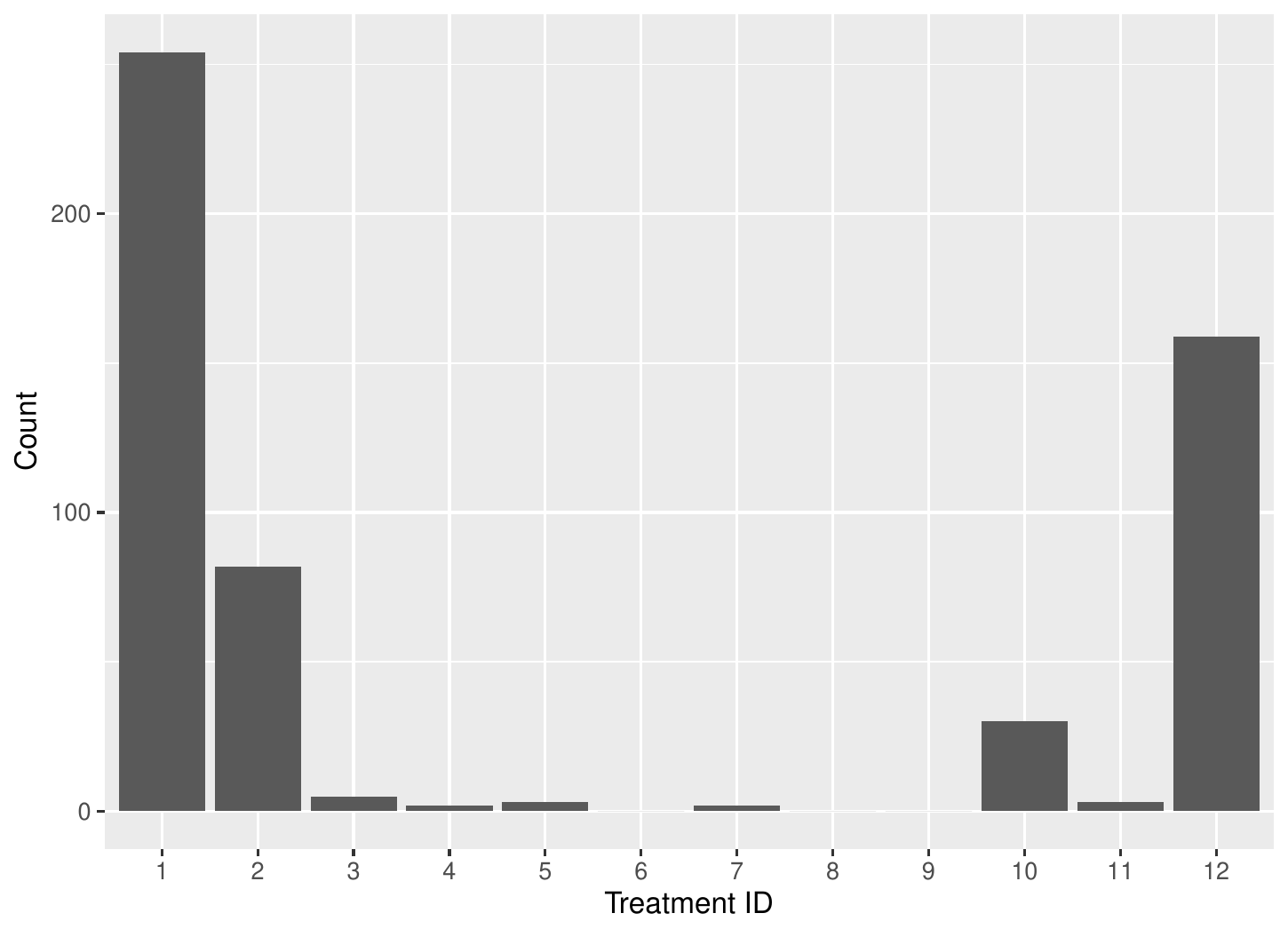}
  \caption{The histogram of the actions which Greedy GQ-learning takes in the $15$ simulation runs.}\label{count_gcq}
\endminipage\hfill
\minipage{0.45\textwidth}%
  \includegraphics[width=\linewidth]{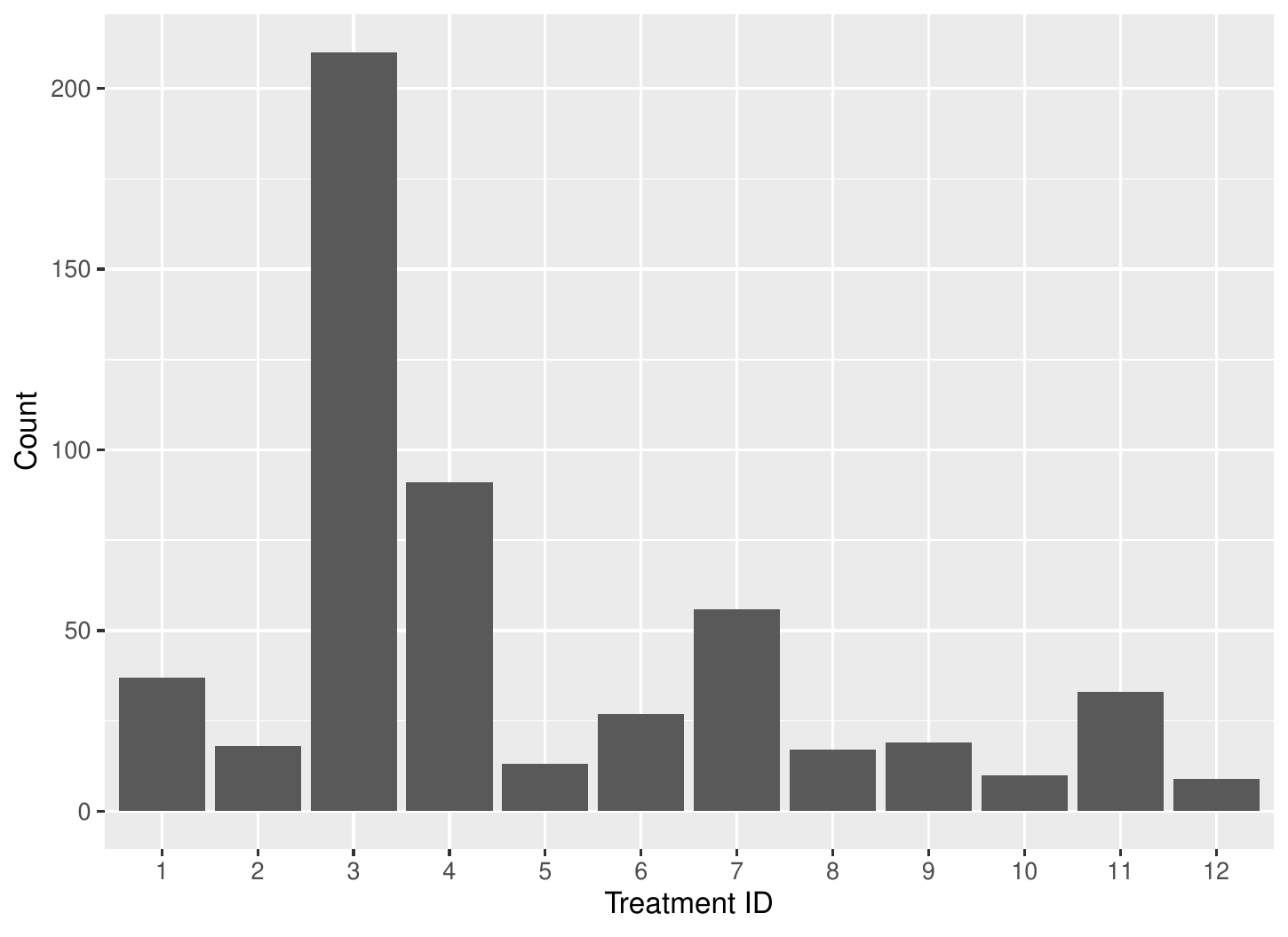}
  \caption{The histogram of the actions which pT-Learning takes in the $15$ simulation runs.}
  \label{count_pt}
\endminipage
\end{figure}
As illustrated in Figure \ref{gcq_avg}, the estimated policies by Greedy GQ-learning are suboptimal or non-optimal in $11$ of $15$ total simulation runs. One of the main reasons could be the poor exploration ability of Greedy GQ-learning. In examining the actions the algorithms explored, Figure \ref{count_gcq} shows that Greedy GQ-learning lacks full exploration of the action choices. For example, actions $6, 8$ and $9$ are not taken at all, and actions $3,4,7$ and $11$ are only taken a few times. In comparison, Figure \ref{count_pt} shows that the pT-Learning explores all action choices with certain chances. The pT-Learning follows a (sparse) stochastic policy model and thus better utilizes the environment and avoids suboptimality. Figure \ref{pt_avg} shows that the cumulative average rewards of the pT-Learning in $15$ simulation runs are all optimal.

\medskip

A deterministic policy will fail to suggest an alternative, near-optimal rule as a backup choice in an unexpected situation. In contrast, a stochastic policy is more robust by selecting actions with a probability distribution, thus providing backup options with performance guarantees. This robustness is a great advantage when facing unexpected situations \citep{ziebart2010modeling}, e.g., unavailability of optimal action at the current stage. We illustrate this point through the estimated policy distribution of pT-Learning at some states in Simulation Example 2. 

\begin{figure}[H]
\centering
\includegraphics[scale=0.45]{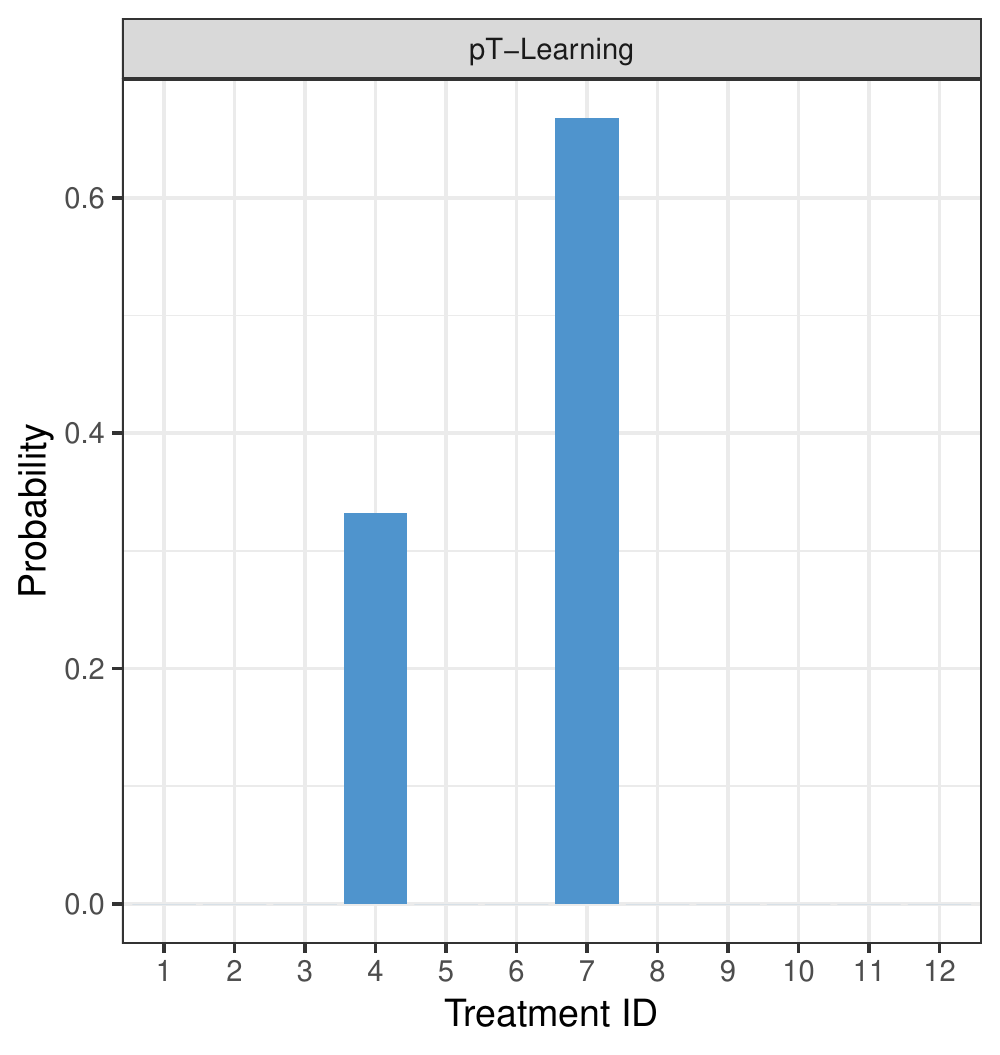}
\caption{The estimated optimal policy distribution of pT-Learning at some states. Treatment $7$ is the optimal treatment, and treatment $4$ is a near-optimal treatment; while the other treatments are sub-optimal or non-optimal.}
\label{stoch_pt}
\end{figure}
When the optimal treatment $7$ is not available due to temporal shortage or restrictions at the current stage, a sparse stochastic policy model is advantageous because it uses the next near-optimal action. This provides a quantitative index to measure the degree of confidence to implement the near-optimal by a specific probability. In Figure \ref{stoch_pt}, treatment $4$ is a potential substitute for treatment $7$. It is near-optimal, identified by the near-optimal support set of pT-Learning. Also, the probability of choosing treatment $4$ is around $0.35$, indicating to select treatment $4$ with a certain probability. In the real data analysis on the OhioT1M dataset, the robustness property of the sparse stochastic policy model is appealing. In practice, when an insulin dose is scarce, or the patient's budget constraint restricts him/her from adopting a sufficient dosage at each decision stage \citep{rehg2017mobile, marling2020ohiot1dm}, the sparse stochastic policy distribution can suggest a smaller dose level. 

\section{A Connection Between pT-learning and Off-policy TD-Algorithms}\label{supp_td}

In Section \ref{proposed_method}, we illustrated how to use an RKHS to capture the optimal critic function. In the following, we present a connection between the pT-learning estimator and off-policy TD algorithms. When $h(\cdot)$ in \eqref{minimax_loss} is restricted to a linear space, for example, $h \in \mathcal{H}_{\text{linear}} \coloneqq \{h: h(s,a) = \alpha_{\text{linear}}^\T \varphi(s,a)\}$, where $\varphi(s,a)$ is a set of basis functions and $ \alpha_{\text{linear}}$ are corresponding weights. The solution of the minimax problem \eqref{minimax_loss} is equivalent to
\$
\mathbb E_{S^{t}, A^{t}, S^{t+1}} \Big[ \big(
\widetilde{\mathcal{T}}_{\pi_\lambda}(S^{t},A^{t},V^{\pi_\lambda}_{\lambda}) - {\Psi}(S^{t})  + \psi(A^{t}|S^{t})  - V^{\pi_\lambda}_{\lambda}(S^{t}) \big) \varphi (S^{t},A^{t}) \Big]  = 0.
\numberthis \label{linear_TC_solu_maintext}
\$
The derivation is provided in the proof of Corollary S.1. In particular, equation \eqref{linear_TC_solu_maintext} can be treated as a generalization of the State-Action-Reward-State-Action (SARSA) algorithm \citep{sutton1996generalization} and on-policy TD algorithms \citep{dann2014policy} for better incorporating off-policy data via replacing the TD error with the pT-error. On the other hand,  equation \eqref{linear_TC_solu_maintext} is also related to the feature importance weighted variants of off-policy TD algorithms \citep{sutton2008convergent, ertefaie2018constructing} except for using the pT-error in \eqref{linear_TC_solu_maintext}. 

\section{pT-learning Algorithm with Experience-Replay}

The samples within an individual trajectory may have strong correlations if the corresponding time points are close. In this case, the strong correlation makes empirical risk minimization more difficult. 
To break the data temporal dependency within a trajectory, we could also consider a modified algorithm that leverages the idea of the experience replay \citep{lin1992self,mnih2015human}. Specifically, we first shuffle the observed data $\{\mathcal{D}_{i}\}^{n}_{i=1}$ into a pool $\mathcal{M}$ such that the pool contains the total $nT$ number of mixed samples. At each iteration, we sample a mini-batch data $\mathcal{M}_0$ from $\mathcal{M}$ to minimize the loss $\widehat{\mathcal{L}_{U}}(\theta,\omega)$ following the same routine in Lines 5-8 in Algorithm \ref{SGD Algorithm}. The probability of selecting an arbitrary mini-batch $\mathcal{M}_0$ of size $n_0$ is $\mathbb{P}(\mathcal{M}_0)=1/\binom {nT}{n_0}$, and an arbitrary transition pair $(S_i^{t},A_i^{t},R_i^{t},S_i^{t+1}) \in \{\mathcal{D}_{i}\}^{n}_{i=1}$ is selected with the probability $n_0/(nT)$. Let us define an event $\textbf{I} = \{\text{samples are completely independent in minibatch}\}$, then the probability of $\textbf{I} $ occurs is 
\$
\mathbb{P}(\textbf{I} ) = 1-\frac{n_0(n_0-1)(T-1)}{{2(nT-1)}} = 1- \mathcal{O}\left(\frac{1}{n}\right).
\$
Hence, the computed gradients are $(n-1)/n$-nearly independent at each iteration, breaking the temporal dependency when the number of trajectories is sufficient. As a byproduct, sampling a mini-batch from the shuffled pool $\mathcal{M}$ may induce non-stationarity in the mini-batch samples, so that some statistical properties or asymptotic results can not very difficult to be studied. For computation costs, this modified algorithm requires $\mathcal{O}(n_0^2T^2q)$ time complexity, which is more computational intensive compared to $\mathcal{O}(n_0T^2q)$ Algorithm \ref{SGD Algorithm}. We present this modified algorithm with experience replay in the following, which is probably an alternative remedy for learning on data with high sample dependency. 

\begin{algorithm}[H]
\setstretch{1.1}
	\caption{pT-learning algorithm with experience-replay}
	\label{SGD Algorithm with ER}
	\begin{algorithmic}[1]
			\STATE \textbf{Input} observed data $\mathcal{D}_{1:n}$ as the transition pairs format $\{(S_i^{t},A_i^{t},R_i^{t},S_i^{t+1}): t=1,...,T\}^n_{i=1}$.
			\STATE \textbf{Initialize} the primary and auxiliary parameters $(\theta,\omega) = (\theta_0,\omega_0)$, the mini-batch size $n_0$, the learning rates $\alpha_{\theta} = \alpha_{\theta}^0 , \alpha_{\omega} = \alpha_{\omega}^0$, the scale parameter $\zeta =\zeta_0$, the adjustment factors $(\kappa_{e}=1,\kappa_{\pi}=1,\kappa_{\alpha}=1 )$, the sparsity parameter $\lambda = \lambda_0$, the kernel bandwidth $\textit{bw}$ = $\textit{bw}_0$, and the stopping criterion $\varepsilon$. 
			\STATE \textbf{Shuffle} the input $\mathcal{D}_{1:n}$ into a pool $\mathcal{M}$ which contains $nT$ number of mixing transition pairs. 	
	\STATE \textbf{For} $k=1$ to $k = \textit{max.iter}$
						\STATE \; \ Randomly sample a mini-batch $\{(S_i^{t},A_i^{t},R_i^{t},S_i^{t+1}): t=1,...,T\}^{n_0}_{i=1}$ from the pool $\mathcal{M}$.
						\STATE \; \ Compute the stochastic gradient with respect to $\theta$ as
	\begin{center}
										\; \;		 $
					\bar{\Delta}_{\theta} = \frac{\zeta_0}{n_0}\sum\limits^{n_0}_{i=1} [D_{i}(\theta) - P_{i}(\theta) + W_{i}(\theta) -Z_{i}(\omega)]^\T \Omega_i [\kappa_{e} \nabla_{\theta} D_{i}(\theta)  + \kappa_{\pi}  \nabla_{\theta} P_{i}(\theta) + \kappa_{\alpha} \nabla_{\theta} W_{i}(\theta) ].
					$
						\end{center}
					\STATE \;  \ Compute the stochastic gradient w.r.t. $\omega$ as 
										\begin{center}
										\; \;			$			\bar{\Delta}_{\omega} = \frac{\zeta_0}{n_0}\sum^{n_0}_{i=1}  [D_{i}(\theta) - P_{i}(\theta) + W_{i}(\theta) -Z_{i}(\omega)]^\T \Omega_i \nabla_{\omega} Z_{i}(\omega)$.
						\end{center}
				\STATE \;  \ Decay the learning rate $\alpha_{\theta}^{k} = \mathcal{O}(k^{-1/2})$, $\alpha_{\omega}^{k} = \mathcal{O}(k^{-1})$.
				\STATE \;  \ Update the parameter of interests as
										\begin{center}
										\; \		 $\theta^{k} \leftarrow \theta^{k-1} - \alpha_{\theta}^{k}\bar{\Delta}_{\theta}, \quad \omega^{k} \leftarrow \omega^{k-1} - \alpha_{\omega}^{k}\bar{\Delta}_{\omega}.
					$
							\end{center}
									\STATE \;  \ Stop if $\|\theta^{k} - \theta^{k-1} \| \leq \varepsilon$.
			  \STATE \textbf{Return}	 $\widehat \theta = \theta^{k} $.
				\end{algorithmic}
\end{algorithm}

\section{Non-stationary Learning and Explore-exploit Tradeoff}

V-learning considers a fully stochastic policy class that assigns non-zero probability to all actions. Thus it achieves a high exploration rate. In contrast, pT-Learning restricts the action set to a small near-optimal set, meaning it possesses a higher exploitation rate but a lower exploration rate. \\
Under stationary settings, pT-Learning may maintain more balanced exploitation and exploration trade-off by controlling $\lambda$ well. Therefore, pT-Learning is likely to yield better model performance than V-learning. However, under non-stationary learning settings where the environment varies over time, a relatively high exploration rate \citep{puterman2014markov} is typically preferred and results in a better policy estimation in the long run. This is particularly applicable for online learning problems, where the estimated optimal policy can be updated in a streaming manner over stages. Consequently, V-learning potentially gains more benefits than pT-Learning due to its high exploration rate. To increase the exploration ability of pT-Learning, one can consider the $\varepsilon$-greedy strategy \citep{sutton1998introduction} to assign the action following the estimated optimal policy with probability $1-\varepsilon$ and with $\varepsilon$ probability for random pick-up. 


\section{State-varying Sparsity and Smoothness Parameter}

In the current framework, the sparsity and smoothness parameter $\lambda$ is homogeneous over states. One possible extension is to allow the parameter $\lambda$ to be varying over states, i.e., 
	\$
	\lambda(s) \coloneqq g(s) \ \text{for} \, s \in \mathcal{S},
	\$ 
	where $g(\cdot)$ is an optimization-friendly smoothed function. Let the parameter $\lambda$ be state-varying, several advantages can be achieved. \\
First, according to the construction of the support set for a fixed $s$,
\$
\mathcal{K}(s) &=\left\{a_{(i)} \in \mathcal{A}: Q^{\pi^{*}_{\lambda}}_{\lambda}(s, a_{(i)}) > \frac{1}{i} \sum_{j=1}^{i} Q^{\pi^{*}_{\lambda}}_{\lambda}(s, a_{(j)}) - \frac{\lambda}{i}\right\} \\
\implies \quad  \mathcal{K}(s) &= \left\{a_{(i)} \in \mathcal{A}: \sum_{j=1}^{i} Q^{\pi^{*}_{\lambda}}_{\lambda}(s, a_{(j)}) - iQ^{\pi^{*}_{\lambda}}_{\lambda}(s, a_{(i)}) < \lambda  \right\},
\$
the parameter $\lambda$ essentially controls the margin between the largest action value and the others in the support set. When the scale of the rewards or the action values on different $s$ vary too much, state-varying $\lambda(s)$ can provide a more flexible and heterogeneous solution to construct the support set. This is especially potentially useful when some groups favor a policy with strong sparsity while others prefer relatively weak ones. \\
Secondly, state-varying $\lambda(s)$ can introduce a flexible smoothness effect. Note that the degree of the non-smoothness for the standard Bellman operator $\mathcal{B}$ may vary for different state regions. In some regions, the number of non-differentiable points introduced by the $\max$ operator is small. In these regions, a small $\lambda$, i.e., a smoothing magnitude,  may be preferred to avoid unnecessary smoothing bias. Otherwise, a large degree of smoothness is favorable. From this point of view, state-varying $\lambda(s)$ can make adjustments over different $s$, and allow the proximal Bellman operator to provide a more adaptive smoothing approximation for the standard Bellman operator. 
This could potentially avoid largely unnecessary smoothing bias.  \\
Thirdly, from the exploration-exploitation balance point of view, a large value of $\lambda$ encourages exploration, and a small value of $\lambda$ leads to high exploitation. For some particular states, a large exploration rate may improve the algorithms. On the other hand, taking more exploitation allows the algorithm to converge faster and reduce computational burden \citep{sutton2018reinforcement}. Therefore, a state-varying $\lambda(s)$ is able to balance the explore-exploit trade-off over different groups on the state space. \\
In conclusion, if we can find a suitable function form for the state-varying $\lambda$, and handle the additional computational costs well, i.e., learning the $\lambda(s)$ if it is unknown, the state-varying $\lambda(s)$ may bring unique the advantages. We conjecture that our theoretical results still hold if $\lambda(s)$ is positive and differentiable over $s$.

\end{document}